\documentclass[12pt]{iopart}
\usepackage{graphicx}

\begin{document}

\title{The Progenitors of Short Gamma-Ray Bursts}

\author{William H. Lee}

\address{Instituto de Astronom\'{\i}a, Universidad Nacional
Aut\'{o}noma de M\'{e}xico, Apdo. Postal 70-264, Cd. Universitaria,
M\'{e}xico D.F. 04510:wlee@astroscu.unam.mx}

\author{Enrico Ramirez-Ruiz}

\address{Institute for Advanced Study, School of Natural Sciences,
  Princeton, NJ 08540, USA: enrico@ias.edu}

\begin{abstract}
Recent months have witnessed dramatic progress in our understanding of
short $\gamma$-ray burst (SGRB) sources. There is now general
agreement that SGRBs -- or at least a substantial subset of them --
are capable of producing directed outflows of relativistic matter with
a kinetic luminosity exceeding by many millions that of active
galactic nuclei. Given the twin requirements of energy and
compactness, it is widely believed that SGRB activity is ultimately
ascribable to a modest fraction of a solar mass of gas accreting onto
a stellar mass black hole or to a precursor stage whose inevitable end
point is a stellar mass black hole. Astrophysical scenarios involving
the violent birth of a rapidly rotating neutron star, or an accreting
black hole in a merging compact binary driven by gravitational wave
emission are reviewed, along with other possible alternatives
(collisions or collapse of compact objects). If a black hole lies at
the center of this activity, then the fundamental pathways through
which mass, angular momentum and energy can flow around and away from
it play a key role in understanding how these prime movers can form
collimated, relativistic outflows. Flow patterns near black holes
accreting matter in the {\it hypercritical} regime, where photons are
unable to provide cooling, but neutrinos do so efficiently, are
discussed in detail, and we believe that they offer the best hope of
understanding the {\it central engine}. On the other hand, statistical
investigations of SGRB niches also furnish valuable information on
their nature and evolutionary behavior. The formation of particular
kinds of progenitor sources appears to be correlated with
environmental effects and cosmic epoch. In addition, there is now
compelling evidence for the continuous fueling of SGRB sources. We
suggest here that the observed late flaring activity could be due to a
secondary accretion episode induced by the delayed fall back of
material dynamically stripped from a compact object during a merger or
collision.  Some important unresolved questions are identified, along
with the types of observation that would discriminate among the
various models. Many of the observed properties can be understood as
resulting from outflows driven by hyperaccreting black holes and
subsequently collimated into a pair of anti-parallel jets. It is
likely that most of the radiation we receive is reprocessed by matter
quite distant to the black hole; SGRB jets, if powered by the hole
itself, may therefore be one of the few observable consequences of how
flows near nuclear density behave under the influence of strong
gravitational fields.
\end{abstract}

\maketitle
%section 1 
\section{Introduction}\label{introduction}

\subsection{Prologue}
In the sections which follow, we shall be concerned predominantly with
the theory of short $\gamma$-ray bursts\footnote{The literature on
this subject has become quite large, and to keep the references to a
manageable size, we have in general referred to the most recent
comprehensive article in a given topic. In particular, we give only
representative references for those topics on which helpful reviews
articles already exist. It is thus possible for the reader to trace
the early work on which some of these conclusions are based. We
apologize to those colleagues whose work is either omitted or not
fully represented.}. If the concepts there proposed are indeed
relevant to an understanding of the nature of these sources, then
their existence becomes inextricably linked to the {\it metabolic
pathways} through which gravity, spin, and energy can combine to form
collimated, ultra relativistic outflows. These threads are few and
fragile, as we are still wrestling with understanding non-relativistic
processes, most notably those associated with the electromagnetic
field and gas dynamics. If we are to improve our picture-making we
must make more and stronger ties to physical theory.  But in
reconstructing the creature, we must be guided by our eyes and their
extensions. In this introductory chapter we have therefore attempted
to briefly summarize the observed properties of these ultra-energetic
phenomena\footnote{The reader is referred to \cite{nakarrev} for an an
excellent review of the observations.}. There are five sections: \S
\ref{sec:his} gives a brief account of their history from birth to
present-age; \S \ref{sec:metabolics} is devoted to their metabolism --
in other words, to their gross energetics, spectra and time
variability; \S \ref{sec:aft} describes the attributes of the
afterglow signals, which, as fading beacons, mark the location of the
fiery and brief $\gamma$-ray event.  These afterglows in turn enable
the measurement of redshift distances, the identification of host
galaxies at cosmological distances, and provide evidence that many
short $\gamma$-ray bursts are associated with old stellar populations
and possibly with no bright supernova. These threads will be woven in
\S \ref{sec:field}. Finally, \S \ref{sec:stage} gives a compendium of
the observational {\it facts}.

\subsection{Burst of Progress}\label{sec:his}

The manifestations of SGRB activity are extremely diverse. SGRBs are
observed throughout the whole electromagnetic spectrum, from GHz radio
waves to 10 MeV $\gamma$-rays, but until recently, they were known
predominantly as bursts of $\gamma$-rays, largely devoid of any
observable traces at any other wavelengths.

\begin{figure}
\centering \includegraphics[width=4.0in]{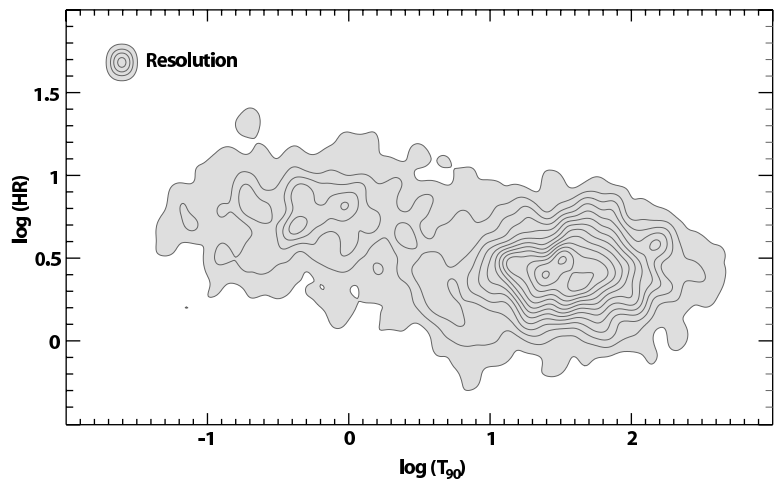}
\caption{Hardness ratio (i.e., ratio of total counts in the
  100-300~keV range, to those in the 50-100 keV range) versus duration
  for BATSE bursts. The hardness ratio is a measure of the shape of
  the spectrum: larger values correspond to harder spectra. Different
  line levels denote number density contours. The bimodality of the
  distribution of their duration is confirmed by the associated
  spectral shape. }
\label{fig:hard}
\end{figure}

Before 2005, most of what we knew about SGRBs was based on
observations from the Burst and Transient Source Experiment (BATSE) on
board the {\it Compton Gamma Ray Observatory}, whose results have been
summarized by Fishman \& Meegan \cite{fishman95}. BATSE, which
measured about 3000 events, detected approximately one burst on a
typical day. While they are on, they outshine every other source in
the $\gamma$-ray sky, including the sun.  Although each is unique, the
bursts fall into one of two rough categories. Bursts that last less
than two seconds are classified as short, and those that last longer
-- the majority -- as long \cite{kouveliotou93}. The two categories
differ spectroscopically, with short bursts having relatively more
high-energy $\gamma$-rays than long bursts do. Figure~\ref{fig:hard}
shows the hardness ratio as a function of the duration of the
emission. It is a measure of the slope of the spectrum, where larger
values mean that the flux at high energies dominates.

Arguably the most important result from BATSE concerned the spatial
distribution of bursts. Both long and short events occur isotropically
-- that is, evenly over the entire sky with no dipole and quadrupole
components, suggesting a cosmological distribution. This finding cast
doubt on the prevailing wisdom, which held that bursts came from
sources within the Milky Way. The uniform distribution instead led
most astronomers to conclude that the instruments were picking up some
kind of cosmological event. Unfortunately, $\gamma$-rays alone did not
provide enough information to settle the question for sure. The
detection of radiation from bursts at other wavelengths would turn out
to be essential. Visible light, for example, could reveal the galaxies
in which the bursts took place, allowing their distances to be
measured. Attempts were made to detect these burst counterparts, but
they proved fruitless.

Observations of burst counterparts \cite{vanparadijs00} were
restricted to the class of long duration bursts \footnote{because {\it
BeppoSAX} is mainly sensitive to bursts longer than about 5 to 10 s.}
until, in 2005, the {\it Swift} spacecraft succeeded in obtaining
high-resolution X-ray images \cite{gehrels05,bloom06} of the fading
afterglow of GRB 050509B -- so named because it occurred on May 9,
2005. This detection, followed by a number of others at an approximate
rate of 10 per year, led to accurate positions, which allowed the
detection and follow-up of the afterglows at optical and longer
wavelengths \cite{hjorth05,covino06,berger05}. This paved the way for
the measurement of redshift distances, the identification of candidate
host galaxies, and the confirmation that they were at cosmological
distances
\cite{berger05,fox05,barthelmy05,gehrels05,bloom06,prochaska06,villasenor05}. {\it
Swift} is equipped with $\gamma$-ray, X-ray and optical detectors for
on-board follow-up, and capable of relaying to the ground arc-second
quality burst coordinates within less than a minute from the burst
trigger, allowing even mid-size ground-based telescopes to obtain
prompt spectra and redshifts.

\subsection{Metabolics}\label{sec:metabolics}

SGRBs are brief flashes of radiation at soft and hard $\gamma$-ray
energies that display a wide variety of time histories.  They were
first detected at soft $\gamma$-ray energies with wide field-of-view
instruments, with peak soft $\gamma$-ray fluxes reaching hundreds of
photons cm$^{-2}$ s$^{-1}$ in rare cases. The BATSE instrument was
sensitive in the 50-300 keV band, and provided the most extensive data
base of SGRB observations during the prompt phase.

\begin{figure}
\centering \includegraphics[width=3.5in]{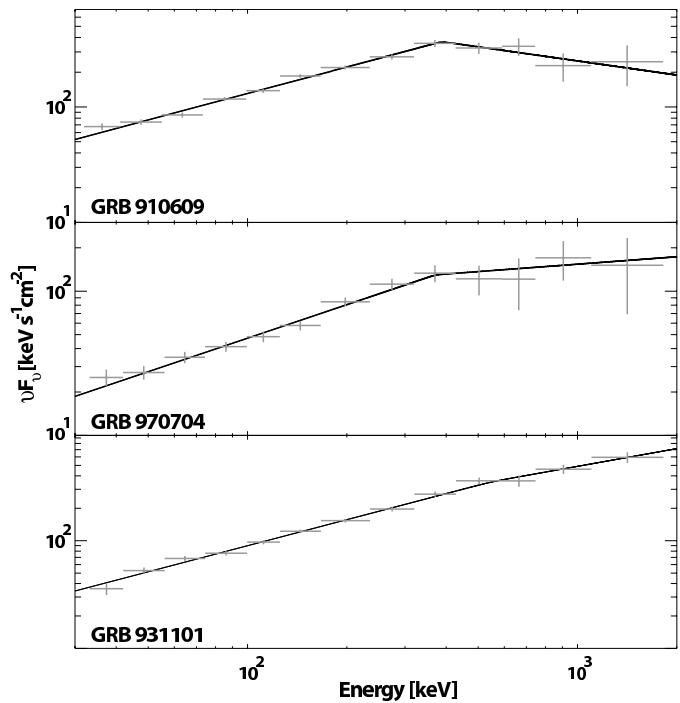}
\caption{Representative spectra $\nu F_\nu \propto \nu^2 N(\nu)$ of
    various SGRBs. The SGRB spectrum is non thermal, the number of
    photons varying typically as $N(\nu) \propto \nu^{-\alpha}$, where
    $\alpha \sim 1$ at low energies changes to $\alpha \sim 2$ to 3
    above a photon energy $\sim$ 1 MeV.}
\label{fig:spec}
\end{figure}
SGRBs typically show a very hard spectrum in the soft to hard
$\gamma$-ray regime. The photon index breaks from $\approx -1$ at
energies $E_{\rm ph}\leq 100$ keV, to a $-2$ to $-3$ spectrum at
$E_{\rm ph} \geq$ several hundred keV \cite{band93}.  Consequently,
the peak photon energies, $E_{\rm pk}$, of the
time-averaged $\nu F_\nu$ spectra of BATSE SGRBs are typically found
in the 500 keV - several MeV range
\cite{mallozzi95,ghirlanda04,kaneko06}. The general trend is that the
spectrum softens, and $E_{\rm pk}$ decreases, with time. More precise
statements must, however, wait for larger area detectors.

In Figure~\ref{fig:spec} representative spectra are plotted, in the
conventional coordinates $\nu$ and $\nu F_\nu$, the energy radiated
per logarithmic frequency interval. Some obvious points should be
emphasized. We measure directly only the specific luminosity $D^2 \nu
I_\nu \equiv (1/4\pi)\nu L_\nu$ (the energy radiated in the direction
of the earth per second per steradian per logarithmic frequency
interval by a source at luminosity distance $D$), and its
dimensionless distance-independent ratios between two frequencies
(BATSE triggers, for example, are based on the count rate between 50
keV and 300keV). The apparent bolometric luminosity $4 \pi
\int^{\infty}_{-\infty} D^2 \nu I_\nu d({\rm In}\nu)$ may be quite
different from the {\it true} bolometric luminosity $\int_{4\pi}
\int^{\infty}_{0}D^2 I_\nu d\nu d\Omega$ if the source is not
isotropic. The GRB spectra shown in Figure~\ref{fig:spec} are those of
GRB 910609, GRB 970704, and GRB 931101 which were observed by
BATSE. The time integrated spectrum on those detectors ranges from 25
keV to 10MeV \cite{kaneko06}.

\begin{figure}
\centering \includegraphics[width=4.5in]{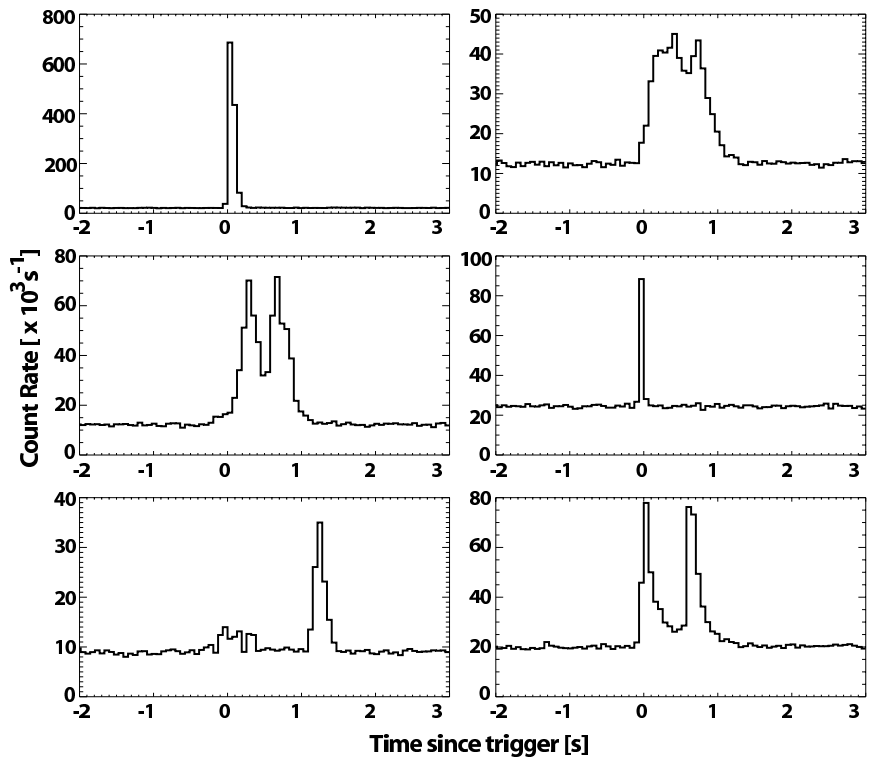}
\caption{BATSE lightcurves of various individual bursts. The {\it y}
  axis is the photon count rate in the 0.05 to 0.5 MeV range; the {\it
  x} axis is the time in seconds since the burst trigger. Both before
  and after the burst trigger, no $\gamma$-rays are detected, above
  background, from the same direction.}
\label{fig:lc}
\end{figure}

A {\it typical} SGRB -- if there is such a thing -- lasts for a
fraction of a second. Observed durations vary, however, by three
orders of magnitude, from several milliseconds \cite{fishman95} to a
few seconds \cite{kouveliotou93}. The shortest BATSE burst had a
duration of 5ms with a 0.2ms structure \cite{bhat92}. Similarly to
long bursts, SGRBs have complicated and irregular time profiles which
vary drastically from one burst to another \cite{nakar02}. They range
from smooth, fast rise and quasi-exponential decay, through curves
with several peaks, to variable curves with many peaks. Various
profiles, selected from the BATSE catalog, are shown in
Figure~\ref{fig:lc}.

The duration of a SGRB is defined by the time during which the middle
50\% ($t_{50}$) or 90\% ($t_{90}$) of the counts above background are
measured. A bimodal duration distribution is measured, irrespective of
whether the $t_{50}$ or $t_{90}$ durations are considered
\cite{kouveliotou93}. About two-thirds of BATSE GRBs are long-duration
GRBs with $t_{90}\geq 2$ s, with the remainder comprising the
SGRBs. The integral size distribution of BATSE SGRBs in terms of peak
flux $\phi_{\rm p}$ is very flat below $\sim 1$ ph cm$^{-2}$ s$^{-1}$,
and becomes steeper than the expected power law with index $-3/2$ of a
Euclidean distribution of sources, at $\phi_{\rm p} > 5$ ph cm$^{-2}$
s$^{-1}$ (see Figure 12 in \cite{fishman95}). This follows from a
cosmological origin of GRB sources, with the decline in the number of
faint bursts due to cosmic expansion. Follow-up X-ray observations
with {\it Swift} and {\it HETE-II} have permitted redshift
determinations that firmly establish the distance scale to the sources
of SGRBs
\cite{donaghy06,gehrels05,bloom06,berger05,barthelmy05,villasenor05,prochaska06,soderberg06,gorosabel06,bloom06b,roming06}.

\begin{figure}
\centering \includegraphics[width=5.0in]{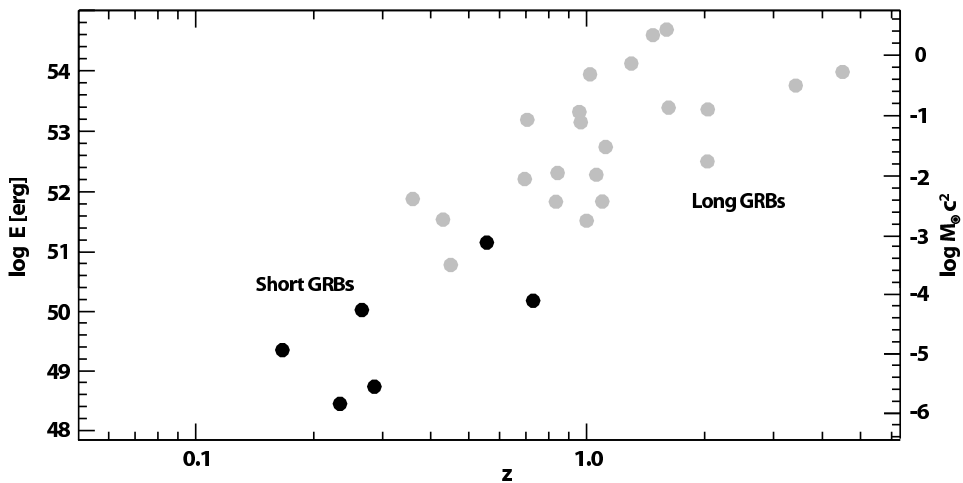}
\caption{Apparent isotropic $\gamma$-ray energy as a function of
redshift. The energy is calculated, assuming isotropic emission, for 6
SGRBs with estimated redshifts. For comparison, the isotropic
$\gamma$-ray energies for 22 long GRBs are also plotted using the
compilation of \cite{bloom01}.}
\label{fig:energ}
\end{figure}

The redshifts of about half a dozen SGRBs are now known (late 2006),
with the median $\langle z \rangle \sim 0.3$ and the largest {\it
spectroscopically} inferred redshift at $z = 0.71$. It should be noted
that the inference of redshift distances relies at present on the
statistical connection to a putative host galaxy and spectroscopy of
the host, and that none are based on the absorption-line systems seen
in the spectra of the afterglows. The corresponding distances imply
apparent isotropic $\gamma$-ray energy releases $E_{\gamma, {\rm iso}}
\approx 10^{48}$-$10^{51} (\Omega_\gamma/ 4\pi)$ ergs, where
$\Omega_\gamma$ is the solid angle into which the $\gamma$-rays are
beamed (Figure~\ref{fig:energ}). For a solar-mass object, this implies
that an unusually large fraction of the energy is converted into
$\gamma$-ray photon energy. This spread in the inferred luminosities
obtained under the assumption of isotropic emission may be reduced if
most GRB outflows are jet-like. A beamed jet would alleviate the
energy requirements, and some observational evidence does suggest the
presence of a jet (see \S \ref{sec:aft}).

\subsection{A Warm Afterglow}\label{sec:aft}

Among the first SGRBs localized by {\it Swift} was GRB 050509B. The
satellite slewed rapidly in the direction of the burst, enabling its
narrow-field X-ray telescope to pinpoint to within 10 arc seconds a
source of faint X-ray emission less than a minute after the trigger.
The source decayed with a power-law behaviour, $\phi_X\propto
t^{\chi}$, with $\chi \sim -1.3$ \cite{gehrels05,bloom06}. Despite
intense follow-up in the optical and radio, no counterparts were
discovered \cite{hjorth05,castro-tirado05,bloom06}. However, the lack
of a detectable afterglow at other wavelengths is not surprising
considering, at face value, the existing GRB afterglow theory
\cite{leeramirez-ruizgranot05}. It took about two months before the
observation of a second short burst, GRB 050709, pinpointed by {\it
HETE-II} \cite{villasenor05}. Other detections have followed, at an
approximate rate of 10 per year, and permitted the observation and
follow-up of afterglows at optical and longer wavelengths.  GRB 050709
was the first SGRB from which an optical counterpart was observed
\cite{hjorth05,covino06}, and GRB 050724 was the first SGRB for which
a radio afterglow was measured \cite{berger05}. GRB 051221A is
probably the best-sampled SGRB to date \cite{soderberg06}.

\begin{figure}
\centering \includegraphics[width=3.7in]{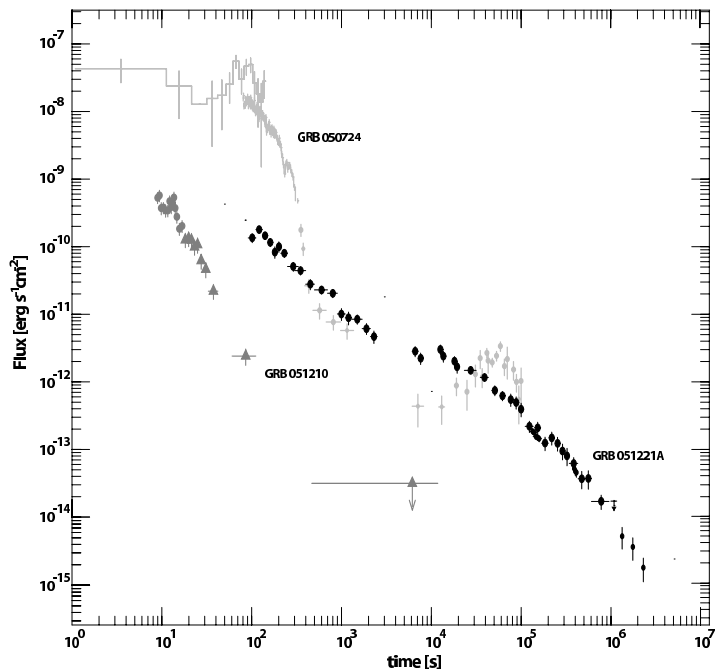}
\caption{ {\it Swift} lightcurves of various individual bursts: GRB
050724, GRB 051210, and GRB 051221A. The {\it y} axis is the flux in
the 2 to 10 keV range; the {\it x} axis is the time in seconds since
the burst trigger. }
\label{fig:flares}
\end{figure}

Of the few bursts localized by the {\it Swift} XRT, only four were
bright enough to permit detailed study: GRB 050724, GRB 051221A, GRB
051210, and GRB 051227 \cite{grupe06,burrows06,laparola06}.  Various
X-ray light curves selected from the {\it Swift} catalog, are shown in
Figure~\ref{fig:flares}. {\it Chandra} follow-up observations obtained
for GRB 050724 and GRB 051221A provide the most constraining
observations for SGRB jets to date \cite{grupe06,burrows06}. The lack
of an observed early downturn in the X-ray light curves has been
interpreted as supporting a mild degree of anisotropy in SGRBs
\cite{burrows06,grupe06,soderberg06,watson06,panaitescu06}.
Anisotropy in the burst outflow and emission affects the light curve
at the time when the inverse of the bulk Lorentz factor equals the
opening angle of the outflow. If the critical Lorentz factor is less
than 3 or so (i.e. the opening angle exceeds 20$^\circ$) such a
transition might be masked by the change from ultra relativistic to
mildly relativistic flow. It would then be generically difficult to
limit the late-time afterglow opening angle in this way if it exceeds
20$^\circ$.  Since some afterglow light curves are unbroken power laws
for over 20 days (e.g., GRB 050724), if the energy input were indeed
just a simple impulsive shell the opening angle of the late-time
afterglow at long wavelengths would probably be greater than
25$^\circ$, i.e. $\Omega_{\rm opt} \geq 0.1$ \cite{grupe06}. However,
even this still means that the energy estimates from the afterglow
assuming isotropy could be $>10$ times too high. The beaming angle for
the $\gamma$-ray emission,$\Omega_\gamma$, could be smaller, and is
much harder to constrain directly.

An observation that attracted much attention was the discovery of long
duration ($\sim 100$ s) X-ray flares following several SGRBs
\cite{villasenor05,barthelmy05} after a delay of $\sim 30$ s (see,
e.g., GRB 050724 in Figure~\ref{fig:flares}). There is also
independent support that X-ray emission on these time scales is
detected when lightcurves of many bursts are stacked
\cite{lazzati01,montanari05}. These observations may indicate that
some sources display continued activity (at a variable level) over a
period of minutes \cite{barthelmy05,panaitescu06}. Additional
structure in the light curves has also emerged from a continued
analysis of some of these objects down to the faintest X-ray flux
levels. In some bursts (e.g., GRBs 050724 and 051221A), the X-ray
light curve exhibits evidence for a late "energy refreshment'' to the
blast wave, on time scales comparable to the afterglow time scale. A
reason for the late rise would be, say, if slower moving ejecta
catches up with the main shock front injecting a substantial amount of
energy and momentum \cite{rees98,ramirez-ruiz01}. However, there are
other mechanisms that can produce rising afterglow fluxes
\cite{fan06,dai06}.

\subsection{Galactic Hosts, Supernova Family Ties and Cosmological Setting}\label{sec:field}

\subsubsection{Demography}

Starting with the detection of GRB 050509B, a growing body of evidence
has suggested that SGRBs are associated with an older and
lower-redshift galactic population than long GRBs and, in a few cases,
with large ($>$ 10 kpc) projected offsets from the centers of their
putative host galaxies (Table~\ref{tab:hosts};
\cite{bloomprochaska06}).

\begin{table}[ht]
\centering
\caption[]{Basic properties of the host galaxies of
SGRBs (adapted from \cite{bloomprochaska06}).\protect\label{tab:hosts}}
\begin{indented}
\item[]\begin{tabular}{lccccc} \mr 
GRB &$z$ & host& in a galaxy & SFR & Metallicity \\
&& classification& cluster &($M_\odot$ yr$^{-1}$)&($Z/Z_\odot$) \\
\mr
050509B & 0.225 & E & Yes & $<$0.1 & $\sim$1 \\
050709  & 0.160 & Irr/late-type dwarf & No  & $>0.3$ & 0.25 \\
050724  & 0.258 & early (E+S0) & No & $ < 0.05$ & 0.2 \\
050813  & 0.722?(1.8) & E?    & Likely & $<$0.2  & $\sim$1\\
051221A & 0.5459& late-type dwarf? & ? & $\sim$1.5  & $\sim$1\\
060502B & 0.287&  E? & ? & $<$0.4& $\sim$0.2\\
\mr
\end{tabular}
\end{indented}
\end{table}

The discovery of GRB 050509B and its fading X-ray afterglow
\cite{gehrels05} led to the first redshift and host galaxy association
\cite{bloom06} for a SGRB.  A chance association with such a galaxy
was deemed unlikely even under conservative assumptions and stood in
stark contrast with the lines-of-sight of long GRBs, for which no
association with an early-type host was ever made. Further {\it Swift}
and {\it HETE-II} detections of SGRBs have continued to support this
hypothesis, though SGRBs are not ubiquitously found at large offsets
and associated with early-type galaxies. GRB 050724
\cite{berger05,prochaska06,gorosabel06,berger06}, GRB 050813
\cite{prochaska06}, and GRB 060502B like 050509B, were found to be in
close association with old, red galaxies. GRB 050724 had optical and
radio afterglow emission that pinpointed its location to be within its
red host, making the association completely unambiguous, though the
association of GRB 050813 and GRB 060502B with any single host remains
tentative \cite{bergerrev}. The absence of observable H$\alpha$ and [O
II] emission constrains the unobscured star formation rates in these
galaxies to $< 0.2 M_\odot$ yr$^{-1}$, and the lack of Balmer
absorption lines implies that the last significant star forming event
occurred $>1$ Gyr ago. Based on positions of the afterglows, two
bursts (050509B and 050813) are very likely associated with clusters
of galaxies \cite{bloom06,pedersen05}.

Not all hosts lack active star formation: GRB 050709
\cite{villasenor05,hjorth05,fox05,covino06} and GRB 051221A
\cite{soderberg06} both had optical afterglows and were associated
with galaxies showing evidence for current, albeit low, star
formation. The host of GRB 051221A, moreover, exhibits evidence for an
evolved stellar population. Despite the availability of both X-ray and
optical afterglow locations, no nearby host has successfully been
identified for either GRB 060121 or GRB 060313. These observations
indicate that these SGRBs occurred during the past $\sim 7$ Gyr of the
universe ($z < 1$) in galaxies with diverse physical characteristics.

In contrast to what is found for SGRBs, all of the confirmed long GRB
host galaxies are actively forming stars with integrated, unobscured
SFRs $\sim 1- 10 M_\odot$ yr$^{-1}$
\cite{christensen04,trentham02,tanvir04}. These host galaxies have
small stellar masses and bluer colors than present-day spiral galaxies
(suggesting a low metallicity; \cite{lefloch03}). The ages implied for
the long-soft GRBs are estimated to be $<$ 0.2 Gyr
\cite{christensen04}, which is significantly younger than the minimum
ages derived for the early-type galaxies found to be associated with
SGRBs. The cluster environments of at least two SGRBs contrast
strikingly with the observation that no well-localized long-soft GRB
has yet been associated with a cluster \cite{bornancini04}.

On the whole, the hosts of SGRBs, and by extension the progenitors,
are not drawn from the same parent population of long GRBs. SGRBs
appear to be more diffusely positioned around galaxies, and their
associated hosts contain a generally older population of stars.

\subsubsection{Soft Gamma-Ray Repeaters in Nearby Galaxies}

It has been noted \cite{hurley05,palmer05} that the giant flare (GF)
observed from the putative galactic magnetar source SGR1806-20 in
December 2004 \cite{hurley05,palmer05} could have looked like a
classical SGRB had it occurred much farther away, thus making the
tell-tale periodic signal characteristic of the neutron star rotation
in the fading emission undetectable. The two previously recorded GFs
of this type, one each from SGR 0520-66 on 5 March 1979
\cite{fenimore96} and SGR 1900 + 14 on 27 August 1998 \cite{hurley99},
would have been detectable by existing instruments only out to $\sim$
8 Mpc, and it was therefore not previously thought that they could be
the source of SGRBs. The main spike of the 27 December event would
have resembled a short, hard GRB if it had occurred within $\sim$ 40
Mpc, a distance scale encompassing the Virgo cluster \cite{palmer05}.
However, the paucity of observed giant flares in our own Galaxy has so
far precluded observationally based determinations of either their
luminosity function or their rate.

In the magnetar model, SGRs are isolated neutron stars with teragauss
exterior magnetic fields and even stronger fields within
\cite{duncan92}, making them the most magnetized objects in the
Universe. The large-scale reorganization of the magnetic field is
thought to produce the observed GFs. The formation rate of magnetars
is expected to track that of stars, which is $\sim 0.013\; M_\odot\;
{\rm Mpc}^{-3}\;{\rm yr}^{-1}$ in our Galaxy \cite{palmer05}. This
suggests that BATSE would have triggered on such events as SGRBs at a
rate of $30 (\dot{N}_{\rm Gal}/0.01\;{\rm yr}^{-1})\;{\rm yr}^{-1}$,
compared with the all--sky BATSE rate of about 150 ${\rm
yr}^{-1}$. Here $\dot{N}_{\rm Gal}$ is the average rate of GFs in the
Galaxy similar to the 27 December event. The observed isotropic
distribution of short BATSE GRBs on the sky and the lack of excess
events from the direction of the Virgo cluster suggests that only a
small fraction, $\leq$ 5\%, of these events can be SGR GFs within 40
Mpc, implying that $\dot{N}_{\rm g} \leq 3 \times 10^{-3}\; {\rm
yr}^{-1}$ on average for a Galaxy like our own \cite{palmer05}.

Before {\it Swift} detected SGRBs, searches for nearby galaxies within
narrow Inter Planetary network (IPN) error boxes revealed already that
only up to $\simeq $ 15\% of them could be accounted for by magnetars
capable of producing GFs \cite{nakar06b}.  Finally, a search for SGRBs
with spectral characteristics similar to that of GFs (thermal spectra
with $kT \simeq 100$~keV) in the BATSE catalog has concluded that a
small fraction (up to a few percent) of them could have originated
from magnetars in nearby galaxies \cite{lazzati05}.
  
Magnetars are thought to be formed during core-collapse events inside
massive stars, and because of their relatively short lifetimes as
observable sources, $\sim 10^{4}$~yr, would naturally be located in
predominantly star-forming galaxies, while essentially none should be
seen in ellipticals. One possible distinction of these from the
classic SGRB population may well come from radio observations, because
their radio afterglows should not be detectable beyond $\sim$ 1 Mpc
\cite{gaensler05,granotrr06}. The fraction of SGR events among what
are now classified as short GRBs may not be dominant, but it should be
detectable and can be tested with future {\it Swift} observations.

\subsubsection{Supernova Partnership}

Current observational limits \cite{hjorth05b,bloom06,castro-tirado05}
indicate that any supernova-like event accompanying SGRBs would have
to be over 50 times fainter than normal Type Ia SNe or Type Ic
hypernovae, 5 times fainter than the faintest known Ia or Ic SNe, and
fainter than the faintest known Type II SNe. These limits strongly
constrain progenitor models for SGRBs.

\begin{figure}
\centering \includegraphics[width=3.7in]{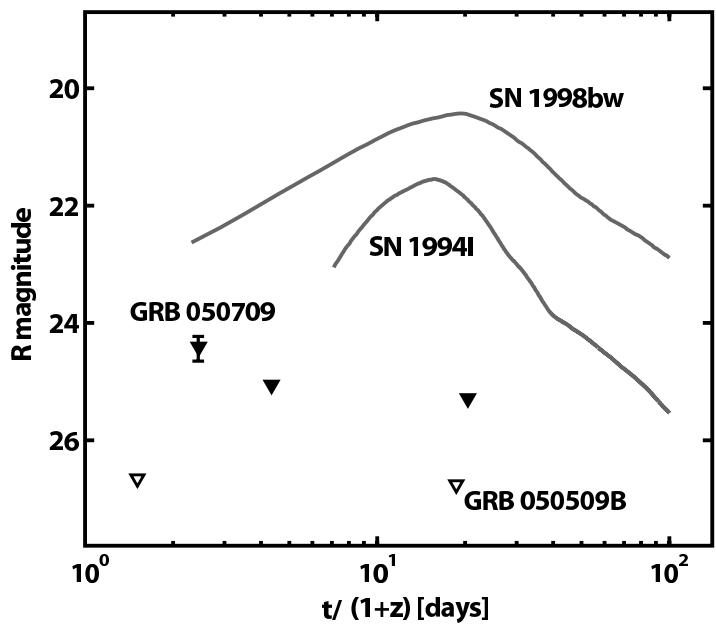}
\caption{Magnitudes and upper limits of SGRB afterglows, GRB 050709
(filled triangles) and GRB 050509B (empty triangles).  For comparison,
two Type Ic supernovae are also shown: SN 1998bw was associated with
the long GRB 980425.}
\label{fig:snlim}
\end{figure}

The limits derived for GRB 050709 (filled triangles) and GRB 050509B
(empty triangles) are plotted in Figure~\ref{fig:snlim} along with two
SN light curves as they would appear at z = 0.225.  The Type Ic SNe
plotted are the very energetic Type Ic SN 1998bw associated with the
long GRB 980425 \cite{galama98} and the faint, fast-rise Type Ic SN
1994I \cite{richmond96}.

The absence of a SN severely rules out models predicting a normal SN
Ia associated with SGRBs. Likewise, current observations disfavor a
progenitor uniquely associated with massive stellar
collapse. Observations of long GRBs at $z < 0.7$ are consistent with
all having SN bumps \cite{zeh04}. The situation is less clear at $z <
0.4$ where all but two long GRBs have had SN features: GRB 980425
\cite{galama98}; GRB 031203 \cite{malesani04}; GRB 030329
\cite{hjorth03}; GRB 011121 \cite{garnavich03}; but no SN was detected
in GRBs 060505 and 060614 down to a flux limit at least hundreds of
times fainter than SN 1998bw \cite{fynbo06}.

\subsubsection{Lifetimes}
The distribution of time delays between progenitor formation and
explosion is not yet well understood. It can be constrained using, for
example, the SGRB rate as a function of redshift
\cite{guetta05,guetta06,nakar06,bloomprochaska06}. If SGRB explosions
lag star formation by some considerable amount of time, then the
intrinsic redshift distribution of SGRBs would be significantly skewed
to lower $z$ when compared to the universal star formation rate (SFR).
A cursory comparison of the redshift distribution of SGRBs with the
SFR reveals what appears to be a significant time delay of a few
Gyr. It should be noted that there are inherent biases in the
discovery of a GRB at a given redshift which are often difficult to
quantify, such as complex trigger efficiencies and non-detections.

An alternative and perhaps less restrictive approach may be to use the
rates of SGRBs in different types of galaxies
\cite{gal-yam06,zheng06}.  On average, early-type galaxies have their
stars formed earlier than late-type galaxies, and this difference,
together with the time delay between progenitor formation and SGRB
outburst, inevitably leads to different burst rates in the two types
of galaxies. For instance, the morphological types for the bursts in
Table~\ref{tab:hosts} reflect a higher incidence of early-type
galaxies than type Ia supernovae and suggest progenitor lifetimes
significantly exceeding 6 Gyr \cite{zheng06}. One difficulty with this
idea is that star formation processes in these two types of galaxies
might not be identical. For example, elliptical galaxies can form by
the merging of two gas-rich galaxies. Many globular clusters can form
in the merging process, which could enhance, for example, the fraction
of binary progenitors and also change the lifetime distribution.

Obviously, these estimates are only sketchy and should be taken as an
order of magnitude estimate at present. However, it should improve as
more hosts are detected.  On the other hand, a large progenitor
lifetime would help explain the apparent high incidence of galaxy
cluster membership. This can be naturally explained by the fact that
galaxies in over dense regions form earlier in hierarchical
cosmologies and thus make a substantial contribution to the local
stellar mass inventory \cite{zheng06}. Detailed observations of the
astrophysics of individual GRB host galaxies may thus be essential
before stringent constraints on the lifetime of SGRB progenitors can
be placed. If confirmed with further host observations, this tendency
of SGRB progenitors to be relatively old can help differentiate
between various ways of forming a SGRB.\\

This concludes our compendium of the {\it facts}. For ease of
reference in the chapters that follow, they have been assembled here
with a minimum of speculative interpretation.

\subsection{Setting the Stage}\label{sec:stage}
SGRBs have been observed assiduously throughout the electromagnetic
spectrum only recently and although we now know much about their
collective and individual properties, we are still long way from being
sure how they operate.

These SGRB sources involve energies that can exceed $10^{50}$ ergs,
the mass equivalent of 1/10,000 of a sun. Compared with the size of
the sun, the seat of this activity is extraordinarily compact, as
indicated by rapid variability of the radiation flux on time scales as
short as milliseconds. It is unlikely that mass can be converted into
energy with better than a few (up to ten) percent efficiency;
therefore, the more powerful SGRB sources must ``process'' upwards of
$10^{-3}M_\odot$ through a region which is not much larger than the
size of a neutron star (NS) or a stellar mass black hole (BH). No
other entity can convert mass to energy with such a high efficiency,
or within such a small volume.

Well-known arguments connected with opacity, variability time scales
and so forth require highly relativistic and variable outflow
\cite{cavallo78, paczynski86,goodman86,shemi90}. Best-guess numbers
are Lorentz factors $\Gamma$ in the range $10^2$ to $10^3$, allowing
rapidly-variable emission to occur at radii in the range $10^{12}$ to
$10^{14}$ cm \cite{rees94,sari97,ramirez-ruiz02,kobayashi97}. Because
the emitting region must be several powers of ten larger than the
compact object that acts as trigger, there is a further physical
requirement: the original energy outflowing in a wind would, after
expansion, be transformed into bulk kinetic energy. This energy cannot
be efficiently radiated as $\gamma$-rays unless it is
re-randomized. The emitted energy is an observable diagnostic of the
microphysical processes of particle acceleration and cooling occurring
within the bulk flow \cite{granot02,sari98,ramirez-ruiz05}. It is
beyond the scope of this article to describe the properties of these
outflows and the physical processes occurring within, and we shall
confine our attention to properties directly relevant to their origins
(An excellent account of the structure and energetics of SGRB outflows
is given in \cite{nakarrev}.). This review is therefore not complete
and in its emphasis reflects the biases of the authors.

Based on current models, the simplest hypothesis -- that the afterglow
is due to a relativistic expanding blast wave -- seems to agree with
the present data
\cite{leeramirez-ruizgranot05,panaitescu06,bloom06,soderberg06}. The
complex time-structure of some bursts suggests that the central engine
may remain active for up to 100 seconds
\cite{nakar02,ramirez-ruiz00,barthelmy05}. However, at much later
times all memory of the initial time-structure would be lost:
essentially all that matters is how much energy and momentum has been
injected, its distribution in angle and velocity.  However we can at
present only infer the energy per solid angle; there are reasons to
suspect that the afterglow is not too narrowly beamed; on the other
hand the constraints on the angle-integrated $\gamma$-ray energy are
not strong \cite{burrows06,grupe06,soderberg06,watson06,panaitescu06}.

As regards the trigger, there remain a number of key questions. What
is the identity of their progenitors? What is the nature of the
triggering mechanism, the transport of the energy and the time scales
involved?  Does it involve a black hole orbited by a dense torus? And,
if so, can we decide between the various alternative ways of forming
it? The presence of SGRBs in old stellar populations helps rule out a source
uniquely associated with recent star formation, while the lack of an
accompanying supernova is strong evidence against a core-collapse
origin. There is now a stronger motivation to develop models in fuller
detail. This article outlines some of these issues.\\

\newpage

%Section 2
\section{Basic Ingredients}\label{sec:basic}

In this section, we present a partial summary of some general ways in
which gravity, angular momentum and the electromagnetic field can
couple to power ultra relativistic outflows, along with a review of the
most popular current models for the central source. There are four
sections: \S \ref{sec:general} gives a brief account of the arguments
in favor of a gravitationally fueled origin; \S \ref{sec:bestiary}
describes the attributes of the most widely favored and conventional
progenitors; the various modes of energy extraction from such systems
are then discussed in \S \ref{sec:pathways}; finally, \S \ref{sec:obs}
gives a compendium of the types of observation that might help
discriminate among the various progenitor models.

\subsection{General Considerations}\label{sec:general}

Shortly after the discovery of quasars in 1963, it was suggested that
accretion of gas onto a compact massive body was responsible for their
enormous energy output \cite{zeldovich64,salpeter64}, which can be
hundreds of times larger than that of entire galaxies. Such a body is
essentially a very efficient converter of gravitational binding energy
into radiation. The deeper the gas can fall into the potential well
before the radiation is converted, the more efficient the process,
hence the appealing nature of compact objects. For black holes
approximately $ \Delta \epsilon \sim GM/R_{\rm ms} \sim 0.1c^2 \equiv
10^{20}$~erg g$^{-1}$ (where $R_{\rm ms}$ is the radius of the
marginally stable orbit) can be released, and even more if the hole is
endowed with a large angular momentum. This efficiency is over a
hundred times that traditionally associated with thermonuclear
reactions (Hydrogen burning releases $0.007c^{2} \sim 6 \times
10^{18}$~erg g$^{-1}$). Since the 1960s, we have also learned about
stellar--mass black holes, where $M \sim 5-10$~M$_\odot$, in Galactic
binary systems and ultra-luminous X-ray sources which, with less
confidence, we also associate with black holes, primarily on energetic
grounds.

In these objects, accretion (and the accompanying radiation) is
usually thought to be limited by the self--regulatory balance between
Newtonian gravity and radiation pressure. A fiducial luminosity is
the {\em Eddington} limit associated with quasi-spherical accretion, at
which radiation pressure balances gravity. If Thomson scattering
provides the main opacity and the relevant material is fully ionized Hydrogen,
then this luminosity is
\begin{equation}
L_{\rm Edd}=\frac{4\pi GM c m_{p}}{\sigma_{\rm T}}=1.3 \times 10^{38}
\left( \frac{M}{M_{\odot}}\right) \mbox{erg~s$^{-1}$},\label{eq:Edd}
\end{equation}
$\sigma_{\rm T}$ being the Thomson cross-section. This may be
converted to a mass accretion rate if one considers that the
accretion luminosity is $L_{\rm Edd}=L_{\rm acc}=GM\dot{M}/R$, giving
\begin{equation}
\dot{M}=10^{18} \left( \frac{R_\ast}{10{\rm km}} \right)\mbox{g~s$^{-1}$}
=1.5 \times 10^{-8} \left( \frac{R_\ast}{10{\rm km}}
\right)\mbox{M$_{\odot}$~yr$^{-1}$}
\end{equation}
where $R_\ast$ is the radius of the compact object. Now clearly this
applies strictly only in a quasi-spherical configuration and a steady
state. Relaxing these assumptions allows for greater luminosities in
transient events, or for configurations in which the energy release is
somehow collimated. However they will not be greater than the
expression given above by more than a factor of a few.

The photon luminosity, for the duration of a typical short burst (a
few seconds at most), is thousands of times larger than that of any
active galactic nucleus (thought to involve supermassive black holes),
and is 12 orders of magnitude above the limit (\ref{eq:Edd}). The
total energy, however, is not very far off from that of other
phenomena encountered in astrophysics, and is in fact reminiscent of
that released in the core of a supernova. The Eddington photon limit
(\ref{eq:Edd}) is circumvented if the main cooling agent is emission
of neutrinos rather than electromagnetic waves.  The associated
interaction cross section is then many orders of magnitude smaller,
and the allowed accretion rates and luminosities are correspondingly
higher. For example, using the cross section for neutrino pair
production, the Eddington limit can be rewritten as
\begin{equation}
L_{{\rm Edd},\nu}=8 \times 10^{53} \left({E_\nu \over 50 {\rm
MeV}}\right)^{-2} (M/M_{\odot})\; {\rm erg~s^{-1}}, \label{eq:Eddnu}
\end{equation}
with an associated accretion rate, assuming unit efficiency for
conversion of mass into neutrino energy,
\begin{equation}
\dot{M}_{{\rm Edd},\nu}= 0.4 (M/M_{\odot}) \left({E_\nu \over 50 {\rm
MeV}}\right)^{-2} \; M_\odot {\rm~s^{-1}}.
\end{equation}
The time it would take an object to radiate away its entire rest--mass
energy in this way is a mass-independent {\em Eddington time} given by
\begin{equation}
t_{{\rm Edd},\nu} = {M \over \dot{M}_{{\rm Edd},\nu}} \sim 2.5
\left({E_\nu \over 50 {\rm MeV}}\right)^{2}\; {\rm~s},
\end{equation}
while the time scale over which an accretion-driven source would
double its mass is $\sim (L/L_{{\rm Edd},\nu})^{-1} \times ({\rm
efficiency})^{-1}\times t_{{\rm Edd},\nu}$. The dynamical time scales
near black holes are modest multiples of $R_{\rm g}/c$, where $R_{\rm
g}$ is the characteristic size of the collapsed object 
\begin{equation}
R_{\rm g} =
GM/c^2 \sim 1.5 \times 10^5 (M/M_\odot)\;{\rm cm},
\end{equation}
and are therefore
much shorter than $t_{{\rm Edd},\nu}$. A fiducial Eddington density,
characteristic near the horizon when the hole accretes at the
Eddington rate, is
\begin{equation}
\rho_{{\rm Edd},\nu} = {\dot{M}_{{\rm Edd},\nu} \over 4\pi R_{\rm
g}^{2} c} \sim 10^{11} (M/M_{\odot})^{-1} \left({E_\nu \over 50
{\rm MeV}}\right)^{-2} \;{\rm g~cm^{-3}}.
\end{equation}
It should be noted that the typical Thomson optical depth under these
conditions is 
\begin{equation}
\tau_{\rm T} \sim n_{{\rm Edd},\nu}^{1/3} R_{\rm g} \sim
10^{16}
\end{equation}
and so, as expected, photons are incapable of escaping and
constitute part of the fluid. For completeness, we can also define an
Eddington temperature, as the black body temperature if a luminosity
$L_{{\rm Edd},\nu}$ emerges from a sphere of radius $R_{\rm g}$,
\begin{equation}
T_{{\rm Edd},\nu}=\left({L_{{\rm Edd},\nu} \over 4\pi R_{\rm g}^2
\sigma_{\rm SB}}\right)^{1/4}\sim 5 \times
10^{11}\;(M/M_{\odot})^{-1/4} \left({E_\nu \over 50 {\rm
MeV}}\right)^{-1/2}\;{\rm K},
\end{equation}
or 
\begin{equation}
k T_{{\rm Edd},\nu} \sim 45 \;(M/M_{\odot})^{-1/4} \left({E_\nu \over 50 {\rm
MeV}}\right)^{-1/2}\;{\rm MeV},
\end{equation}
and an Eddington magnetic field strength
\begin{equation}
B_{{\rm Edd},\nu} = \left({L_{{\rm Edd},\nu} \over R_{\rm
g}^2c}\right)^{1/2}\sim 3\times 10^{16} (M/M_{\odot})^{11/2} \left({E_\nu \over
50 {\rm MeV}}\right)^{-1} \;{\rm G}.
\end{equation}
Finally, for comparison, we define $T_{\rm th}$ as the temperature
the accreted material would reach if its gravitational potential
energy were turned entirely into thermal energy. It is given by
\begin{equation}
T_{\rm th} = {G M m_{\rm p} \over 3kR_{\rm g}}\sim 3 \times
10^{12}\;{\rm K}.
\end{equation}
Some authors use the related concept of the virial temperature,
$T_{\rm vir} =T_{\rm th}/2$, for a system in mechanical and thermal
equilibrium. In general, the radiation temperature is expected to be
$\leq T_{\rm th}$. In deriving the above estimates we have assumed that the
radiating material can be characterized by a single temperature. This
may not apply, for example, when a hot corona deforms the neutrino
spectrum away from that of a cooler thermal emitter
\cite{ramirez-ruiz06}.

Although we have considered here the specific case of neutrino pair
creation, the estimates vary little when one considers for example
coherent scattering of neutrinos by nuclei and/or free nucleons
(except for the energy scaling). Similar overall fiducial numbers also
hold for neutron stars, except that the simple mass scalings obtained
here are lost.

It is thus clear from the above estimates that when mass accretes onto
a (stellar-mass) black hole or neutron star under these conditions,
the densities and temperatures are so large ($\rho \simeq
10^{11}$~g~cm$^{-3}$, $T\simeq 10^{11}$~K) that: (i) photons are
completely trapped; and (ii) neutrinos, being copiously emitted, are
the main source of cooling since they can mostly escape. This regime,
which requires correspondingly large accretion rates, is termed
hypercritical accretion, and was considered for SN1987A shortly after
the explosion \cite{chevalier89,houck91}. Such high accretion rates
are never reached for black holes in XRBs or AGN, where the luminosity
remains well below the photon Eddington rate (\ref{eq:Edd}). They can,
however, be achieved in the process of forming neutron stars and
stellar--mass black holes during the collapse of massive stellar
cores.  Note that at sufficiently high accretion rates, the density
reaches the threshold for optical thickness even to neutrinos, which
cannot then simply stream out (this occurs at $\rho\simeq
10^{11}$g~cm$^{-3}$).

\subsection{Bestiary}\label{sec:bestiary}
The current view is that SGRBs arise in a very small fraction
(approximately $\sim 10^{-6}$) of stars which undergo a catastrophic
energy release event toward the end of their evolution. One
conventional possibility is the coalescence of binary neutron stars
\cite{paczynski86,goodman86,goodman87,eichler89,paczynski91,narayan92,meszaros92,mochkovitch93,jaroszynski93,jaroszynski96}. Double
neutron star binaries, such as the famous PSR1913+16
\cite{hulse75}, will eventually coalesce due to angular momentum and
energy losses to gravitational radiation. The resulting system could
be top--heavy and unable to survive as a single neutron star. However,
a black hole would be unable to swallow the large amount of angular
momentum present. The expected outcome would then be a spinning hole,
orbited by a torus of NS debris.

Other types of progenitor have been suggested - e.g. a NS-BH merger
\cite{lee95,kluzniak98,janka99}, where the neutron star is tidally
disrupted before being swallowed by the hole; the merger of a White
Dwarf (WD) with a black hole \cite{fryerwd99}; the coalescence of
binary WDs \cite{paczynski86,goodman87,katz96,levan06}; or accretion
induced collapse (AIC) of a NS \cite{vietri99,macfadyen06}, where the
collapsing neutron star has too much angular momentum to collapse
quietly into a black hole. In an alternative class of models, it is
supposed that the compact objects are contained within a globular
cluster, and that the binary system will evolve mainly through
hardening of the binary through three--body interactions
\cite{hansen98, grindlay06} or physical star-star collisions
\cite{lee06b} rather than by pure gravitational wave
emission. Finally, violent reconfigurations of the magnetic field on
magnetars \cite{duncan92} may offer another
possibility. Table~\ref{tab:rates} provides a summary of the various
rate estimates for some of these possible SGRB progenitors. Aside from
the rate of SNe events, the rate of SGRBs and plausible progenitors in
Table~\ref{tab:rates} are highly imprecise. The main sources of
uncertainty are related to supernova kicks, the mass ratio
distribution in binaries, mass limits for black hole formation,
stellar radii and common envelope evolution
\cite{fryerrates99,clark79,lipunov87,hills91,phinney91,tutukov93,lipunov95,portegies-zwart96,fryer98,bethe98,belczynski99,bloomdns99,fryerrates99,belczynski01,belczynski02,belczynski06}.

\begin{table}[ht]
\centering
\caption[]{Estimated progenitors of SGRBs and their plausible rates
 \cite{guetta06,levan06,fryerrates99,phinney91,bethe98,izzard04} (in
 yr$^{-1}$ Gpc$^{-3}$). The rates for these various occurrences are
 plagued by a variety of uncertainties in the supernova explosion
 mechanism, stellar evolution, and binary evolution, and should be
 taken as an order of magnitude estimate only. The oberved rates of
 Type (Ia + Ib/c) events given here provide a generous upper limit.
 \protect\label{tab:rates}}
\begin{indented}
\item[]\begin{tabular}{lc} \mr
Progenitor &  Rate ($z=0$)\\
\mr
NS--NS & 1-800\\
BH--NS & 0.1--1000\\
BH--WD & 0.01-100\\
NS AIC & 0.1 - 100\\
WD-WD & 3000\\
SN Ib/c & 60000\\
SN Ia & 150000\\
SGRBs & 10$(4\pi/ \Omega)$\\
\mr
\end{tabular}
\end{indented}
\end{table}

The rate of gravitational mergers among BH-BH and BH-NS binaries
depends on the orbital periods after spiral--in, and hence on the
ill--understood details of the process. A binary with total mass
$10M_1M_\odot$, mass ratio $q$, and period $P_{\rm d}$ (in days) in a
nearly circular orbit will merge in 
\begin{equation}
\tau_{\rm m}=1 \times 10^9 P_{\rm d}^{8/3}M_1^{-5/3}(1+q)(1+1/q)
\;{\rm yr}.
\end{equation}
Thus for a coalescence to occur within a Hubble time, spiral--in must
have reduced the orbital period to $\leq$ 1 day. Since a black hole
can have a mass exceeding that of its companion helium core,
spiral--in need not occur, as opposed to the case of a lower mass
object like a neutron star (except in very non conservative mass
transfer). If the companion has a strong wind, spiral--in might never
begin. If it does, envelope ejection might be less efficient than that
by a neutron star, since black hole accretion could have a very low
efficiency. If spiral--in does occur, however, the subsequent
evolution may be simpler than in the double neutron star case. Helium
stars less massive than 3$M_\odot$ expand their envelopes dramatically
during core carbon burning \cite{habets86}, requiring a second
spiral-in to make a NS-NS binary which could merge in a Hubble
time. The more massive He cores which could leave black hole remnants
do not expand much, so a second spiral-in is not needed.

Thus BH-NS binaries form at rates comparable to the NS-NS rate, but
the fraction which merge depends on the miasma of mass transfer and
spiral-in which determine the final period distribution. Gravitational
waveforms of any merging systems will allow the masses of the merging
bodies to be accurately determined \cite{schutz86,cutler94}, shedding
light on the underlying physics, as well as on the minimum black-hole
mass determined by post-collapse infall. The much more certain NS-NS
mergers will also allow the mass distribution of neutron stars to be
determined.

How might such a progenitor generate a relativistic outflow or a
sudden release of electromagnetic energy? We now address various
possible routes.
 
\begin{figure}[h!]
\centering \includegraphics[width=4.0in]{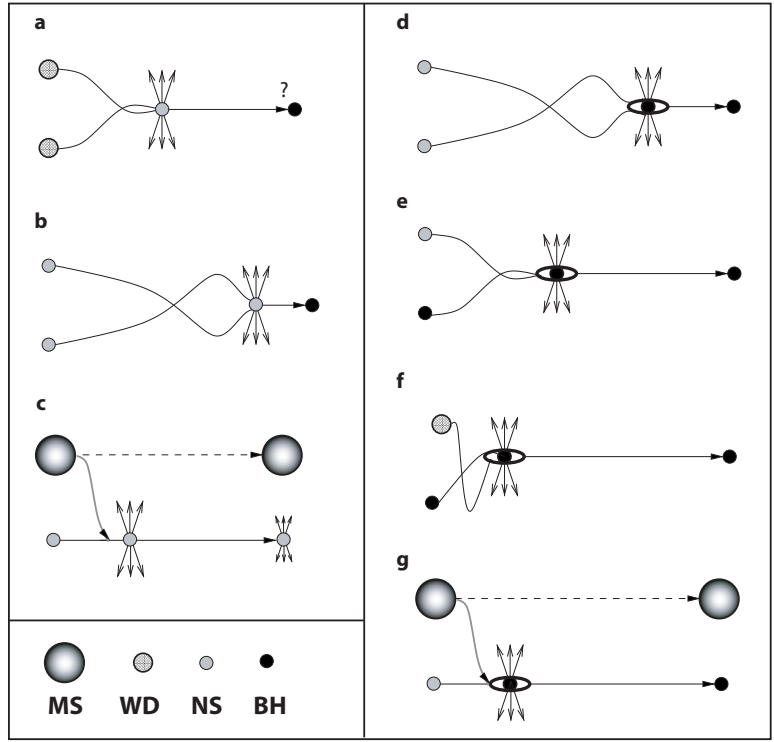}
\caption{Schematic scenarios for plausible SGRB progenitors: neutron
star mergers (b,d); neutron star (e) or white dwarf (f) disruption by
a black hole; white dwarf mergers (a); accretion induced collapse of
neutron star (g); and a recycled magnetar (c). The dominant production
channel for each scenario is depicted, where MS denotes the primary
main sequence star. The (rough) relative in-spiral times due to
gravitational radiation for compact mergers are shown (not to
scale). (d,e,f,g) eventually lead to the formation of a black hole
with a debris torus around it. Physical star-star collisions or
three--body gravitational interactions rather than gravitational
mergers could lead to similar conditions. A second way that the
progenitor system may liberate its binding and rotational energy is to
develop a large magnetic field and act as a magnetar (a,b,c). Panels
adapted from \cite{bloomphd}.}
\label{fig:prog}
\end{figure}

\subsection{Metabolic Pathways}\label{sec:pathways}

It has become increasingly apparent in the last few years that most
plausible SGRB progenitors suggested so far (e.g. NS-NS or NS-BH,
WD-BH mergers or physical collisions, and NS AIC) are expected to lead
to a black hole plus debris torus system (Figure~\ref{fig:prog}).  In
this case there is no external agent feeding the accretion disk, and
thus the event is over roughly on an accretion time scale (which would
be on the order of one second). A possible exception includes the
formation from accretion \cite{spruit99} or from a WD-WD
\cite{levan06} or NS-NS merger \cite{rosswog03b,rosswogrr04,price06}
of a rapidly rotating (in some cases very massive) neutron star with
an ultrahigh magnetic field (Figure~\ref{fig:prog}). If there is an
ordered field $B$, and a characteristic angular velocity $\omega$, for
a spinning compact source of radius $R_\ast$, then the magnetic dipole
moment is $\sim BR_\ast^3$. General arguments suggest
\cite{pacini67,gold68} that the non thermal magnetic-dipole-like
luminosity will be $\propto B^2R_\ast^6\omega^4/ c^3$, and simple
scaling from these familiar results of pulsar theory require fields of
order $10^{15}$~G to carry away the rotational or gravitational energy
(which is $\sim 10^{53}$~erg) in a time scale of seconds
\cite{usov94,thompson94}.

One of the most perceptive theoretical discoveries that was made about
black holes was that, when they spin, a fraction of their mass can be
ascribed to rotational energy and is, in principle, extractable
\cite{penrose71}. This is most convincingly demonstrated by observing
that there exist orbits of test particles with negative total energy,
(including their rest mass), within the ergosphere. If an infalling
plasma cloud is attached to magnetic field lines anchored at a large
distance (e.g., in an accreting torus), and the field drags the cloud
backwards relative to the rotation of the hole placing it on an orbit
with negative energy, the work that has been performed can be thought
of as energy that has been effectively extracted from the spin of the
black hole \cite{ruffini75}.

The binding energy of the orbiting debris, and the spin energy of the
BH are thus the two main reservoirs for the case of a black hole
central engine: up to 42\% of the rest mass energy of the torus, and
29\% of the rest--mass energy of the black hole itself can be
extracted for a maximal black hole spin. SGRB activity can be powered
by the black hole only as long as there is interaction with the
surrounding gas, which will probably need to be centrifugally
supported (although even for relatively low angular momentum a
considerable amount of energy can still be extracted \cite{lee06}).

The angular momentum is quite generally a crucial parameter, in many
ways determining the geometry of the accretion flow. Even a little
rotation can make a big difference, breaking the spherical symmetry
and producing accretion {\em disks} instead of {\em radial} inflow, as
envisaged originally by Bondi \cite{bondi52}. If the gas has no
angular momentum and the magnetic field is dynamically unimportant,
there will be essentially radial inflow. Spherical accretion onto
black holes is relatively inefficient despite the deep potential well,
because the gas is compressed, but not shocked, and thus cannot easily
convert gravitational to thermal energy. The flow pattern changes
dramatically if the inflowing gas has a small amount of angular
momentum. The quasi-spherical approximation breaks down when the gas
reaches a radius $R_{\rm circ} \sim l^2/GM$, where $l$ is the angular
momentum per unit mass, and if injection occurs more or less
isotropically at large radii, a familiar accretion disk will form (as
occurs in X--ray binaries and white dwarf binary systems). The matter
will instead dissipate its motion perpendicular to the plane of
symmetry and form a differentially rotating disk, the rotational
velocity at each point being approximately Keplerian, and then
gradually spiral inwards as viscosity transports its angular momentum
outwards.

If the emission process is very efficient, the disk is dynamically
cold and geometrically thin, in the sense that locally, $kT \ll
GMm_{\rm p}/R$ and the pressure scale height $H(R) \ll R$. If gas
passing through a thin disk reaches a radius within which the internal
pressure builds up -- either because it is unable to cool in an inflow
time or because the radiation pressure force is competitive with
gravity -- the disk will become geometrically thick, with $H \sim
R$. In a thick disk or torus the pressure provides substantial support
in the radial as well as the vertical direction, and the angular
momentum distribution (now as a function of height as well as radius)
may be far from Keplerian. Both types of configurations have been
studied extensively and the agreement in some cases between
observation and theory is extremely good, assuming a fundamentally
empirical recipe for angular momentum transport
\cite{shakura73,ichimaru77,narayan94,abramowicz95}.

As mentioned above in \S \ref{sec:general}, in principle flow onto a
compact object can liberate gravitational potential energy at a rate
approaching a few tenths of $\dot{M}c^2$, where $\dot{M}$ is the mass
inflow rate. Even for such high efficiencies the mass requirements of
the more luminous SGRB sources are rather high, with
\begin{equation}
\dot{M} \sim 3 \times 10^{-3}\left({0.1 \over
\epsilon}\right)\left({L_{\rm SGRB} \over 10^{51}\;{\rm
erg\;s^{-1}}}\right) M_\odot\;{\rm s^{-1}}, \label{eq:mdot}
\end{equation} 
where $\epsilon$ is the overall efficiency. The inner regions of disks
with mass fluxes in this range are generally able to cool by neutrinos
on time scales shorter than the inflow time. If
$\dot{m}=\dot{M}/\dot{M}_{{\rm Edd},\nu} \leq 1$, then the bulk of the
neutrino radiation comes from a region only a few gravitational radii
in size, and the physical conditions can be scaled in terms of the
{\it Eddington quantities} defined in \S ~\ref{sec:general}. The
remaining relevant parameter, related to the angular momentum, is
$v_{\rm inflow}/v_{\rm freefall}$, where $v_{\rm
freefall}\simeq(2GM/R)^{1/2}$ is the free fall velocity. The inward
drift speed $v_{\rm inflow}$ would be of order $v_{\rm freefall}$ for
supersonic radial accretion. When angular momentum is important, this
ratio depends on the mechanism for its transport through the disk,
which is related to the effective shear viscosity. For a thin disk,
the factor $(v_{\rm inflow}/v_{\rm freefall})$ is of order $\alpha
(H/R)^{2}$, where $H$ is the scale height at radius $R$ and $\alpha$
is the phenomenological viscosity parameter.

Suppose that a given accretion rate yields a luminosity $L_\nu$ with
an efficiency $0.1$. Then the characteristic density, at a distance
$R$ from the hole, with account of the effects of rotation, is
\begin{equation}
\rho\sim \dot{m}(R/R_{\rm g})^{-3/2}(v_{\rm inflow}/v_{\rm
freefall})\rho_{{\rm Edd},\nu},
\end{equation}
and the maximum magnetic field, corresponding to equipartition with
the bulk kinetic energy, would be
\begin{equation}
B_{\rm eq}\sim \dot{m}^{1/2}(R/R_{\rm g})^{-5/4}(v_{\rm inflow}/v_{\rm
freefall})^{1/2}B_{{\rm Edd},\nu}.
\end{equation}
Any neutrinos emerging directly from the central {\it core} would have
energies of a few MeV. Note that, as mentioned above, $kT_{{\rm
Edd},\nu}$ is far below the {\it virial} temperature $kT_{\rm
vir}\simeq m_p c^2(R/R_{\rm g})$.

The flow pattern when accretion occurs would be then determined by the
value of the parameters $L_\nu/L_{{\rm Edd},\nu}$, which determine the
importance of radiation pressure and gravity, and the ratio $t_{\rm
cool}/t_{\rm dynamical}$, which fixes the temperature if a stationary
flow pattern is set up. The preceding general discussion of
neutrino--cooled accretion flows thus provides a basis for general
models. So far we have discussed the power output that might be
generated by the accretion process, but we have made no attempt to
describe in detail the flow of the accreting gas.  A hint that its
dynamics may not be straightforward is provided by the existence of
the Eddington limit for neutrinos, clearly illustrating that for the
high accretion rates expected in SGRBs, forces other than gravity may
be important. In addition, in many cases (and probably in most) the
accreting matter possesses considerable angular momentum per unit mass
which it has to lose somehow in order to be accreted at all. The
reader is referred to \S \ref{sec:disks} for fuller details.  We now
concentrate briefly on the origins of the jets which provide the power
for SGRB sources.

\begin{figure}[h!]
\centering \includegraphics[width=5in]{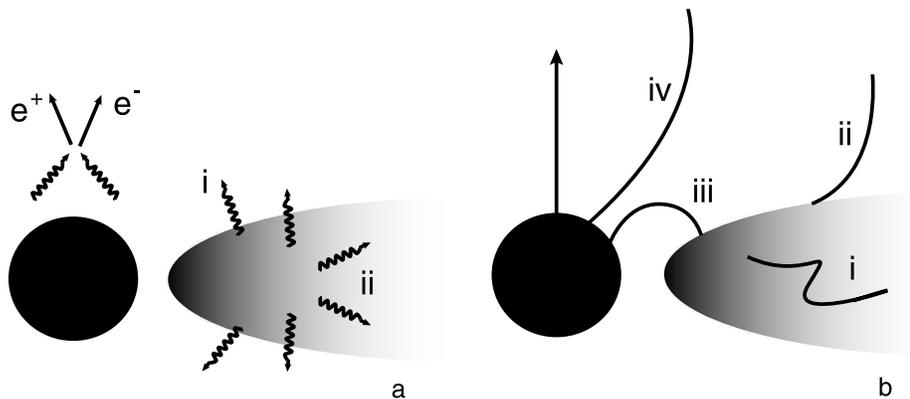}
\caption{Metabolic pathways for energy extraction. {\it Panel
a:} Energy released as neutrinos is reconverted via collisions outside
the dense core into e$^{\pm}$ pairs or photons (i). The
neutrinos that are emitted from the inner regions of the debris
deposit part of their energy in the outer parts of the disk (ii),
driving a strong baryonic outflow. This wind may be responsible for
collimating the jet. {\it Panel b:} Strong magnetic fields anchored in
the dense matter can convert the binding and/or spin energy into a
Poynting outflow. A dynamo process of some kind is widely believed to
be able to operate in accretion disks, and simple physical
considerations suggest that fields generated in this way would have a
canonical length-scale of the order of the disk thickness (i). Open
field lines can connect the disk outflow and may drive a hydromagnetic
wind (ii). The above mechanism can tap the binding energy of the
debris torus, while a rapidly rotating hole could contain an even
larger energy reservoir, extractable in principle through MHD coupling
to the exterior (iii;iv). Panel {\it b} adapted from \cite{blan02}.}
\label{fig:diag}
\end{figure}

Two ingredients are necessary for the production of jets: first, there
must be a source of material with sufficient free energy to escape the
gravitational field of the compact object; second, there must be a way
of imparting some directionality to the escaping flow. Our eventual
aim must be to understand the overall flow pattern around a central
compact object, involving accretion, rotation, and directional outflow
but we are still far from achieving this.  Most current works who have
discussed outflow and collimation have simply invoked some central
supply of energy and material. A self-consistent model incorporating
outflow and inflow must explain why some fraction of the matter can
acquire a disproportionate share of energy (i.e., a high enthalpy).  A
brief summary of the various metabolic pathways is presented in
Figure~\ref{fig:diag}.

The neutrino luminosity emitted when disk material accretes via
viscous (or magnetic) torques on a time scale $\Delta t \sim 1$ s is
roughly
\begin{equation}
L_\nu \sim 2 \times 10^{52} \left({M_{\rm disk} \over 0.1
M_\odot}\right)\left({\Delta t \over 1 {\rm s}}\right)^{-1}
\mbox{erg~s$^{-1}$}
\end{equation}
for a canonical radiation efficiency of 0.1. One fairly direct
solution is to reconvert some of this energy via collisions outside
the disk into electron-positron pairs or photons
\cite{mochkovitch93,rosswogrr02}. If this occurs in a region of low baryon density
(e.g., along the rotation axis, away from the equatorial plane of the
disk) a relativistic pair-dominated wind can be produced. An obvious
requirement for this mechanism to be efficient is that the neutrinos
escape (free streaming, or diffusing out if the density is high
enough) in a time scale shorter than that for advection into the black
hole. The efficiency for conversion into pairs (scaling with the square
of the neutrino density) is too low if the neutrino production is too
gradual, so this can become a delicate balancing act. Typical
estimates suggest a lower bound of $L_{\nu\bar\nu} \sim 10^{-3}
L_{\nu}$ when the entire surface area emits close to a single
temperature black--body. The efficiency may be significantly larger if
dissipation takes place in a corona--like environment
\cite{ramirez-ruiz06}.

One attractive energy extraction mechanism that could circumvent the
above restriction in efficiency is a relativistic magneto hydrodynamic
(MHD) wind \cite{usov92,thompson94}. Such a wind carries both bulk
kinetic energy and ordered Poynting flux, and it is possible that
gamma-ray production occurs mainly at large distances from the source
\cite{duncan92,usov94,thompson94,meszaros97,spruit01,lyutikov03}. A rapidly
rotating neutron star (or accretion disk) releases energy via magnetic
torques at a rate
\begin{equation}
L_{\rm em} \sim 10^{49} \left({B \over 10^{15}\;{\rm G}}\right)^2
\left({P \over 10^{-3}\;{\rm s}}\right)^{-4} \left({R\over
10^6\;{\rm cm}}\right)^6\;{\rm erg\;s^{-1}},
\end{equation}
where $P$ is the spin period, and $B$ is the strength of the poloidal
field at a radius $R$. The last stable orbit for a Schwarzschild hole
lies at a coordinate distance $R=6R_{\rm g}=9(M/M_\odot)$ km, to be
compared with $R_{\rm g}=3/2(M/M_\odot)$ km for an extremal Kerr
hole. Thus the massive neutron disk surrounding a Schwarzschild black
hole of approximately $2 M_{\odot}$ should emit a spin-down luminosity
comparable to that of a millisecond neutron star. A similar MHD
outflow would result if angular momentum were extracted from a central
Kerr hole via electromagnetic torques \cite{blandford77}.  The field
required to produce $L_{\rm em} \geq 10^{51}\;{\rm erg\;s^{-1}}$ is
enormous, and may be provided by a helical dynamo operating in hot,
convective nuclear matter with a millisecond period \cite{duncan92}. A
dipole field of the order of $10^{15}$~G appears weak compared to the
strongest field that can in principle be generated by differential
rotation ($\sim 10^{17}[P/1\;{\rm ms}]^{-1}\;{\rm G}$), or by
convection ($\sim 10^{16}\;{\rm G}$), although how this may come about
in detail is not resolved. Note, however, that it only takes a
residual torus (or even a cold disk) of $10^{-3}\ M_\odot$to confine a
field of $10^{15}$~G. Orbiting debris with such large magnetic seed
fields and turbulent fluid motions will give rise to a plethora of
electromagnetic activity (Figure \ref{fig:diag}). The topic lies
beyond the scope of this paper and we refer the reader to the
excellent review by Blandford \cite{blan02}.

A potential death-trap for such relativistic outflows is the amount of
entrained baryonic mass from the surrounding medium. For instance, a
Poynting flux of $10^{52}$ erg could not accelerate an outflow to
$\Gamma \geq 100$ if it had to drag more than $\sim 10^{-5}M_\odot$ of
baryons with it. A related complication renders the production of
relativistic jets even more challenging, because the high neutrino
fluxes are capable of ablating baryonic material from the surface of
the disk at a rate \cite{qian96}
\begin{equation} 
\dot{M}_\eta \sim 5 \times 10^{-4}\left({L_\nu \over 10^{52}\;{\rm
erg\;s^{-1}}}\right)^{5/3} M_\odot {\rm s^{-1}}.
\label{eq:ablation}
\end{equation}
Thus a rest mass flux $\dot{M}_\eta$ limits the bulk Lorentz factor of
the wind to
\begin{equation} 
\Gamma_\eta={L_{\rm wind} \over \dot{M}_\eta c^2}=10 \left({L_{\rm
wind} \over 10^{52}\;{\rm erg\;s^{-1}}}\right) \left({\dot{M}_\eta
\over 5 \times 10^{-4} M_\odot {\rm s^{-1}}}\right)^{-1}.
\end{equation}
Assuming that the external poloidal field strength is limited by the
vigor of the convective motions, the spin-down luminosity scales with
neutrino flux as $L_{\rm wind} \approx L_{\rm em}\propto B^2\propto
v_{\rm con}^2\propto L_\nu^{2/3}$, where $v_{\rm con}$ is the
convective velocity. The ablation rate given in equation
(\ref{eq:ablation}) then indicates that the limiting bulk Lorentz
factor $\Gamma_\eta$ of the wind decreases as $L_\nu^{-1}$. Thus the
burst luminosity emitted by a magnetized neutrino cooled disk may be
self-limiting. Mass loss could, however, be suppressed if the
relativistic wind were somehow collimated into a jet. This suggests
that centrifugally driven mass loss will be heaviest in the outer
parts of the disk, and that a detectable burst may be emitted only
within a relatively small solid angle centered on the rotation axis.

\subsection{Observational Tests}\label{sec:obs}
In the preceding sections, we have endeavored to outline some of the
basic physical processes that are believed to be of most relevance to
interpreting SGRB sources, not so much because we believe that there
is strong evidence in favor of them, but instead because it provides a
framework in which to discuss the observations and to demonstrate
that, as yet, SGRBs pose no threat to conventional physics. Of course,
we are conscious that most observations tell us less about the primary
source than the about secondary reprocessing of this power in the
circumburst and galactic environment -- from which we can learn about
the central engine only by a chain of uncertain inferences. Unless SGRBs are eventually found to be accompanied by telltale emission features like
the supernovae of long-duration GRBs, the only definitive
understanding of the progenitors will come from possible associations to direct gravitational or neutrino signals.

Spectral investigations will nonetheless be important as probes of the
basic energetics and microphysical parameters of relativistic shocks
\cite{leeramirez-ruizgranot05,panaitescu06}. We are more sanguine
about the possibility of deriving accurate particle densities from
afterglow observations. However, there is the possibility that
occurrence in an unusually low density environment could cause the
external shock to occur at much larger radii and over a much longer
time scale than in usual afterglows \cite{panaitescu01,perna02}. The
X-ray afterglow intensity could be then below the threshold for
triggering.  This may be the case for SGRBs arising from compact
binaries that are ejected from the host galaxy into an external
environment that is much less dense that the ISM assumed for usual
models. Another possibility for an unusually low density environment,
made up only of very high energy but extremely low density electrons,
is if the GRB goes off inside a pulsar cavity inflated by one of the
neutron stars in the precursor binary \cite{lee02,konigl02}. Such
bubbles can be as large as fractions of a parsec or more, giving rise
to a deceleration shock months after the SGRB with a consequently much
lower brightness that could avoid triggering and detection. The
difference between the low-density and high-density environment cases
could be tested if future observations of afterglows reveal a
correlation with the degree of galaxy clustering or with individual
galaxies.

The association of SGRBs with both star-forming galaxies and with
ellipticals dominated by old stellar populations, to some researchers,
suggested an analogy to type Ia supernovae, as it indicated a class of
progenitors with a wide distribution of delay times between formation
and explosion. Similarly, just as core-collapse supernovae are
discovered almost exclusively in late-time star-forming galaxies, so
too are long GRBs. Indeed, a detailed census of the types of host
galaxies, burst locations and redshifts should help decide between the
various alternatives suggested in \S~\ref{sec:bestiary}. This is
because if the progenitor lifetime is long and the systemic kick is
small, then the bursts should correspond spatially to the oldest
populations in a given host galaxy. For early-type galaxies, the
distribution would presumably follow the light of the host
\cite{bloomprochaska06}. In contrast, a neutron star binary could take
millions of years to spiral together, and could by then, especially if
given a substantial kick velocity on formation, have moved many
kiloparsecs from its point of origin. The burst offsets would then
presumably be larger for smaller mass hosts. As new redshifts,
offsets, and host galaxies of SGRBs are gathered, the theories of the
progenitors will undoubtedly be honed.\\

\newpage

%Section 3
\section{Compact Object Mergers and Accretion Disk Assembly}\label{trigger}

SGRB activity manifests itself on many different scales. The simplest
hypothesis, however, is that the central {\it prime mover} or {\it
trigger} is qualitatively similar in all events. The primary energy
may be reprocessed in different ways, depending on the details of the
circumburst (and galactic) environment. A brief summary of the various
evolutionary pathways that may be involved in producing a SGRB is
given in \S \ref{sec:bestiary}. The important message is that most
current alternatives involve either an accreting stellar mass black
hole (neutron star) or a precursor stage whose inevitable end point is
such an object. Furthermore, as we discussed in \S \ref{sec:general},
a black hole (neutron star) embedded in infalling matter offers a more
efficient power source than any other conceivable progenitor; it is
our firm prejudice that SGRBs are energized by mechanisms that involve
either neutron stars or black holes.  For this reason, in this
section, we turn out attention to the formation of the {\it trigger}~:
the actual astrophysical system that is capable of, and presumably
powers the observed burst. It is beyond the scope of this article to
describe completely all evolutionary pathways, and we shall confine
our attention to binary mergers and physical collisions. The study of
compact binaries, particularly those involving pulsars, has evolved
dramatically over the past thirty years, and carries a great deal of
historical baggage. Consequently, most theoretical work has been
directed towards describing these binary encounters. Our discussion
will reflect this bias \footnote{The reader is referred to the
excellent review by Rosswog \cite{rosswog06} for an alternative but
complementary description of these source models.}. There are eight
sections: \S~\ref{sec:mass} gives a summary of observed compact object
masses; \S~\ref{sec:couples} gives an overall picture of the merger
scenario as we best understand it now; then in \S~\ref{sec:early} we
describe early considerations of these events; this is followed in
\S~\ref{sec:dynstab} by a summary of dynamical stability
considerations in such systems; the merger itself is considered in
\S~\ref{sec:merger}; the effects of General Relativity and the fate of
the remnant core are addressed in \S~\ref{sec:GReffects} and
\S~\ref{sec:corefate} respectively; finally, the outcome of compact
object collisions is studied in \S~\ref{sec:colls}.

\subsection{The Mass of the Progenitor}\label{sec:mass}
The mass of the compact object dictates a characteristic length and
luminosity scale for SGRB activity, so we must consider what we know
(and what we do not) from observations of systems containing compact
objects in a more leisurely state of affairs.

\begin{figure}
\centering \includegraphics[width=4.5in]{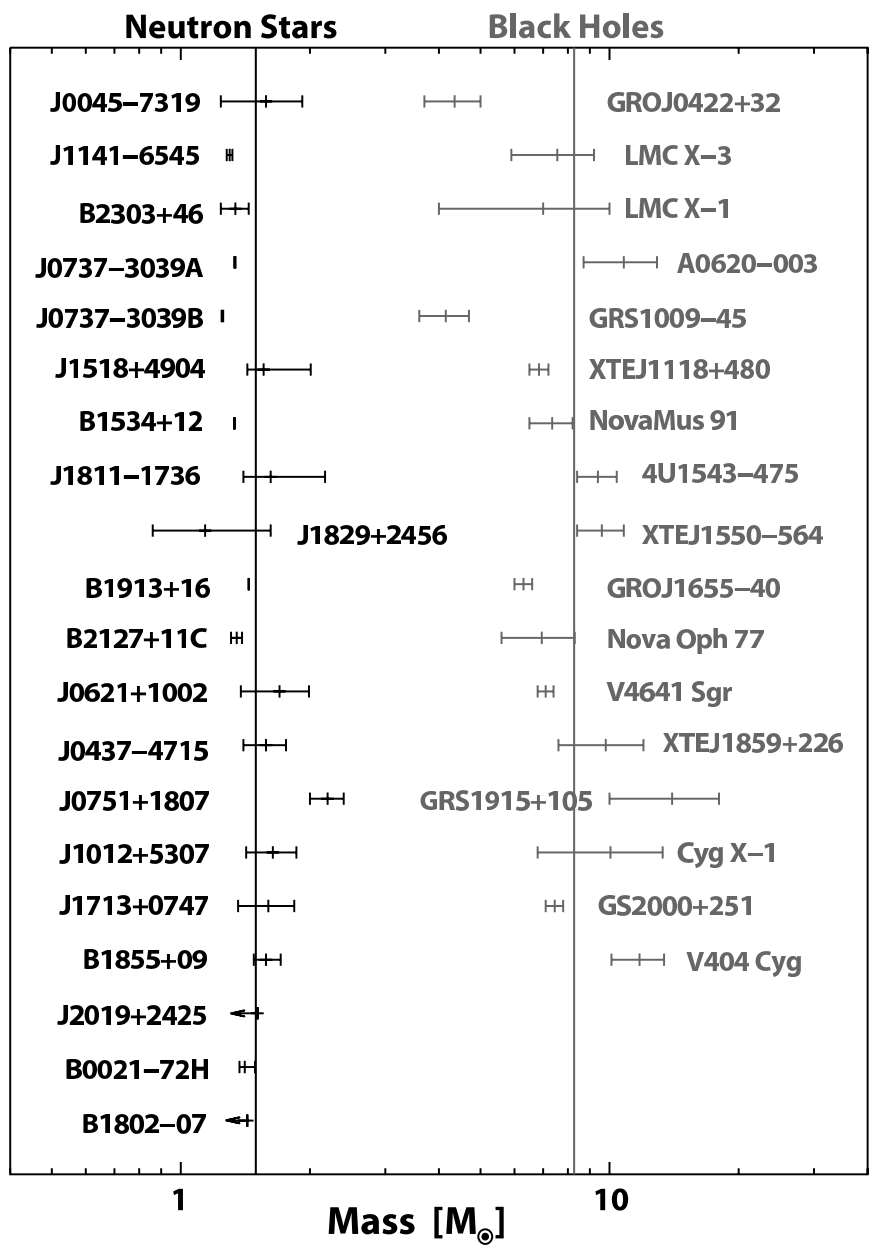}
\caption{Measured masses for neutron stars and black holes in
binaries. The average values are $M_{\rm NS}=1.49M_{\odot}$ and
$M_{\rm BH}=8.27M_{\odot}$ (where the scatter is much larger). The
data are taken from \cite{stairs04} for neutron stars, and
\cite{mcclintock06} for black holes.}
\label{masses}
\end{figure}

Measuring stellar masses, although a considerable observational
challenge, is not nearly as difficult as measuring their radii (which
is why the equation of state at nuclear densities remains one of the
primary open questions in compact object and nuclear physics
studies). When they occur in binaries, accurate mass determinations
are possible (Figure~\ref{masses}), and for cases where there are two
compact objects in tight orbits, the constraints can be truly
spectacular (PSRB1913+16 \cite{weisberg03}, PSRB1534+12 \cite{stairs02},
and PSRJ0737-3039A,B \cite{lyne04,kramer06}). Neutron star masses are tightly
clustered around the Chandrasekhar mass, at 1.4$M_{\odot}$. Those that
deviate farthest from this value towards high masses are found in Low
Masss X--ray Binaries \cite{tauris06}, where the slow but steady mass
transfer from a companion has produced top--heavy objects (the most
dramatic example being J0751+1807 at 2.1M$_{\odot}$ \cite{nice05}). At
such an extreme, it is possible that conversion to strange (quark)
matter \cite{bodmer71,witten84,fahri84,alcock86} may have occurred (see
Jaikumar et al. for a review \cite{jaikumar06}). Incredibly, it is
possible that observations of superbursts (unusually long Type I
X--ray bursts) may be able to discern if this is actually the case
\cite{page05,cumming06}.

Regarding black holes, the scatter is considerably greater, probably
reflecting the fact that a threshold mass is necessary to create one
in a core--collapse event, but no obvious hard limit exists at high
masses, and a degree of variation is expected \cite{heger00}. The
constraints are also less stringent in most cases, due to a
combination of effects, but nevertheless a boundary can be drawn at
roughly 3M$_{\odot}$, separating the most massive neutron star from
the least massive black hole. This is also close to the maximum mass
allowed for a neutron star in hydrostatic equilibrium in General
Relativity, based only on the condition that the equation of state
must preserve causality \cite{rhoades74}. One would thus expect that
encounters between objects of this type would have mass ratios of
about unity if two neutron stars are involved, and 0.2-0.3 if one of
each interacts.

Although white dwarfs are not usually considered as direct progenitors
of GRBs, one should keep an open mind about their possible role,
particularly in the light of unexpected discoveries made by {\it
Swift} concerning late time variability and the inferences about the
parent population of SGRBs. They may be responsible for the violent
birth of neutron stars through accretion induced collapse or the
merger of two dwarfs in a
binary \cite{mochkovitch89,mochkovitch90,nomoto91,segretain97,fryer99,king01,dessart06},
and it is possible that this remnant could power a
GRB \cite{paczynski86,goodman87,katz96,levan06}. The mass of a white
dwarf is somewhat lower that that of a neutron star, but they are
larger, by a factor of at least $f\sim 10^{3}$, leading to a natural
(dynamical) time scale that is greater than for neutron stars and
black holes by a factor $f^{3/2}\sim 3 \times 10^{4}$.

\subsection{Waltzing Couples}\label{sec:couples}

\begin{figure}
\centering \includegraphics[width=5.0in]{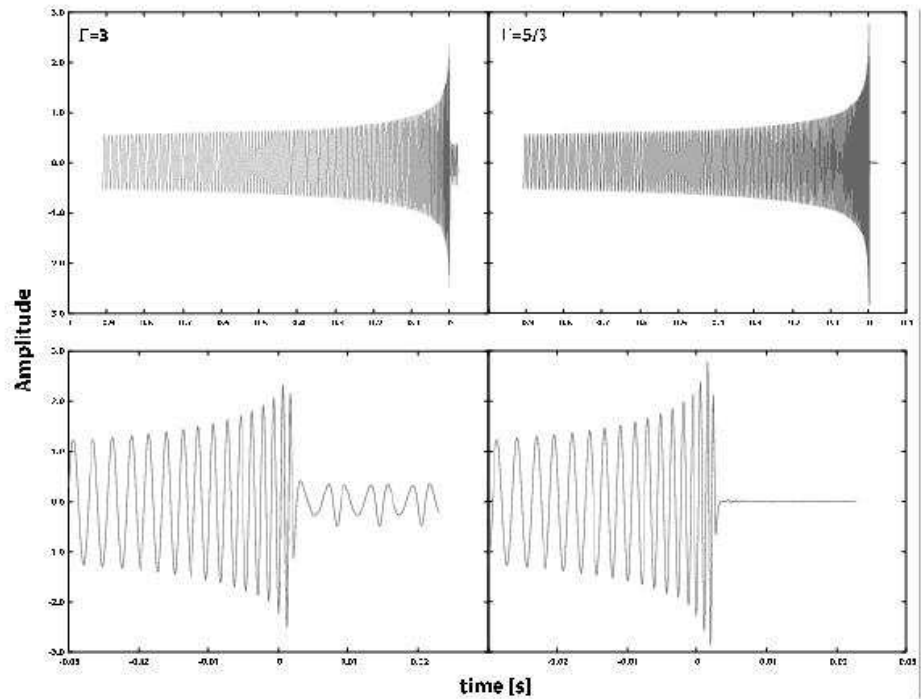}
\caption{Gravitational wave signals for the final stages of spiral-in
of compact binaries consisting of a black hole and a neutron star with
initial mass ratio $q=M_{\rm NS}/M_{\rm BH}=0.3$ and adiabatic index
$\Gamma=3$ (left column \cite{lee01}) and $\Gamma=5/3$ (right column
\cite{lee00}). In this example, the observer is directly above the
orbital plane of the binary along the rotation axis. The rise in
amplitude and frequency shows the characteristic accelerated phase of
decay, followed by a rapid transition to a ring--down after the
violent tidal disruption of the star at $t\approx 0$. In the bottom
panels, details of the merger and tidal disruption waveform are shown,
from whose spectrum one could in principle extract valuable
information regarding the equation of state at neutron star
densities. Note that the neutron star with the more compressible
equation of state is fully and rapidly disrupted, resulting in the
total disappearance of the gravitational wave signal, whereas the one
with a stiff equation of state leaves behind a low-mass orbiting
remnant, which produces persistent emission at late times (see also
Figure~\ref{profiles}).}
\label{gravwaves}
\end{figure}

As a stellar binary revolves, gravitational waves carry away its
energy and angular momentum. The loss rates depend on the system mass,
semi--major axis, $a$, and eccentricity, and rapidly increase as
orbital decay progresses. In the stage where $a \gg R_\ast$, where
$R_\ast$ is the typical stellar radius, the evolution can be computed
to high accuracy in the weak--field limit, using post--Newtonian
expansions for point
masses \cite{kidder92,cutler93,blanchet95,buonanno03,blanchet04,blanchet05,pati00,pati02,will05,blanchet06}. The
characteristic decay time for circular orbits of period $P$ is given
by
\begin{equation}
\frac{t_0}{P}\simeq 10^{5} \left( \frac{P}{1 \;\mbox{\rm s}}\right
)^{5/3},
\end{equation}
which, when applied to the binary pulsars PSR1913+16 \cite{hulse75}
and PSRJ0737-3039 \cite{burgay03}, gives $t_0 \simeq 300$~Myr and
$85$~Myr, respectively. These waveforms, calculated for the last few
minutes of the binary system's life, will be required as templates for
detection with interferometric gravitational wave detectors such as
LIGO \cite{abramowici92} and VIRGO \cite{bradaschia90}. For more
common binaries with much longer periods, the decay rate will be
negligible and the emission may be observed with the space--based
mission LISA. For systems reaching small separations, the decay rate
will increase catastrophically and a collision will ensue, followed by
relaxation to a final, nearly steady state in which, in many cases,
only a black hole will remain. This is the ring--down phase, in which
the emission of gravitational waves and the configuration of the
fossil black hole can be studied through perturbative methods on a Kerr
background (see \cite{kokkotas99} for a review). A composite
gravitational radiation signal is shown in Figure~\ref{gravwaves} as
an example. The decay at early times ($t<0$) was computed using the
weak-field limit approximation for point masses, while the merger
waveform itself ($t>0$) is the result of a three-dimensional numerical
simulation for the disruption of a neutron star by a point mass. The
initially slow decay accelerates (while the amplitude and frequency of
the signal increase) and comes to an abrupt end when the separation is
of the order of the stellar radius, giving a complex signal which
depends on the details of the neutron star structure. In principle,
observation of such radiation and its frequency power spectrum could
help constrain the mass--radius relationship for neutron stars
\cite{faber02}.

The middle ground, during which a violent redistribution of angular
momentum and energy takes place, will produce a burst of
gravitational, neutrino, and electromagnetic energy, perhaps in the
form of a classical GRB \cite{rees99}. There is no approximate,
analytical solution to be found in this regime, and multidimensional
numerical simulations are clearly required to address the behavior of
the system. The process is highly dynamical, with thermodynamics and
emission processes mattering little, and gravitational dynamics ruling
the evolution (with the possible exception of systems containing white
dwarfs).

Even before high--resolution numerical simulations were carried out in
essentially Newtonian gravity, it was assumed that the dynamical
merger would indeed result in the formation of an accretion disk with
enough mass and internal energy to account for the energy budget of a
typical GRB, either through the tidal disruption of the neutron star
in the former, or post--merger collapse of most of the central core in
the latter. Compact binaries were thus seriously suggested as possible
progenitors for GRBs at cosmological distances, with possible maximum
power in excess of 10$^{50}$~erg~s$^{-1}$
\cite{paczynski86,goodman86,eichler89,paczynski91,narayan92,mochkovitch93,jaroszynski93,jaroszynski96}.

The characteristics of the problem, and the qualitatively different
nature of the questions that can be of interest (or addressed) has
dictated a range of approaches when searching for a solution. On the
one hand, there is the question of generation of gravitational waves,
clearly deserving of substantial effort in its own right, and tackled
as such by several researchers. On the other, the generation of an
electromagnetic signal (as a GRB, or another observable transient)
requires the consideration of a different kind of input physics,
mainly in the domain of thermodynamics, the equation of state under a
wide array of conditions, nuclear physics, weak interactions and
magnetohydrodynamics. Efforts in the field have generally proeeded
along these largely orthogonal axes, with little overlap between them
until quite recently, and it is only now that numerical simulations
with any realistic microphysics are beginning to employ the full
machinery of general relativity. It is thus only natural that results
midway along this path have not always been in agreement, and indeed
are not in some aspects even now. The recent and spectacular advances
in the successfull evolution of black hole binaries in numerical
relativity give us today a glimpse of what may be possible to study in
greater detail in the coming years (see the recent overview by
J. Centrella \cite{centrella06}). 

%3.3 
\subsection{First Inroads}\label{sec:early}

To our knowledge, the merger of compact objects (i.e., neutron stars
and/or black holes) was first considered by James Lattimer and David
Schramm \cite{lattimer74,lattimer76}. They initially studied the
dynamics and disruption of a neutron star by a black hole in a bound
system, and the discovery of the Hulse--Taylor pulsar in 1974
\cite{hulse75} soon made it clear that such encounters were an
inevitable, if rare in terms of the lifetime of a typical galaxy,
consequence of binary stellar evolution pathways. The computations
were carried out in full General Relativity, with certain
simplifications to make them tractable and their interest originally
laid in the possibility of such events ejecting neutron--rich material
to the interstellar medium, which might subsequently undergo
r--process nucleosynthesis and contribute to the observed abundances of
heavy elements \cite{burbidge57,wallerstein97}. GRBs had only recently
been made known to the astrophysical community \cite{klebesadel73},
and the likelihood of them being of Galactic origin made Lattimer \&
Schramm consider such events to be unrelated. In their words:
``Although the frequency of these events is too small to be important
as far as currently observed anti--neutrino, $\gamma$--ray, or optical
bursts are concerned, enough neutron star material may be ejected to
be of nucleosynthetic importance.''\cite{lattimer76}. Over thirty
years after their pioneering work, coalescing compact binaries stand
today as one of the leading runners in models for SGRBs. It is
noteworthy that only now are detailed computer simulations of such
collisions beginning to fully consider the effects of General
Relativity in the dynamical merger phase of tidal disruption (John
A. Wheeler's ``tube of toothpaste'' mechanism
\cite{wheeler71,mashhoon73}).

%3.4
\subsection{Dynamical Stability of Close Binaries}\label{sec:dynstab}

\begin{figure} 
\centering \includegraphics[width=4.0in]{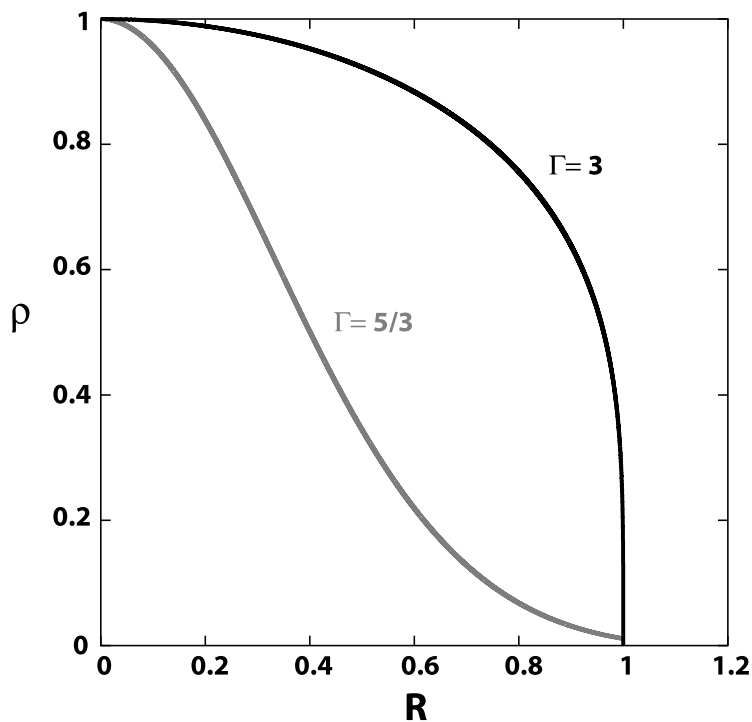} 
\caption{Normalized radial density profiles for spherically symmetric
configurations of a given mass in hydrostatic equilibrium, for
different compressibilities (parametrized through the adiabatic index
$\Gamma$). The varying degree of central condensation makes the
stiffer stars (with higher values of $\Gamma$) more susceptible to
tidal effects when placed in a binary system.}
\label{profiles}
\end{figure}

The stiffness of the nuclear equation of state, or equivalently, the
compressibility, is a parameter derived from microphysics that has a
predominant role in the final outcome of a merger. It is most easily
expressed by writing the pressure--density relation as
\begin{equation}
P \propto \rho^{\Gamma}, 
\end{equation}
where $\Gamma$ is the adiabatic index. Figure~\ref{profiles} shows the
effect of its variation on the structure of a star by plotting the
density profiles of static, spherically symmetric configurations in
hydrostatic equilibrium. For $\Gamma \gg 1$, the result is a
nearly--constant density (a stiff equation of state), while for
$\Gamma \rightarrow 4/3$ (lower values will not produce a stable
solution in Newtonian gravity) the star is highly centrally condensed
(a soft equation of state). It is immediately obvious that when placed
in the vicinity of a massive companion, the star made of more
compressible material will feel tidal effects to a much lesser
degree. The issue of how matter responds to pressure at extremely high
densities is of course far from being solved. For example, while the
stiffness of ordinary matter may be quite high at $\rho\simeq
\rho_{\rm nuc}$, with $\Gamma \simeq 2-3$, the presence of exotic
condensates at supra--nuclear densities could soften the equation of
state considerably \cite{glendenning00}, leading to potentially
different behavior.

\begin{figure}
\centering \includegraphics[width=4.0in]{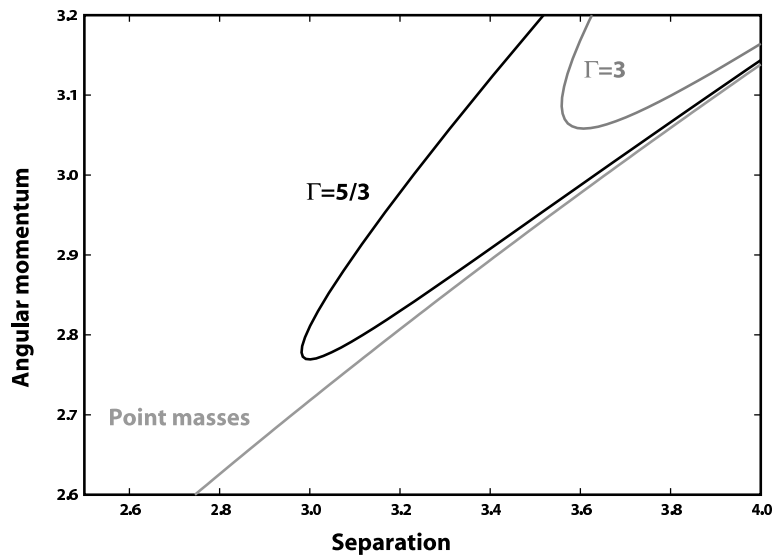}
\caption{Total dimensionless angular momentum as a function of
separation (in units of the neutron star radius) for equilibrium
configurations of BH-NS binaries with mass ratio $q=0.31$ in a
Newtonian approximation. The neutron star is modeled as a
non--spinning tri--axial Roche--Riemann ellipsoid (see \cite{lai93b})
of varying adiabatic index (curves for $\Gamma=3; 5/3$ are
shown). Tidal effects increase the steepness of the effective
interaction potential and produce a minimum in $J_{\rm eq}(R)$,
indicating the onset of a dynamical instability. The effect of the
equation of state is clear, with the less compressible neutron star
encountering instability, which will induce a dynamical plunge, at a
greater separation than its more compressible counterpart. For
reference, the equilibrium curve for point masses is also shown. As
expected, it is a monotonically increasing function of radius and
exhibits none of the turning points characteristic of the appearance
of an unstable branch.}
\label{jvsr}
\end{figure}

The equilibrium configurations for an incompressible fluid, and their
stability, were considered by Chandrasekhar \cite{chandra69}, both for
single and binary stars, in combinations that could represent double
neutron stars or neutron stars in orbit around point masses. This work
was extended later by Lai and collaborators
\cite{lai93a,lai93b,lai94a,lai94b,lai94c} to compressible
configurations by the use of polytropic relations as equations of
state. In a series of papers, they showed that purely Newtonian
hydrodynamic effects can de--stabilize close binaries if the equation
of state is sufficiently stiff ($\Gamma \geq 2$). This can occur
before contact, when the separation is typically $\sim 3$ stellar
radii. The effective binary potential becomes so steep that a
dynamical (i.e., on an orbital time scale) plunge occurs, forcing the
two stars to come together. This is independent of any consideration
of General Relativity, and clearly shows just how powerful the
departures from point--mass behavior can be when the combination of a
finite stellar radius and compressibility are considered. For
illustrative purposes, we show in Figure~\ref{jvsr} the equilibrium
curves of total orbital angular momentum for a Newtonian neutron
star--point mass binary for two compressibilities, as well as the
point mass result (which is simply $J_{\rm eq}(R) \propto
R^{1/2}$). This has important implications for the initiation of mass
transfer from one star to the other, the occurrence of tidal
disruption and the final configuration of the system once it comes
into near contact (see e.g., Chapter 4 in \cite{frank92}).

The compressibility also impacts directly upon the mass radius
relation of a star. For polytropes, we have 
\begin{equation}
R_\ast \propto M^{\Gamma-2 \over 3\Gamma-4}, 
\end{equation}
so the star will expand upon mass loss (or contract) for $\Gamma < 2$
(or $\Gamma > 2$). Thus it can further overflow its Roche lobe once
mass transfer begins, in which case the process itself is unstable
leads to complete disruption, or retreat from the Roche surface and,
in principle, shut off mass transfer in the absence of external
driving (which will not occur, because gravitational radiation
emission will remove orbital angular momentum continuously). In either
case, the details of the dynamics will depend on the global adiabatic
index at high densities.

%3.5
\subsection{Investigating the Merger Phase}\label{sec:merger}

The numerical studies of coalescing binaries began in the 1980s and
1990s, with computations of the gravitational wave emission and
determinations of the stability and dynamics of the system at
separations comparable with the stellar radius under various
simplifying approximations
\cite{oohara89,oohara90,oohara92,nakamura89,nakamura91,zhuge94,zhuge96,shibata92,shibata93,shibata97,wilson96,marronetti98,mathews00,lai93b,rasio92,rasio94,lee95,kluzniak98,rosswog99,rosswog00,rosswog02}
. The thermodynamical evolution of the fluid during merger, with the
potential for electromagnetic energy release and GRBs was also
considered by various groups
\cite{davies94,janka96b,ruffert96,ruffert97,janka99,ruffert01,rosswog03a,rosswog03b}. These
studies used Newtonian or post-Newtonian gravity with additional terms
included in the equations of motion to mimic gravitational radiation
reaction (which is the ultimate agent driving the binary evolution),
and in some cases approximations to the equations of General
Relativity.

\begin{figure}
\centering \includegraphics[width=5.0in]{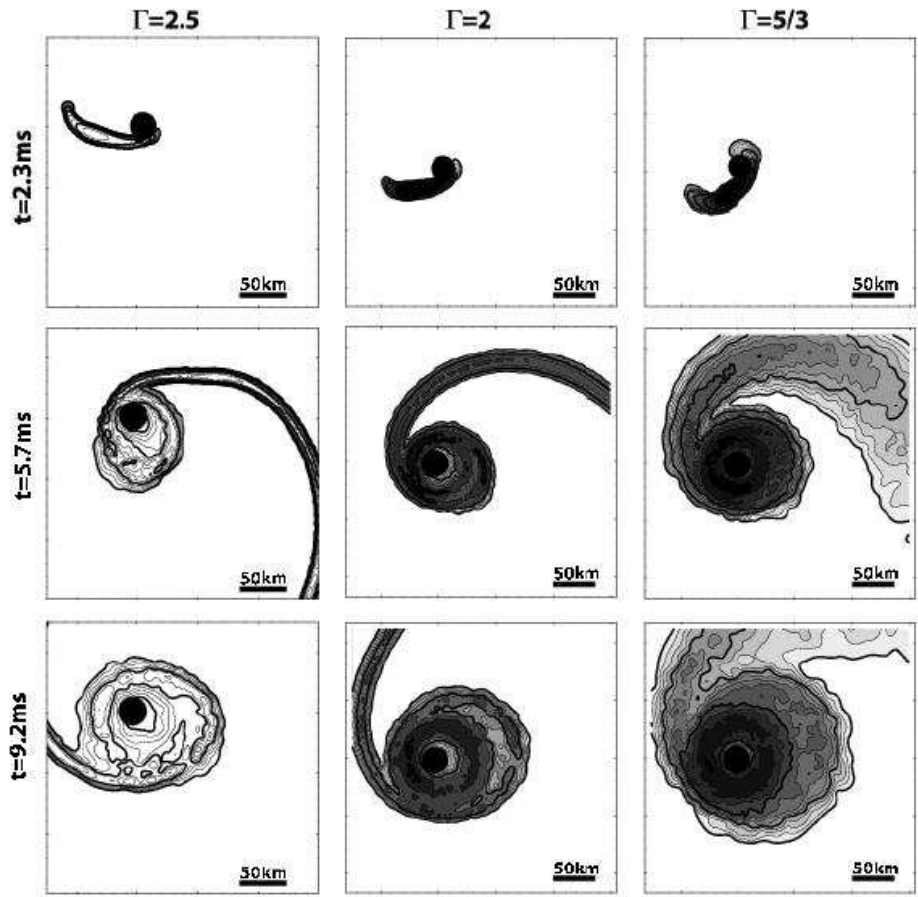}
\caption{The tidal disruption of a neutron star by a black hole in a
binary: Each column (top to bottom in time) depicts the interaction
(computed in three dimensions) for a given value of the adiabatic
index $\Gamma$, going from stiff to soft (left to right). Logarithmic
density contours of density (spaced every quarter decade, with the
lowest one at $\log \rho{\rm [g~cm^{-3}]}=11$), are shown in the
orbital plane, with additional shading for $\Gamma=2; 5/3$. The
distance scale is indicated in each panel, and a black disk of radius
$2R_{\rm g}$ represents the black hole. All three cases initially had
identical mass ratios, $q=0.31$ \cite{lee00,lee01}.}
\label{BHNSmergers}
\end{figure}

Figure~\ref{BHNSmergers} shows the dynamical evolution of a merging
neutron star--black hole binary, during which the neutron star
(modeled as a polytrope) is tidally disrupted and an accretion disk
promptly forms \cite{lee99a,lee99b,lee00,lee01,janka99}. The general
conclusion of this body of work (which includes a mock radiation
reaction force to account for gravitational wave emission) is that
when the orbital separation is of the order of a few stellar radii,
the star becomes greatly deformed due to tidal effects, and the system
becomes dynamically unstable. For stellar components with relatively
stiff equations of state ($\Gamma > 5/2$) the instability is due to a
strong steepening of the effective potential because of tidal effects
(Figure~\ref{profiles}). In the opposite limit, the mass--radius
relationship for the star is such that mass loss leads to an expansion
and Roche lobe overflow, further accelerating the process and leading
to a runaway in which disruption occurs. All of the features
anticipated in the work of Lai and collaborators concerning the
stability of the binary are apparent, and are seen as well in the
numerical simulations of merging binaries performed by Rasio and
collaborators \cite{rasio92,rasio94,rasio95}. The main additional
result in terms of the dynamics is that configurations that are stable
in principle up to Roche lobe overflow are de--stabilized by the mass
transfer process itself, and fully merge as well within a few orbital
periods. It is also quite clear that in the presence of gravitational
radiation reaction (even in its crudest approximation) the system
enters a dynamical infall at small separations that no amount of (even
conservative) mass transfer can revert. Qualitatively, the result is
largely independent of the mass ratio, the assumed initial condition
in terms of neutron star spin (both tidally locked and irrotational
configurations were explored in \cite{lee99a,lee99b,lee00,lee01}) and
the assumed compressibility. Quantitatively, the details are
different, and are primarily reflected in the final disk mass and the
gravitational wave signal (see, e.g., \cite{duez02}).

\begin{figure}
\centering \includegraphics[width=5.0in]{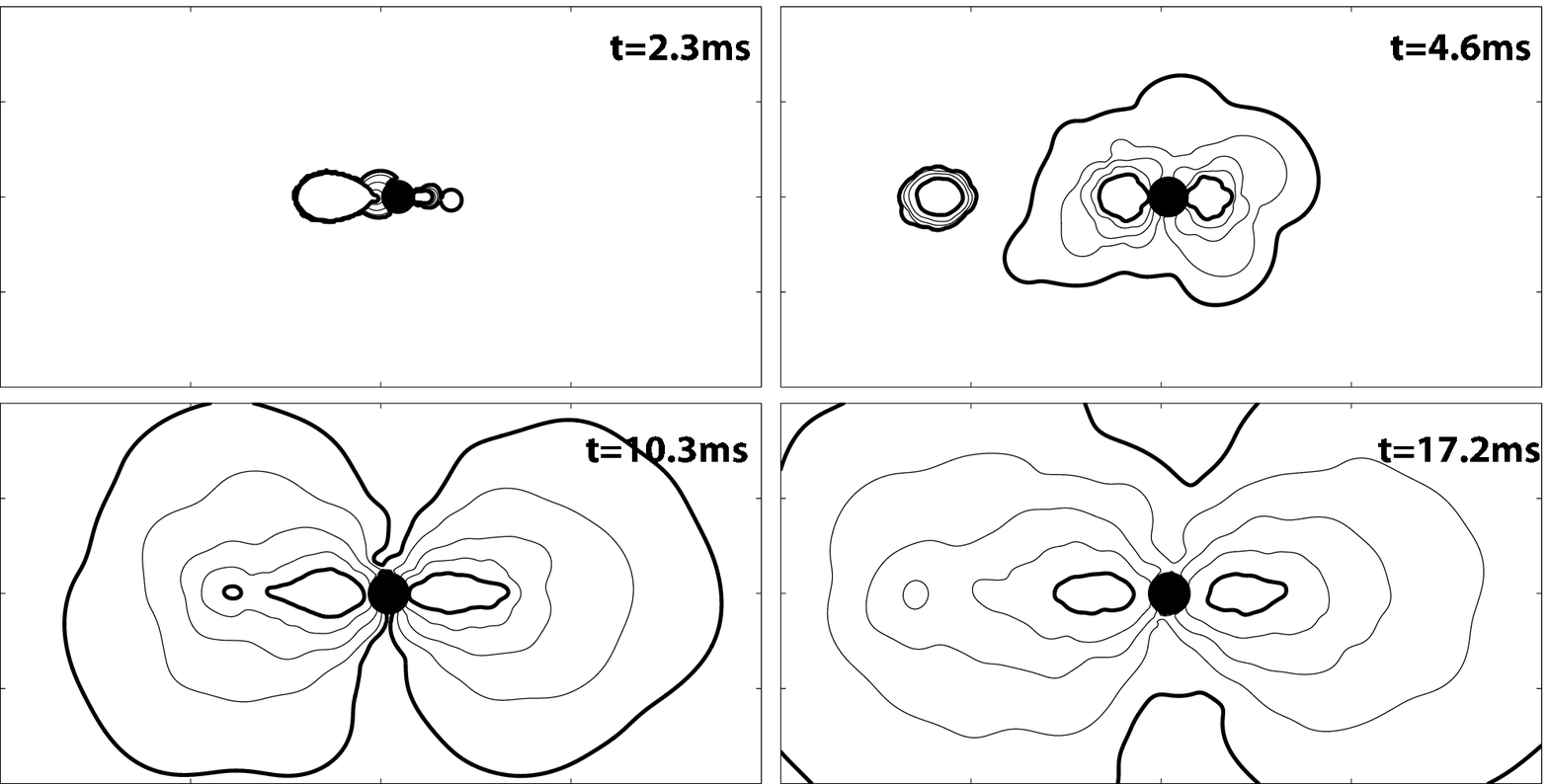}
\caption{During the tidal disruption of a neutron star by a black hole
in a binary, an accretion disk is formed from some of the material
stripped from the neutron star. A sequence of meridional slices
through a three--dimensional simulation is shown, plotting logarithmic
density contours for the density (contours are spaced every decade,
with bold ones at $\log \rho{\rm [g~cm^{-3}]}=$8; 12). In the first
panel, the tidally deformed star is still visible to the left of the
black hole. By 4~ms, the disk is growing around the black hole, and a
slice of the tidal tail (see Figure~\ref{BHNSmergers}) is visible to
the left of the disk. By 17~ms the disk is practically fully formed,
and is draining onto the black hole on a much longer time scale,
determined by angular momentum transport effects. This particular
calculation used $\Gamma=5/3$, and the box shown covers $ 400 \times
200 $~km \cite{lee01}.}
\label{assembly}
\end{figure}

After a few initial orbital periods, a black hole of $M\sim 3-5
M_{\odot}$, surrounded by a thick and hot debris disk ($M_{\rm
disk}\sim 0.01-0.1 M_{\odot}$) approximately $4 \times 10^{7}$~cm
across is all that remains of the initial couple
(Figure~\ref{assembly}).

For double NSs the situation is essentially the same in terms of the
orbital dynamics prior to merger. Synchronization during spiral-in is
impossible because the viscosity in neutron star matter is too small
\cite{bildsten92,kochanek92}, and so they merge with spin frequencies
close to zero, compared with the orbital frequency (such
configurations are termed {\em irrotational}). Contact occurs at
subsonic velocities and a shear layer then develops at the
interface. Modeling this numerically is extremely difficult, because
the layer is unstable at all wavelengths and vortices develop all
along it \cite{faber02b}, possibly amplifying the magnetic field to
extremely large values very quickly \cite{price06}.  The mass ratio in
double NS binaries is quite close to unity, so a fairly symmetrical
remnant would be a natural and expected outcome. However, in general
even small departures from unity can have important consequences in
the inner structure of the final object: the lighter star is disrupted
and spread over the surface of its massive companion, which can remain
largely undisturbed \cite{benz90,rasio94,rosswog00}. The final
configuration now consists of a supra--massive neutron star (i.e., one
with more mass than a cold, non--rotating configuration could support)
surrounded again by a thick shock--heated envelope and hot torus
similar to that formed in BH-NS mergers.  The center of the remnant is
rapidly, and in many cases, differentially rotating, which can have a
profound impact on its subsequent evolution (see below,
\S~\ref{sec:corefate}).

By definition, in either case the merger entails the formation of an
object rotating at the break up limit. Material dynamically stripped
from a star is thus violently ejected by tidal torques through the
outer Lagrange point, removing energy and angular momentum and forming
a large tail. For double neutron star binaries there are two such
structures, and obviously only one in a BH-NS pair. The mass
(typically up to a few tenths of a solar mass), energy and composition
of these structures can be more easily studied by Lagrangian particle
based methods, as they are several thousand kilometers across by the
time the merger is over, and have thus moved off the numerical domain
in grid based codes. By the inherent asymmetry in BH-NS binaries (in
form and mass ratio), the black hole can receive a substantial kick
(up to 10$^{2}$~km~s$^{-1}$ \cite{kluzniak98}) during the process, in
addition to that arising from gravitational wave emission.  Some of
the fluid (as much as a few hundredths of a solar mass) in these flows
is often gravitationally unbound, and could, as originally envisaged
by Lattimer \& Schramm, undergo r--process nucleosynthesis
\cite{rosswog99,frei99}. The rest will eventually return to the
vicinity of the compact object, with possible interesting consequences
for SGRB late time emission (see \S~\ref{sec:prolonged}).

It is important to note that the debris disks are formed in only a few
dynamical time scales (a few milliseconds), which is practically
instantaneous considering their later, more leisurely evolution (which
will be detailed below).

The possibility that a fraction of the neutron star in a BH-NS binary
would survive the initial mass transfer episode and produce cycles of
accretion was explored at long time scales by Davies et
al. \cite{davies05}. This was motivated by the fact that numerical
simulations employing relatively stiff equations of state at high
binary mass ratios \cite{kluzniak98,lee00,rosswog04} showed that this
might actually occur. However, this is most likely not the case,
because: (i) the required equation of state was unrealistically stiff;
and (ii) gravity was essentially computed in a Newtonian
formalism. More recent calculations with pseudo--Newtonian potentials
\cite{rosswog05,leeramirez-ruizgranot05} and the use of General
Relativity \cite{faber06a,faber06b} consistently fail to reproduce
this behavior for a range of compressibilities in the equation of
state. Instead, the neutron star is promptly and fully disrupted soon
after the onset of mass transfer.

%3.6
\subsection{Effects of General Relativity}\label{sec:GReffects}

Performing calculations with full General Relativity is clearly
necessary, but is not an easy task. More importantly, it is not even
clear how to estimate its effects with simple analytical
considerations. Computations of the location of the {\it innermost
stable circular orbit} (ISCO) in black hole systems indicate that
tidal disruption may be avoided completely, with the star plunging
directly beyond the horizon, essentially being accreted whole in a
matter of a millisecond \cite{miller05}. This would preclude the
formation of a GRB lasting 10 to 100 times longer. But we stress that,
just as for stability considerations in the Newtonian case, even if
the location of the ISCO can be a useful guide in some cases, it
cannot accurately describe the dynamical behavior of the system once
mass transfer begins. In pseudo--Newtonian numerical simulations
\cite{rosswog05} and post-Newtonian orbital evolution estimates
\cite{prakash04}, even when a near radial plunge is observed, the star
is frequently distorted enough by tidal forces that long tidal tails
and disk--like structures can form. The outcome is particularly
sensitive to the mass ratio $q=M_{\rm NS}/M_{\rm BH}$ and initial
General Relativity calculations showed that the spin of the black hole
is also very important, with rotating BHs favoring the creation of
disks \cite{rasio05}.  Dynamical calculations of BH-NS systems in
pseudo--Newtonian potentials that mimic General Relativity effects
typically show that for mass ratios $q\simeq 0.25$ it is possible to
form a disk, although of lower mass than previously thought, $M_{\rm
disk}\approx 10^{-2}
M_{\odot}$\cite{rosswog05,leeramirez-ruizgranot05}. A long one--armed
tidal tail is formed as in the Newtonian calculations, with some
material being dynamically ejected from the system. Recently, Faber et
al. \cite{faber06a,faber06b} have presented their first results for
the dynamical merger phase of a BH-NS binary in General Relativity for
low mass ratios ($q \simeq 0.1$). While the the equation of state is
simple (a polytropic relation with $\Gamma=2$ was assumed), the final
outcome is clearly dependent on the complex dynamics of mass
transfer. Even though at one point in the evolution most of the
stellar material lies within the analytically computed ISCO, tidal
torques transfer enough angular momentum to a large fraction of the
fluid, producing an accretion disk with $M_{\rm disk} \simeq 0.1
M_{\odot}$ by the end of the simulation at $t \simeq 70$~ms. The black
hole horizon is a point of no return, but the innermost stable
circular orbit is most definitely {\it not}.

Extending our own work, and further motivated by the first accurate
localizations of SGRB afterglows (starting with GRB050509B,
\cite{bloom06}), we computed the binary evolution of BH-NS systems in
a pseudo--Newtonian potential in an attempt to better estimate the
circumstances under which a disk may form, and its specific
configuration if it does \cite{lee99b,leeramirez-ruizgranot05}. The
most important parameters are the disk mass and size, as these set the
global energy, density, and time scales for its evolution. The
calculations employed the same formalism and code as before
\cite{lee01}, substituting the pseudo-Newtonian potential of
Paczy\'{n}ski \& Wiita \cite{paczynski80}, and using irrotational
binaries as initial conditions (i.e., ones in which the neutron star
spin is negligible). We explored variations in both the mass ratio
($q\sim 0.1-0.3$) and neutron star compressibility ($\Gamma \sim 5/3 -
2$), and found that full tidal disruption takes place as the star
approaches the black hole, with a torus forming around it. It
contains, however, substantially less mass than in the Newtonian case,
because a larger fraction is directly accreted by the black
hole. Additionally, a large portion of the remaining material is on
highly eccentric orbits in the long tidal tails, and, while it will
return to the vicinity of the black hole, it does not constitute
immediately part of the accretion disk. Overall the torus mass is at
least one order of magnitude lower than previously estimated, and
somewhat colder, since it has not been shock--heated to the same
degree by self--interaction of the accretion stream (see
Figure~\ref{BHNSmergers}). Rosswog \cite{rosswog05} has performed high
resolution calculations of this type as well, but with the use of a
realistic equation of state, and finds similar evolution and outcomes.

In the case of merging neutron star pairs, general relativistic
calculations \cite{shibata99,shibata00,shibata03} have progressed even
further, going beyond the adoption of a simple polytropic
pressure--density relation \cite{shibata05,shibata06,oechslin06}. The
outcome is predictably complicated, and depends sensitively on the
total mass of the system and the initial mass ratio. Asymmetric
binaries tend to produce more massive disks than those with identical
components. As mentioned earlier for the case of Newtonian
calculations, a small deviation from a mass ratio of unity is
significant, since it can lead to the nearly complete tidal disruption
of the lighter component, while the more massive star is largely
unaffected. Shibata \& Taniguchi \cite{shibata06} find that the lower
the mass ratio, the more massive the resulting accretion disk,
reaching $M_{\rm disk}=0.03 M_\odot$ for $q=0.7$. Both PSR1913+16 and
PSRJ0737-3039 have $q>0.9$ \cite{weisberg03,burgay03,lyne04,stairs04},
so this mass ratio may be unrealistically low, but when $q \simeq 0.9$
a considerable disk containing $\simeq 10^{-3} M_{\odot}$ still
forms. This can easily account for the $10^{49}-10^{50}$~erg required
for a typical SGRB, simply based on its gravitational binding energy.

We give in Table~\ref{diskmass} a summary of estimates for the disk
properties for different progenitors, based on the calculations of
various groups described above. The characteristic masses, maximum
densities and temperatures are $M_{\rm disk}/M_{\odot} \simeq
10^{-4}-10^{-1}$, $\rho \simeq 10^{10}-10^{12}$g~cm$^{-3}$ and $T
\simeq 10^{10}-10^{11}$K respectively, and provide a starting point
for the more detailed disk evolution calculations we will describe
below in \S~\ref{sec:disks}.

\begin{table}
\caption{\label{diskmass}Remnant disk masses in compact mergers for a
variety of equations of state and various gravity methods: Newtonian
(N); Paczy\'{n}ski \& Wiita (PW); and General Relativity (GR).}
\begin{indented}
\item[]\begin{tabular}{@{}lllll}
\br
Prog. & $M_{\rm disk}/M_{\odot}$ & Gravity \& & Eq. of State 
& Ref. \\
& & Method & & \\
\mr
BH/NS&0.1-0.3&N, SPH & Polytropes& \cite{lee99a,lee99b,lee00,lee01}\\
BH/NS&0.03-0.04&PW, SPH & Polytropes & \cite{leeramirez-ruizgranot05}\\
BH/NS&0.26-0.67&N, Grid & LS\cite{lattimer91}&\cite{janka99}\\
BH/NS&0.001-.01&PW, SPH & Shen\cite{shen98}&\cite{rosswog04,rosswog05}\\
BH/NS&0.001-0.01&GR, SPH & Polytropes&\cite{faber06a,faber06b}\\
\mr 
NS/NS&0.2-0.5&N & SPH, Polytropes&\cite{rasio92,rasio94}\\
NS/NS&0.4&N, SPH & Polytropes&\cite{davies94}\\
NS/NS&0.01-0.25&N, Grid & LS\cite{lattimer91}&\cite{ruffert96,ruffert97,ruffert01}\\
NS/NS&0.25-0.55&N, SPH & LS\cite{lattimer91},
Shen\cite{shen98}&\cite{rosswog99,rosswog00,rosswog02,rosswog03a,rosswog03b,price06}\\
NS/NS&0.05-0.26&GR, SPH & Shen\cite{shen98}&\cite{oechslin06}\\
NS/NS&0.0001-0.01&GR, Grid & APR\cite{akmal98}&\cite{shibata06}\\
\br
\end{tabular}
\end{indented}
\end{table}

\subsection{The Fate of the Central Core}\label{sec:corefate}

In the case of merging neutron stars, the ultimate fate of the central
object is still unresolved, and depends on the maximum mass that a
hot, differentially rotating configuration can support. It has been
known for some time that the maximum allowed mass can be increased by
the effects of rotation (see, e.g., \cite{cook94}), and in particular,
{\em differential} rotation \cite{baumgarte00}. In addition, the
post--merger core is certainly not cold, as shock heating (at least
for a mass ratio of unity) can raise the internal energy
substantially. Shibata \& Taniguchi \cite{shibata06} find that this
threshold for collapse can be estimated as $M_{\rm thres} \simeq
1.35M_{\rm cold}$, where $M_{\rm cold}$ is the corresponding value for
a non--rotating and spherical, cold configuration. Based on the recent
observation of PSRJ0751+1807 \cite{nice05}, $M_{\rm cold} \geq 2.1
M_{\odot}$, so a total mass greater than $\simeq 2.83 M_{\odot}$ is
required for prompt collapse to a black hole. If $M_{\rm cold} <
M_{\rm core}< M_{\rm thres}$, various mechanisms (e.g., emission of
gravitational waves, redistribution of angular momentum, magnetic
fields) could act to dissipate and/or transport energy and angular
momentum, possibly inducing collapse after a delay which could range
from seconds to weeks (this has also been suggested as a two--step
process which would be responsible for a SN/GRB association in the
case of long GRBs \cite{stella98}).

Differential rotation, besides providing additional support against
collapse, could amplify seed magnetic fields in the core to large
values \cite{price06,rosswog03b,rosswogrr04} and turn the it into a
powerful magnetar. Spin--down torques, magnetic field winding and
flaring alone could then in principle power a GRB, independently of
the presence of a torus--like accretion structure
\cite{usov92,kluzniak98}. Note that this would also be one ingredient
of the expected outcome of accretion--induced collapse of a WD, or a
double WD merger. The surrounding environment would be much different,
clearly, and its effect on the generation of a relativistic fireball
would certainly be important (and possibly catastrophic
\cite{fryer99}). The magneto rotational instability (MRI) may also
operate in such an environment, transferring angular momentum to the
debris disk fast enough to keep it from being swallowed by the core
when (and if) it collapses \cite{duez06,shibata06b,duez06b}. An
alternative possibility is that the core may deform into a bar-like
structure, provided the equation of state is sufficiently stiff
\cite{shibata06}. Gravitational torques in the presence of this
asymmetry may then transfer angular momentum to the orbiting debris
and increase the mass of the disk. In either of these last two cases
one would then have a system similar to that occurring in a BH-NS
merger. In general, the details remain highly uncertain, but powering
a GRB is still a possible evolutionary pathway (see Figure 21 in
\cite{shibata06}).

%3.8
\subsection{Colliding Compact Objects}\label{sec:colls}

The collision of unbound compact objects, rather than their merger in
binaries, has not received much attention in the context of
GRBs. Janka \& Ruffert \cite{janka96b} computed the interaction of two
identical neutron stars on parabolic orbits, and found a
post-collision environment so contaminated with baryons that they
concluded it was not viable as a GRB central engine. In addition, they
estimated the event rate within galaxies, and found it much too low to
be of interest. We have reconsidered this problem \cite{lee06b}, and
find that if interactions in Globular Clusters (GC) are taken into account,
the rates can have an important effect on the production of SGRBs, if
the outcome of the collision is a favorable one. Exactly how frequent
a collision takes place depends on the velocity dispersion and the
stellar density in a given GC, as well as the number density of GCs in
a given galaxy \cite{lee06b}.

A similar scenario, the tidal disruption of stars by massive black holes (where the mass
ratio is typically $q=10^{-7}$) was considered as a mechanism for
feeding AGN, thus accounting for their luminosity
\cite{frank78,lacy82}. Simple initial estimates and more detailed
analytical calculations \cite{carter83,rees88} were subsequently
confirmed in their fundamental aspects through numerical
simulations by various groups \cite{evans89,laguna93,ayal00}.

The relevant quantity fixing the strength of the interaction is
\begin{equation}
\eta=\left( \frac{M_{*}}{M_{\rm BH}} \frac{R_{p}^3}{R_{*}^3}
\right)^{1/2},
\label{tidaleta}
\end{equation}
where $M_{*}$ and $M_{\rm BH}$ are the stellar and black hole mass,
$R_{p}$ is the pericenter distance and $R_{*}$ is the stellar
radius. When $\eta \approx 1$ the tidal field is sufficiently strong
to disrupt the star (the energy for this is ultimately extracted from
the orbital motion). For the scenario considered by Rees \cite{rees88},
the star moves essentially in a fixed background metric since $q \ll
1$ (a condition fully exploited in the numerical calculations later
carried out), and the typical pericenter distance is up to one hundred
times the stellar radius. Thus the star does not directly impact its
companion, and the interaction is entirely gravitational. It was found
that approximately half the mass of the star is ejected from the
system (at greater than escape velocity) and the other half remains
bound. Using the Keplerian relation
\begin{equation}
{d\epsilon \over dt} ={1 \over 3}(2\pi G M_{\rm BH})^{2/3}\; t^{-5/3}
\label{eq:Kep}
\end{equation}
for the orbiting material, where $d\epsilon$ is the specific energy,
and the fact that the differential mass distribution with energy,
$dM/d\epsilon$, is practically constant, the rate at which mass
returns to the black hole is given by $dM/dt \propto t^{-5/3}$.

Much can be learned from these calculations, and applied to the case
of colliding compact objects of comparable mass. Much is different as
well, primarily because the mass ratio is of order $q \approx 1-0.1$,
rather than 10$^{-7}$. This means that the pericenter distance is now
comparable to the stellar radius. The initial encounter may thus
resemble more a direct impact, with immediate exchange of mass, and a
corresponding alteration to the mass ratio on a dynamical time
scale. A second effect, as for mergers, is related to the
compressibility of the material, and how the star responds in radius
once mass transfer begins.

\begin{figure}
\centering \includegraphics[width=4.6in]{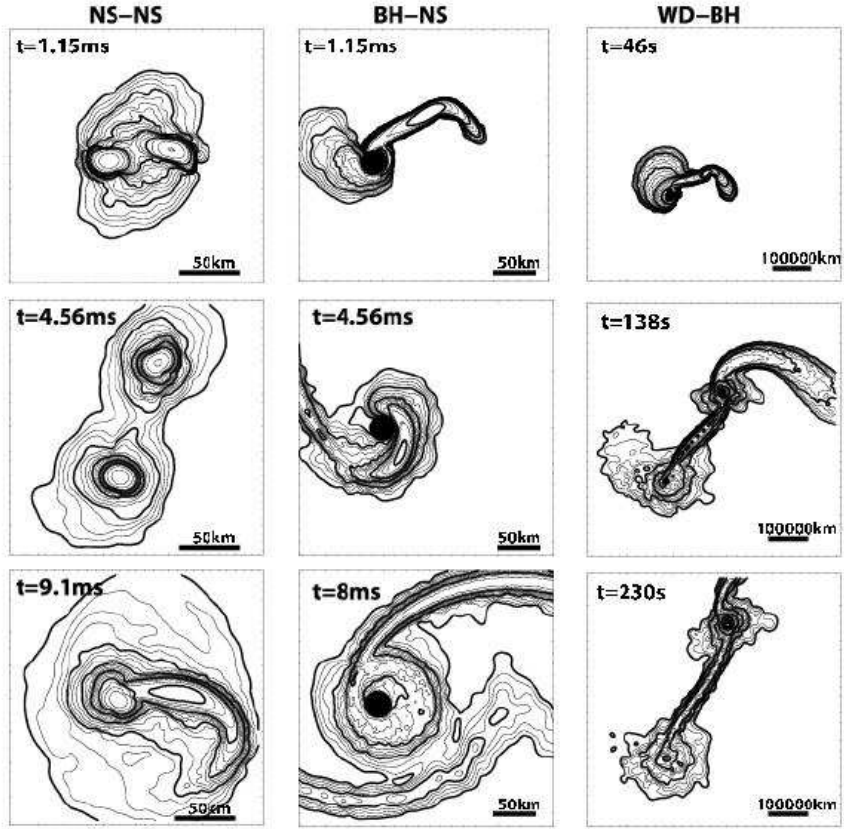}
\caption{The collision of compact objects, and their remnants: The
columns (top to bottom in time) show parabolic encounters between two
neutron stars with masses 1.75M$_{\odot}$ and 1.4M$_{\odot}$ (left), a
neutron star and a black hole with masses 1.4M$_{\odot}$ and
4.5M$_{\odot}$ (middle), and a white dwarf and a massive neutron star
with masses 0.5M$_{\odot}$ and 2.5M$_{\odot}$ (right). All neutron
stars are modeled with adiabatic indices $\Gamma=2$, and the white
dwarf with index $\Gamma=5/3$, appropriate for a low--mass
configuration. The scale is indicated in each panel, with logarithmic
contours of density shown in the orbital plane (spaced every quarter
decade, with the lowest one at $\log \rho{\rm [g~cm^{-3}]}=10; 10; 0$
for each column respectively). The two tidal tails and the accretion
disk are clearly visible in the BH-NS collision. Note the large
difference in spatial and temporal scales for the case involving the
WD. The estimated delay for the WD core to return to the vicinity of
the BH is 1000~s. All encounters have a parameter $\eta=1$, as defined
in equation~(\ref{tidaleta}).}
\label{collisions}
\end{figure}

We have now computed the outcome of parabolic encounters between black
holes, neutron stars and white dwarfs, in various combinations, to
determine the structure of the remnant and investigate its viability
as a possible GRB central engine. The obvious difference between
systems containing a WD and those that do not is that the time scales
are typically longer in the former by a factor $(R_{\rm WD}/R_{\rm
NS})^{3/2}$. It is therefore natural to expect accretion and possible
flares at hundreds, or even thousands of seconds, rather than tens of
seconds after the main interaction. Whether these manifest themselves
in $\gamma$-rays at all is another matter, and one that cannot be
easily tackled numerically yet. The reason is that the WD radius is
typically a thousand times larger than that of its companion, and
hence the dynamical range that needs to be properly resolved is
correspondingly increased.

We find that collisions occur in two stages (see
Figure~\ref{collisions}). During the first periastron passage, an
outward tidal stream is stripped from the low--mass member and ejected
at high velocity. This act binds the surviving core, which rapidly
returns for a second passage and is entirely disrupted, forming a
second tidal tail, as well as an accretion disk around the massive
companion. The duration of the evolution up to this point is
comparable to the final disruption in a binary (about 10~ms). The end
result is the formation of a system similar to that occurring in a
binary merger, in that it contains a massive body (with the same
uncertainties in delay and pathway to collapse to a black hole as
before) surrounded by a debris disk of a few tenths to a few
hundredths of a solar mass. We also considered the case of a white
dwarf colliding with a massive neutron star (or low--mass black
hole). We were unable at this point to follow the entire evolution of
the system, because of the longer time scales involved, but a large
tidal tail is also formed at first encounter, with the surviving core
being bound as a result. Already, however, a disk is apparent, being a
thousand times larger than in the NS/BH encounter. A detailed
dynamical study of this type of collision will be presented elsewhere
\cite{lee06b}.\\

Despite a variety of possible initial configurations, our best guess
for the ultimate prime mover thus consists of a similar astrophysical
object: a stellar mass black hole surrounded by a hot debris torus;
moreover the overall energetics of these various progenitors differ by
at most a few orders of magnitude, the spread reflecting the differing
spin energy in the hole and the different masses left behind in the
orbiting debris\footnote{A possible exception includes the formation
of a rapidly rotating (in some cases very massive) neutron star with
an ultrahigh magnetic field.}.  On the hypothesis that the central
engine involves hyperaccreting black holes, we would obviously expect
the hole mass, the rate at which the gas is supplied to the hole, and
the angular momentum of the hole to be essential parameters. In this
case there is no external agent feeding the accretion disk (excluding
tidal tail interactions, which we will consider in detail in
\S~\ref{sec:prolonged}), and thus the event is over roughly on an
accretion time scale. Any attempt to explain their properties thus
requires that we now consider its evolution and associated energy
release.\\

\newpage

%Section 4
\section{Neutrino Cooled Accretion Flows and Short Gamma-Ray Bursts}\label{sec:disks}
The way in which gas flows onto an accreting object depends largely on
the conditions where it is injected. In this section we address the
general dynamical evolution of neutrino cooled accretion flows under
the conditions expected to occur in SGRB central engines.  There are
five sections: \S~\ref{sec:history} is devoted to general
considerations and a review of previous work addressing neutrino
cooled flows; \S~\ref{sec:thermo} gives a summary of the microphysical
and thermodynamical conditions in the fluid and considers the relevant
equation of state and cooling processes in the disk; detailed
calculations in a pseudo-General Relativistic potential are presented
in \S~\ref{sec:evolution}; the generation of disk--driven winds and
their accompanying outflows is discussed in \S~\ref{sec:winds};
finally, stability considerations are addressed in
\S~\ref{sec:stability}.

\subsection{General Considerations}\label{sec:history}

Despite all of the complications inherent in the study of neutrino
cooled flows (and accretion flows in general), two main ingredients,
easier to qualify, are crucial for understanding the life and death of
the accreting source.  The first is simply the total mass available in
the system at the outset. This fixes the global energy scales, both
thermal and gravitational, and allows for an approximate determination
of the available energy, with
\begin{equation}
E_{\rm gr}\simeq E_{\rm th}\simeq \frac{GM_{\rm BH}M_{\rm
disk}}{R_{\rm disk}}\simeq 10^{52}\left(\frac{M_{\rm BH}}{3M_{\odot}}
\right) \left(\frac{M_{\rm disk}}{0.1 M_{\odot}}\right)
\left({R_{\rm disk} \over 10^7\;{\rm cm}}\right)^{-1} \, {\rm erg}.
\end{equation}
Note that we have estimated the thermal and gravitational energies to
be of the same order. This is different than for classical thin disks,
because the hypercritical accretion flows we are to consider are born
in highly dynamical situations where there is an enormous store of
internal energy (this is true for binary mergers as well as for
collapsar--type scenarios). The fundamental reason is that in either
case, internal energy was the primary source of support prior to the
catastrophic gravitationally--induced collapse or tidal
disruption. 

The second is the nature of the mechanism feeding the inner disk with
mass and energy as a function of time. In the standard picture
following core collapse, normally envisaged for long GRBs
\cite{michel88,chevalier89,macfadyen99}, the inflowing mass supplies
an accretion rate $\dot{M}\sim 10^{-3}M_\odot\;{\rm s^{-1}}$ for a
time $t_{\rm f} \simeq 10$~s and is thereafter fed by injection of
infalling matter at a rate that drops off as $t^{-5/3}$. For the case
of merging compact objects, it is usually assumed that no external
agent feeds the previously assembled disk, and thus that the event is
over roughly on an accretion time scale. This picture can be altered
by the presence of tidal tails formed with material stripped from the
stars during the initial merger phase and ejected on eccentric
orbits. We will address this later. Regardless of the tidal tails, the
physical conditions and instantaneous structure within the flow are
similar in both scenarios, and conclusions drawn from one case may be
used to infer the situation in the other.

The equivalent Eddington luminosity for neutrinos fixes a rough upper
bound for the power output (considering coherent scattering off free
nucleons as a source of opacity) of $L_{{\rm Edd},\nu} \simeq
10^{54}$~erg~s$^{-1}$. Combined with the available energy in the disk,
one finds a characteristic lifetime at this power level of $t_{{\rm
Edd},\nu} \sim t_{\rm burst} \simeq 10$~ms. Two important effects
alter this estimate: (i) the power is in fact much smaller than
$L_{{\rm Edd},\nu}$, by at least two orders of magnitude, and (ii)
accretion onto the black hole drains the disk of mass and energy on a
viscous time scale, $t_{\rm \alpha}$, which can be shorter if the
efficiency of angular momentum transport is high.

Speculation about the structure and evolution of a massive accretion
disk surrounding a new-born black hole began as soon as they were
suggested to be the prime movers of GRB sources
\cite{eichler89,narayan92,mochkovitch93}. Despite the large
theoretical uncertainties, it was clear from the start that indeed
dense, rapidly evolving structures with huge accretion rates (on the
order of one solar mass per second) were to be expected, with natural
time scales (dynamical and viscous) that could in principle account
for GRBs. Whether or not the resulting signal turns into a burst or
not is still a matter of debate, but as Stan Woosley put it
\cite{woosley93} ``If the signature of a $5 M_{\odot}$ black hole
accreting stellar masses of material in a minute is not a gamma--ray
burst, what is it?''. The physical regime at the expected densities
and temperatures places these systems alongside core--collapse
supernovae in the bestiary of astrophysics, and possibly makes them
even more energetic.

The detailed structure of neutrino cooled accretion flows (NDAFs) has
been investigated by a number of groups over the past few
years. Initially only steady-state, azimuthally symmetric and
vertically integrated solutions were considered, with increasing level
of detail mainly in what concerns the thermodynamics and the equation
of state \cite{popham99,narayan01,kohri02,dimatteo02}, but also in the
effects of General Relativity \cite{chen06}. These studies indicated
that the instantaneous power output for plausible accretion rates, if
maintained, could account for the overall energetics of GRBs. Neutrino
cooling, when initiated at the proper temperature, would be an
efficient mechanism for the removal of internal energy, and the disk
would thin rather abruptly. The 1D solutions of Popham et
al. \cite{popham99} were matched in the inner regions quite nicely with
the initial numerical results for collapsar simulations
\cite{macfadyen99}. Narayan et al. \cite{narayan01} considered among
other things, how the radius of matter injection would affect the
flow, and concluded that it needed to be rather small in order for
neutrino cooling to be efficient. Otherwise, a large fraction of the
accreting mass at large radii would, being unable to get rid of its
internal energy, simply be blown away in a large--scale outflow. Kohri
\& Mineshige \cite{kohri02} then considered microphysical effects in
greater detail, both in terms of the equation of state and the cooling
processes. Finally, the effects of General Relativity on the
disk structure have been studied by Chen \& Beloborodov \cite{chen06}
for stationary flows around Schwarzschild black holes.

More recently, time--dependent calculations in one \cite{janiuk04},
two \cite{lee04,lee05a} --assuming azimuthal symmetry-- and three
dimensions \cite{setiawan04,setiawan05} have tackled the question of
time dependence and instabilities. The trade--off between relaxing the
assumption of $\phi$--symmetry or not is reflected in the time
interval than can be reliably modeled at high resolution: $t_{\rm
sim}\simeq 50$~ms in 3D vs. $t_{\rm sim}\simeq 1$~s in 2D. The former
allows for a full exploration of azimuthal modes and instabilities,
while the latter may provide reliable physical estimates on time
scales comparable to the duration of SGRBs.  Both 2D and 3D
simulations have relied on initial conditions taken from 3D binary
merger calculations \cite{lee01,ruffert99}, and as such are a natural
extension of these models. It has become clear from these studies
that the neutrino energy release is rather rapid. As long as the
accretion (or viscous) time scale $t_{\rm acc}\sim M_{\rm
disk}/\dot{M}$ is longer than the cooling time $t_{\rm cool}\sim
E_{\rm int}/L_{\nu}$, the fluid will be able to radiate most of its
internal energy. The power is thus maintained at a fairly constant
level for up to 100~ms, then drops rapidly. However, the density in
the disk remains fairly high for essentially $t_{\rm acc}$, which can
be as long as a few seconds. It is thus in principle able to anchor
strong magnetic fields capable of driving MHD flows. Another important
result is that for high enough accretion rates (or equivalently,
densities), the inner most regions of the disk can become opaque to
neutrinos. This has important consequences for the cooling time scale,
since the internal energy cannot escape immediately, but must diffuse
out. It is also important for the composition, since a negative radial
lepton gradient, akin to that occurring above proto--neutron stars
following core--collapse \cite{janka96}, can be established. Despite
the stabilizing influence of differential rotation \cite{tassoul78},
we have found that the disk can become convectively unstable, which
can have important consequences for the generation of magnetic fields
and nucleosynthetic products in a possible outflow.

General Relativity, as already pointed out, not only plays a crucial
role in the merger dynamics of compact binaries, but may also be
important in determining the evolution and energy output from the
accretion disks thus formed. The disk is small enough that a
substantial amount of material lies in the vicinity of the marginally
stable orbit, where the potential departs significantly from its
Newtonian form. The black hole itself may be rapidly rotating,
depending on the previous binary history, and thus produce inertial
frame--dragging. Finally, the emitted neutrinos are subject to light
bending effects, which can alter the efficiency for annihilation, and
thus the corresponding spectrum. Dynamical simulations of post--merger
disk evolution have been performed without \cite{lee04,lee05a} and
with \cite{setiawan04,setiawan05} account of these differences,
through the use of a pseudo--potential of the form proposed by
Paczy\'{n}ski \& Wiita \cite{paczynski80} and Artemova et
al. \cite{artemova96} for rotating black holes. These effects have
been taken into account in the computation of the energy deposition
rates in the vicinity of the disk
\cite{jaroszynski93,jaroszynski96,salmonson99,asano00,asano01,miller03},
taking realistic configurations derived from compact mergers as input
conditions in recent computations \cite{birkl06}.

\subsection{Physical Conditions and Relevant Processes}\label{sec:thermo}

These disks are typically compact, with the bulk of the mass residing
within $4 \times 10^{7}$~cm of the black hole (which contains
$3-5M_{\odot}$). They are dense ($10^{9}\; {\rm g}\; {\rm cm}^{-3}
\leq \rho \leq 10^{12}\; {\rm g}\; {\rm cm}^{-3}$) and hot ($10^{9}\;
{\rm K} \leq T \leq 10^{11}\; {\rm K}$). The temperature is in fact
high enough that the nuclei are practically fully photodisintegrated
in the inner regions.  Neutronization due to the high densities can
cause the electron fraction to drop substantially below $Y_{e}=1/2$,
and capture of $e^{\pm}$ pairs by free neutrons and protons provides
the bulk of the cooling for $\rho \geq 10^{10}$~g~cm$^{-3}$. The gas
is composed essentially of $\alpha$ particles and free nucleons in
nuclear statistical equilibrium (NSE), $e^{\pm}$ pairs (the relative
importance of positrons is quite sensitive to the degeneracy parameter
$\eta_{e}=\mu_{e}/kT$, where $\mu_{e}$ is the electron chemical
potential), trapped photons and neutrinos of all species ($e^{\pm}$
captures and annihilation produce only electron neutrinos, but
nucleon--nucleon bremsstrahlung and plasmon decays can produce $\mu$
and $\tau$ neutrinos as well). Studying the thermal evolution and the
corresponding energy release thus requires a detailed equation of
state and consideration of weak interaction rates and emission
processes. In addition, the background gravitational field is intense
enough that relativistic effects can come into play, at least in the
inner regions of the flow.  Finally, the whole situation can hardly be
considered to be in a steady state, since it originates, in most
cases, either from the collapse of the Fe-core in a massive star, or
from the merger or collision of two compact objects.

We have thus considered an equation of state in which the total
pressure is given by
\begin{equation}
P=P_{\rm rad}+P_{\rm gas}+P_{\rm e}+P_{\nu},
\end{equation}
where $P_{\nu}$ is the pressure associated with neutrinos, $P_{\rm
rad}=aT^{4}/3$, $P_{\rm gas}=(1+3X_{\rm nuc})\rho k T/(4m_{\rm p})$
and $X_{\rm nuc}$ is the mass fraction of photodisintegrated (and
ideal) nuclei. As an approximate (but quite accurate) solution to the
equations of nuclear statistical equilibrium between free nucleons and
Helium (we do not consider the creation of iron--like nuclei) we use
\begin{equation}
X_{\rm nuc}=22.4 \left( \frac{T}{10^{10}\;{\rm K}}\right)^{9/8}
\left(\frac{\rho}{10^{10}\;{\rm g}{\rm cm}^{-3}} \right)^{-3/4}
\exp \left (-8.2 \; \frac{10^{10}{\rm K}}{T} \right). 
\label{xnuc}
\end{equation}
The degeneracy parameter of Fermions can vary over a large range in
the disk, so it is necessary to use an expression (due to
\cite{bdn96}) that allows for it (but under the condition of
relativity, which translates to $\rho \geq 10^{6}$~g~cm$^{-3}$), namely
\begin{equation}
P_{\rm e}=\frac{1}{12\pi (\hbar c)^{3}} 
\left[ \eta_{e}^{4} + 2 \pi^{2} \eta_{e}^{2} (kT)^{2} 
+ \frac{7}{15} \pi^{4} (kT)^{4}\right],
\end{equation}
and 
\begin{equation}
\frac{\rho Y_{e}}{m_{p}}=n_{-}-n_{+}=\frac{1}{3 \pi^{2} (\hbar c)^{3}}
\left[ \eta_{e}^{3} + \eta_{e} \pi^{2} (kT)^{2} \right]
\end{equation}
for the electron fraction. This formula reduces to the appropriate
limits when the temperature is low ($kT \ll \eta_{e}$, implying $P
\propto \rho^{4/3}$ for a cold Fermi gas) and when it is high ($kT \gg
\eta_{e}$, giving $P \propto T^{4}$ for relativistic $e^{\pm}$
pairs). Note that this automatically takes into account the presence
of pairs in the limit of an ultra relativistic gas, making it
unnecessary to alter the factor $1/3$ appearing in the term for
radiation pressure.

There is an additional effect which alters the equilibrium composition
of the fluid, related to the optical depth. We approximate this by
following Beloborodov~\cite{beloborodov03} and computing the electron
fraction as
\begin{equation}
Y_{e}=\frac{1}{2}+0.487 \left( \frac{Q/2 - \eta_{e}}{kT} \right) 
\end{equation}
in the optically thin and mildly degenerate case
($Q=[m_{n}-m_{p}]c^{2}$), and
\begin{equation}
\frac{1-Y_{e}}{Y_{e}}=\exp \left( \frac{\eta_{e}-Q}{kT} \right)
\end{equation}
in the optically thick case (this follows from setting the neutrino
chemical potential equal to zero in the reaction $e + p
\leftrightarrow n$ and using Maxwell--Boltzmann statistics for the
non--degenerate nucleons). We use an interpolated fit weighted with
factors involving $\exp (-\tau_{\nu})$ to smoothly transition from one
regime to the other.

We have included neutrino energy losses from $e^{\pm}$ pair captures
onto free nucleons by the use of tables \cite{langanke01} and
$e^{\pm}$ pair annihilation from fitting functions
\cite{itoh96}. Additionally plasmon decays \cite{ruffert96} and
nucleon-nucleon bremsstrahlung \cite{hannestad98} losses are
considered, although their contribution to the total luminosity is
negligible. The resulting cooling rates are accurate over a wide range
of temperature and density, and are vary consistently with the
changing composition and conditions within the flow (particularly for
the two most important contributions of $e^{\pm}$ capture and
annihilation).

We approximate the optical depth of the material to neutrinos by
computing the cross-section for scattering off free nucleons and
$\alpha$ particles heavy nucleons. These provide the dominant
contribution, and are given by
\begin{equation}
\sigma_{N}=\frac{1}{4} \sigma_{0} \left( \frac{E_{\nu}}{m_{e}c^{2}}
\right)^{2},
\end{equation}
and
\begin{equation}
\sigma_{\alpha}=\sigma_{0} \left( \frac{E_{\nu}}{m_{e}c^{2}}
\right)^{2} [4 \sin^{2} \theta_{W}]^{2},
\end{equation}
where $\sigma_{0}=1.76 \times 10^{-44}$~cm$^{-2}$, $\theta_{W}$ is the
Weinberg angle and $E_{\nu}$ is the energy of the neutrinos (roughly
equal to the Fermi energy $E_{\rm F}$ of the mildly degenerate
electrons). The optical depth is estimated as $\tau_{\nu}=H/l_{\nu}$,
where $H$ is the local disk scale height and $l_{\nu}$ is the neutrino
mean free path. We find that in the inner regions, $\tau_{\nu}\sim
10^{2}$ in high mass disks.

The cooling is then suppressed by a factor $\exp (-\tau_{\nu})$ to
mimic the effects of diffusion, so the total neutrino luminosity is
\begin{equation}
L_{\nu}=\int \rho^{-1} (\dot{q}_{\rm ff} + \dot{q}_{\rm plasmon} +
\dot{q}_{\rm pair} + \dot{q}_{\rm cap}) \exp (-\tau_{\nu}) dm.
\end{equation}
The transition from a transparent to an opaque fluid occurs at
$\tau_{\nu}\approx 1$, which corresponds roughly to $\rho \approx
10^{11}$g~cm$^{-3}$ (this happens largely in the high--mass disks,
with their low--mass counterparts remaining mostly
transparent). Photodisintegration cools the gas at a rate
$\dot{q}_{\rm phot}=6.8 \times 10^{18} (dX_{\rm
nuc}/dt)$~erg~s$^{-1}$~cm$^{-3}$, which is included in the energy
equation. Finally, angular momentum is transported through the disk by
shear stresses, and we use the $\alpha$ prescription to vary the
magnitude of the effective viscosity (all terms in the stress tensor
are included in the momentum and energy equations \cite{lee02}).

Clearly, although we have attempted to give an accurate thermodynamic
description of the gas, several approximations remain. In the first
place, we have assumed that all reactions considered reach
equilibrium. This is quite accurate for photodisintegration, but not
necessarily true for weak interactions. Second, our prescription for
neutrino optical depth effects is obviously a far cry from the proper
Boltzmann transport used in one--dimensional SN calculations (or even
diffusion), but we believe it qualitatively captures the nature of the
transparent--opaque transition, at approximately the correct
density. Finally, we have {\em not} considered energy deposition from
neutrinos back to the gas through absorption or incoherent
scattering. This could be quite important in the outer regions of the
disk (see \S~\ref{sec:pathways} and \ref{sec:winds}) and relevant for
the baryon loading of a possible GRB--producing outflow.

\begin{figure}
\centering \includegraphics[width=6.0in]{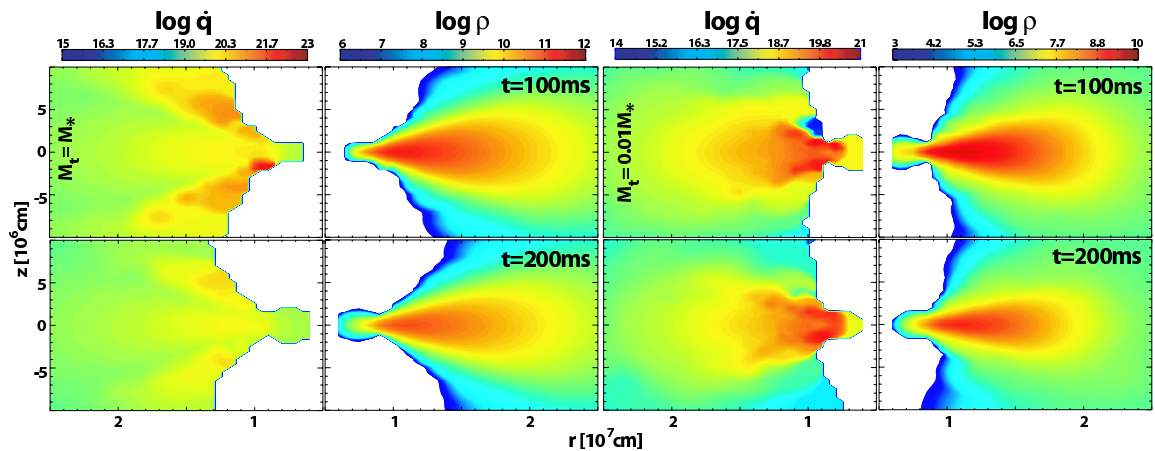}
\caption{Meridional structure and evolution of hyperaccreting flows
around a Schwarszchild black hole, computed with the pseudo--Newtonian
potential of Paczy\'{n}ski \& Wiita \cite{paczynski80}. The density,
$\rho$ (g~cm$^{-3}$) and cooling rate through neutrinos $\dot{q}$
(erg~g$^{-1}$~s$^{-1}$) are plotted 100 and 200~ms after the start of
the calculation, for disks with an initial mass M$_{*}$ (left panels)
and 0.01M$_{*}$ (right panels), where $M_{*}=0.3M_{\odot}$. The
$\alpha$ viscosity coefficient is 10$^{-2}$. Note the different bounds
for the color scale between the two cases, in density as well as
cooling.}
\label{rhodudt}
\end{figure}

\subsection{Dynamical Evolution in a Pseudo-GR Potential}
\label{sec:evolution}

Here we present results for our own 2D calculations in azimuthal
symmetry for the evolution of hypercritical accretion flows around a
black hole following compact object mergers. The initial conditions
are essentially the same as those in \cite{lee05a}, which were taken
from 3D simulations of black hole--neutron star mergers, and are
evolved in the pseudo--Newtonian potential of Pac\'{z}y\'{n}ski \&
Wiita \cite{paczynski80}, which reproduces the existence of a
marginally stable orbit at $r_{\rm ms}=6GM_{\rm BH}/c^{2}$ for
non--spinning black holes. The standard disk mass initially is
$M_{\ast}=0.3M_{\odot}$. This is probably close to the highest mass
that will be found in such disks, and we have also investigated
conditions in lighter disks, with $M_{\rm disk}=0.1, 0.01 M_{*}$. The
microphysics included is the same as in our previous calculations (and
described above in \S~\ref{sec:thermo}) thus allowing for a clear
identification of the effects of strong field gravity.

\begin{figure}
\centering \includegraphics[width=5.0in]{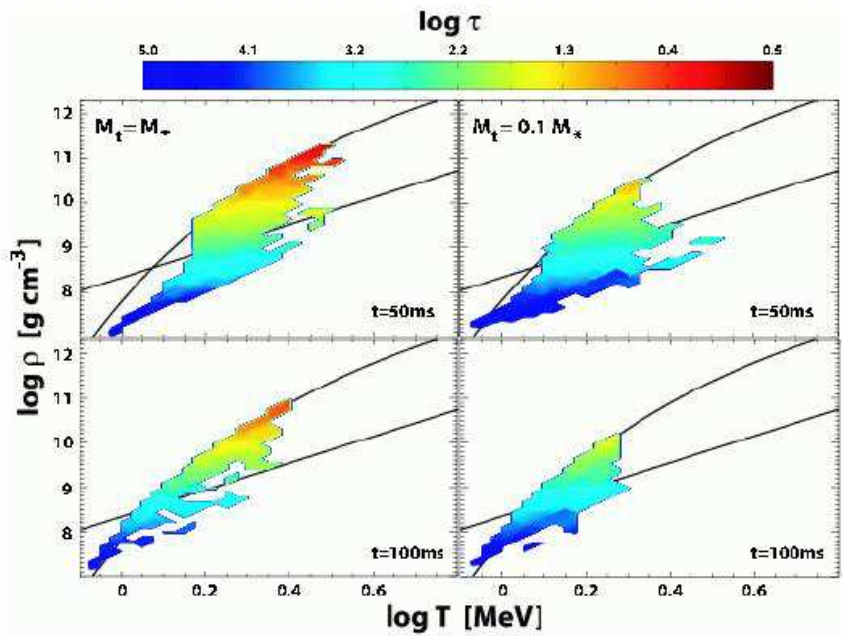}
\caption{Color--coded optical depth in the $\rho$--$T$ plane at 50 and
100~ms for disks with initial mass M$_{*}$ (left panels) and
0.1M$_{*}$ (right panels). The curved solid line marks the
photodisintegration threshold between $\alpha$ particles and free
nucleons (by mass), and the straight solid line marks the electron
degeneracy threshold, given by $kT=7.7 (\rho/ 10^{11}
\mbox{g~cm$^{-3}$})^{1/3}$~MeV. As the disk drains into the black
hole, it becomes more transparent, cooler and less degenerate. The
$\alpha$ viscosity coefficient is 10$^{-2}$.}
\label{rhotau}
\end{figure}

The spatial structure of the disk is shown in Figure~\ref{rhodudt},
where the cooling is also indicated. The overall structure is similar
to that seen in the Newtonian simulations, but the maximum density is
reduced, and only in the high--mass disks and at early times is the
presence of an optically thick region (along the equator, and close to
the black hole) evident. The thermodynamic structure is plotted in
Figure~\ref{rhotau}, where the regions in the density--temperature
plane occupied by the disk are shown, along with the optical
depth. The flow remains almost entirely photo dissociated in the inner
regions, thus giving rise to the intense neutrino emission from pair
capture onto free nucleons. It is also clear that the electron gas is
neither ideal nor fully degenerate, and finite--degeneracy effects
must be considered to compute its evolution properly.

\begin{figure}
\centering \includegraphics[width=3.0in]{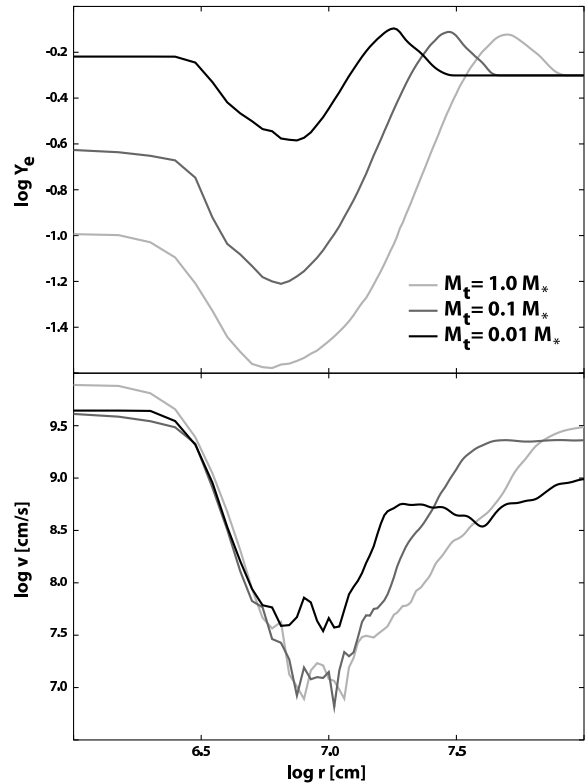}
\caption{Top: Height--integrated radial profiles of the electron
fraction $Y_{e}$ for disks with varying mass in a pseudo--Newtonian
relativistic potential. Bottom: Meridional velocity magnitude,
$(v_{r}^{2}+v_{z}^{2})^{1/2}$, for the same calculation. The
marginally stable orbit is located at $\log (R{\rm[cm]})\approx 6.6$. In the
outer disk, the gas consists mostly of $\alpha$ particles, and
$Y_{e}=1/2$. As the fluid nears the black hole it rises slightly, then
becomes smaller as neutronization takes place. The trend is then reversed and
$Y_{e}$ increases again. The minimum is a sensitive function of the
disk mass (or equivalently, the density).}
\label{Ye}
\end{figure}

In our Newtonian calculations we found that neutronization was
important in the inner regions of the flow. This is still the case for
the relativistic calculations, although the effect is reduced,
particularly for low disk masses. Figure~\ref{Ye} shows one snapshot
of height--integrated radial profiles of the electron fraction in the
disk for different disk masses. The most massive disk achieves higher
densities and neutronization is substantially enhanced. The negative
gradient in $Y_{e}(R)$ at small radii was also observed in Newtonian
calculations, where it was a consequence of the transition to the
optically thick regime. In the relativistic case it occurs even for
low--mass disks, and is related to the appearance of a plunging
region, where the radial velocity rapidly increases, once the fluid
reaches the marginally stable orbit (see the bottom panel in
Figure~\ref{Ye}). The net effect is also to produce an inversion of
the radial lepton gradient, indicating that convection is possible
(differential rotation has a stabilizing effect on this, and the full
Solberg--Hoiland criterion must be considered, see below).

\begin{figure}
\centering \includegraphics[width=3.0in]{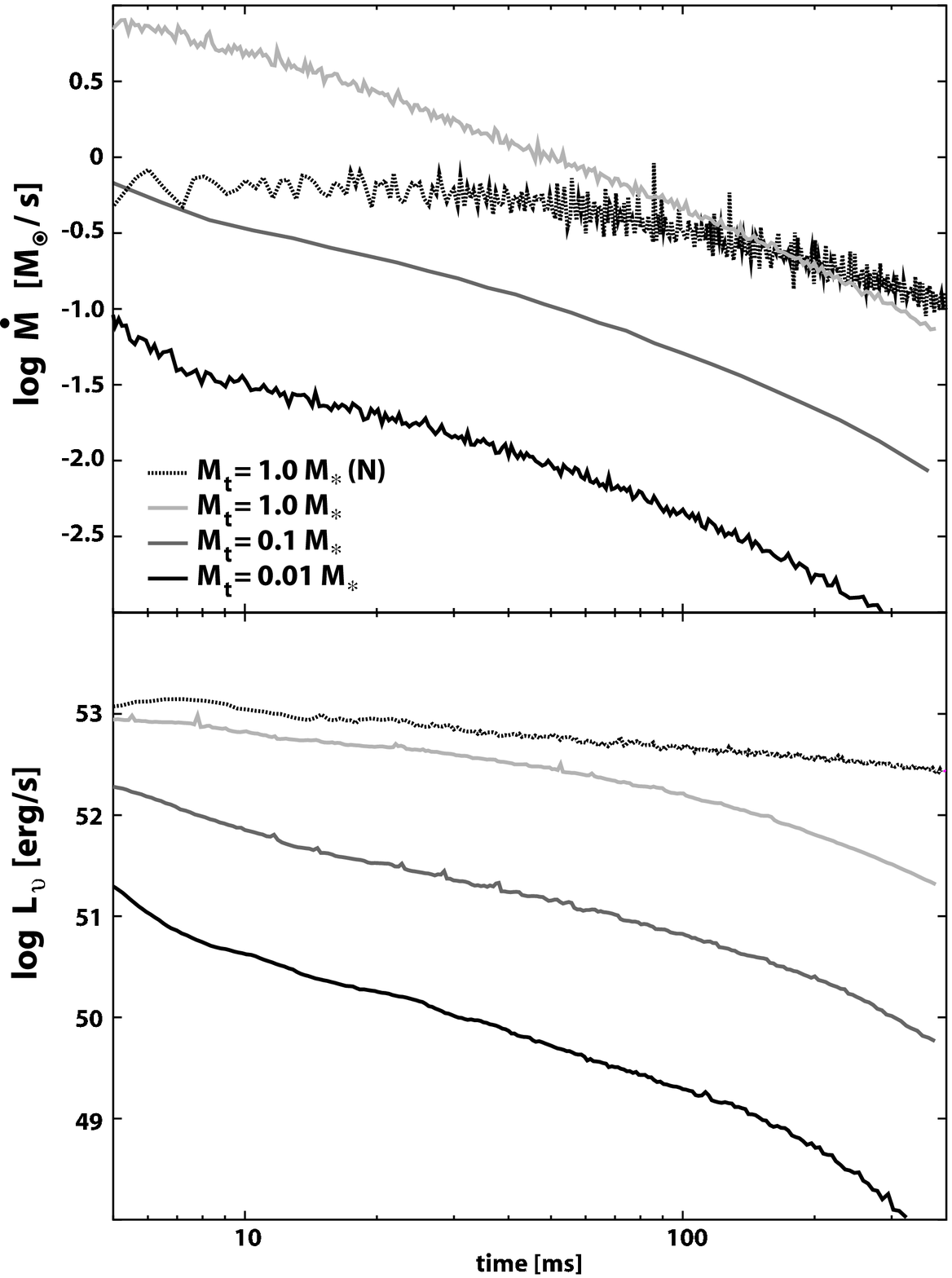}
\caption{The neutrino luminosity decreases as the disk drains into the
black hole on a viscous time scale. The top (bottom) panel shows
$\dot{M}$ ($L_{\nu}$) as a function of time for calculations with
three different initial disk masses, covering two orders of magnitude
($M_{*}=0.3M_{\odot}$) in the pseudo--Newtonian potential of
Pac\'{z}y\'{n}ski \& Wiita. For reference, the curve labeled [N] is
the result of a calculation evolved in a Newtonian $1/R$
potential. The viscosity parameter is set to $\alpha=10^{-2}$ in all
cases.}
\label{Lnu}
\end{figure}

The accretion rate and neutrino luminosity are plotted as functions of
time in Figure~\ref{Lnu} for $\alpha=10^{-2}$ and disk masses covering
two orders of magnitude. For comparison, the result for high disk mass
in the Newtonian case is also shown. Clearly the interval during which
a large luminosity is maintained is lower in the relativistic case,
simply because the disk drains more rapidly onto the black hole. The
accretion rates scale roughly linearly with the disk mass, whereas the
power output in neutrinos shows a slightly steeper dependence (it also
depends sensitively on the temperature), particularly evident at late
times.

One possibility for the driving of a relativistic outflow and the
powering of a GRB is $\nu\overline{\nu}$ annihilation, and another is
the often--quoted magnetic outflow \cite{blandford77}. If we assume a
1\% efficiency for the first case at $L_{\nu}=10^{53}$~erg~s$^{-1}$
(which scales with the square of the neutrino luminosity), and using
the present set of calculations, we infer a scaling with disk mass for
the total energy release as
\begin{equation}
E_{\nu\overline{\nu}}=2 \times 10^{48}\left( \frac{M_{\rm disk}}
{0.03 M_{\odot}}\right)^{2} \mbox{erg}. 
\end{equation} 
This has the same dependence as in the Newtonian case, but with a
reduced intensity, by about a factor of five. For the magnetic
scenario, the total energy release (assuming the magnetic field energy
density is in equipartition with the internal energy density $\rho
c_{s}^{2}$) scales as
\begin{equation}
E_{\rm BZ}= 2 \times 10^{49} \left( \frac{M_{\rm disk}}{0.03
M_{\odot}}\right) \left( \frac{\alpha}{10^{-1}}\right)^{-0.55} \mbox{erg},
\end{equation} 
again in the same way as in the Newtonian case, but also reduced (this
time by about a factor twenty).

\begin{figure}
\centering \includegraphics[width=4.0in]{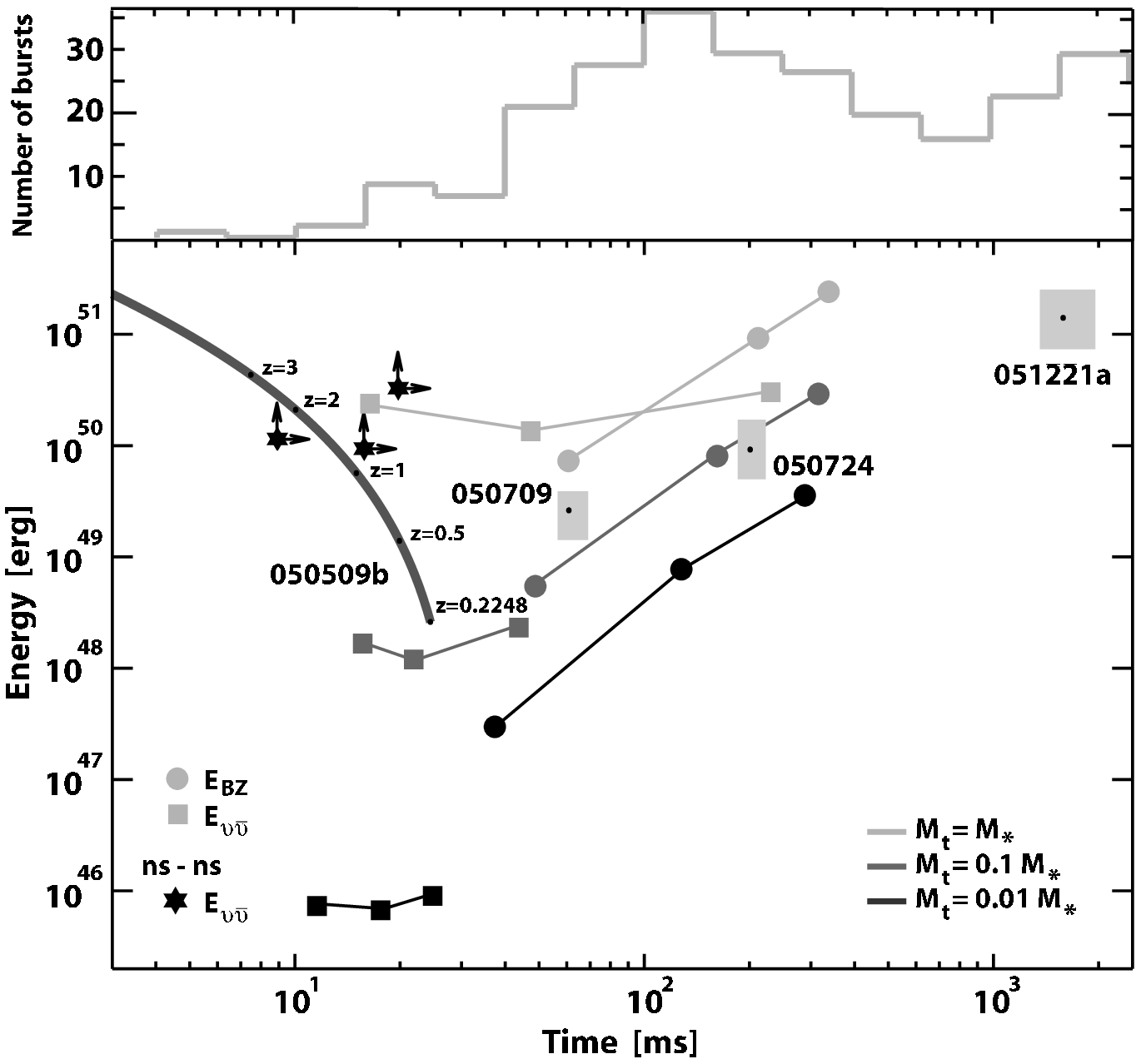}
\caption{Top: Histogram of observed burst durations, taken from
\cite{paciesas99}. Bottom: comparison of energy vs. duration for
GRB959509B (for which the dependence on redshift is indicated by the
gray line), GRB050709, GRB050724, and GRB051221A with estimates from
compact binary mergers. The connected squares and circles show the
total isotropic energy release (assuming collimation into a solid
angle $\Omega=4\pi/10)$ and duration ($t_{90}$) for
$\nu\overline{\nu}$ annihilation and Blandford--Znajek powered events,
respectively, as computed from our 2D disk evolution calculations in a
pseudo--Newtonian potential \cite{paczynski80}. The range in initial
disk masses covers two orders of magnitude, and the effective
viscosities are (from left to right), $\alpha=10^{-1}, 10^{-2},
10^{-3}$. The stars are estimates from $\nu\overline{\nu}$--driven
outflows in double neutron star mergers \cite{rosswog03,rosswogrr03}.}
\label{lum}
\end{figure}

A comparison of these scalings and the actual data for the prompt
emission of four SGRBs detected in 2005 is shown in
Figure~\ref{lum}. Clearly neutrinos are probably not the best way to
produce a burst lasting more than a few tens of milliseconds, unless
the disk mass is quite high. The energy release is somewhat
insensitive to the burst duration in this case. On the other hand,
magnetic energy extraction seems like a plausible mechanism, and
would, at a given disk mass (if this is a standard quantity in any
sense), reproduce the trend that longer bursts seem to have a greater
fluence. It is also apparent that a reduced disk mass is less of a
problem for magnetic mechanisms than for neutrinos as regards the
energetics.

\subsection{Disk Winds and Outflows}\label{sec:winds}

The outer regions and surface of the accretion flow are subject to
various effects that can unbind substantial quantities of mass and
drive powerful winds, with important implications for nucleosynthesis
\cite{qian96,pruet03,pruet04}, GRBs and supernovae \cite{kohri05} and
the possible collimation of relativistic outflows
\cite{rosswogrr02,levinson00,rosswogrr03,aloy05}. One of these is
energy deposition by neutrino heating, another is thermonuclear
burning and the corresponding energy release. The former has been
studied extensively by a number of authors, in the supernova as well
as GRB contexts. Dynamical disk evolution calculations do not always
include this explicitly (our own simulations fall in this category),
but reasonable estimates can nevertheless be derived in some cases. We
include the latter in its crudest approximation, considering only the
transition from $\alpha$ particles to free nucleons, whose mass
fraction is given in equation~(\ref{xnuc}).

\begin{figure}
\centering \includegraphics[width=4.0in]{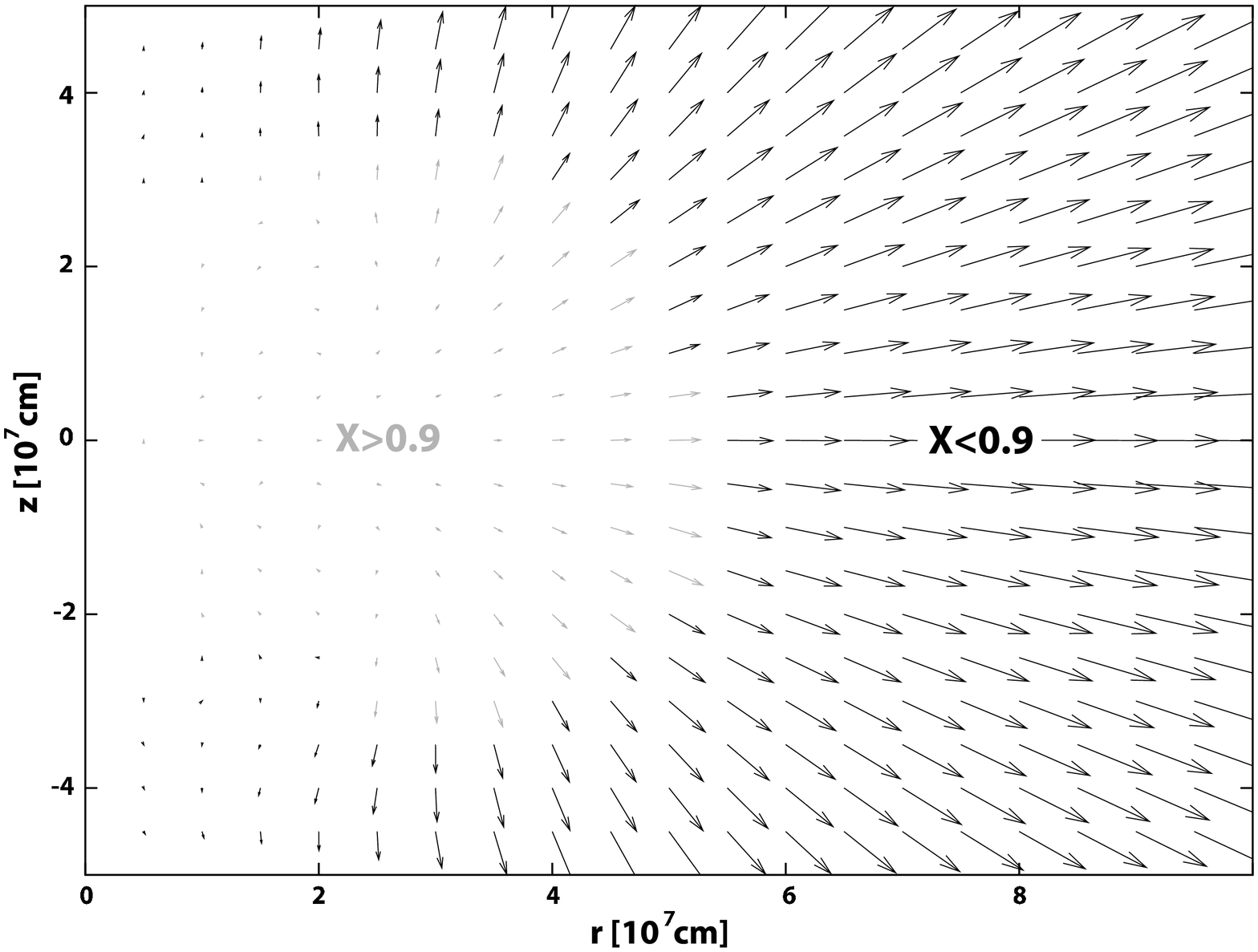}
\caption{The energy released from He synthesis in the outer regions of
the hypercritical accretion flow surrounding the black hole is
sufficient to unbind the gas and produce a strong wind. The meridional
velocity field shows an accelerated outflow in the regions where the
mass fraction $X$ of free nucleons drops appreciably (a threshold of
0.9 has been set in the plot), indicating the creation of $\alpha$
particles.}
\label{xwind}
\end{figure}

Now the gravitational binding energy per nucleon is $E_{\rm gr}\simeq
GM_{\rm BH}m_{p}/R \simeq 5 (10^{8}\mbox{cm}/R) \mbox{MeV}$, and the
production of one $\alpha$ particle releases 7.7~MeV/nucleon into the
flow, so enough energy is available to produce a large--scale wind
where the binding energy becomes sufficiently small. This is in fact
seen in our calculations (Figure~\ref{xwind}), where the outward
(mostly vertical) velocity is $v_{\rm w}\simeq c/10$ and $\dot{M}_{\rm
wind}\simeq 2-5 \times 10^{-2}M_{\odot}$~s$^{-1}$. The corresponding
power is $L_{\rm wind}=\dot{M}_{\rm wind} v_{\rm wind}^{2}/2 \simeq
1-3 \times 10^{50}$~erg~s$^{-1}$ for a disk with $M_{\rm disk} \simeq
0.3M_{\odot}$ initially. The mass outflow from a neutrino--driven wind
in spherical symmetry was estimated by Qian \& Woosley \cite{qian96}
as
\begin{equation} 
\dot{M}_{\rm wind} \simeq 5 \times 10^{-4} \left(
\frac{L_{\nu}}{10^{52} \mbox{erg~s$^{-1}$}}\right)^{5/3}
\mbox{M$_{\odot}$~s$^{-1}$},
\end{equation}
 so at the disk neutrino luminosity of $L_{\nu}\simeq
10^{53}$~erg~s$^{-1}$, it is interesting to note that the power of
each mechanism is comparable (Table~\ref{winds}).

\begin{table}
\caption{\label{winds}Energetics of winds.}
\begin{indented}
\item[]\begin{tabular}{@{}lcccccc}
\br
$\alpha$
& $\dot{M}_{\rm wind}$ 
& $\rho$ 
& $v_{\rm w}/c$
& $L_{\rm wind}$ 
& $M_{\rm disk}$
& Potential$^{\rm a}$ \\
& 10$^{-2}$M$_{\odot}$~s$^{-1}$ 
& g~cm$^{-3}$ & & 10$^{50}$erg~s$^{-1}$ 
& M$_{\odot}$ & \\
\mr
0.1 & 2.5 & $10^{6}$ & 0.08 & 1.5 & 0.3 & N \\
0.01 & 2.5 & $4 \times 10^{5}$ & 0.1 & 2.25 & 0.3 & N \\
0.01 & 1.5 & $4 \times 10^{5}$ & 0.1 & 1.3 & 0.3 & PW$^{\rm a}$ \\
0.001 & 1 & $2 \times 10^{5}$ & 0.1 & 0.9 & 0.3 & N \\
\mr
0.1 & 0.015 & $4 \times 10^{4}$ & 0.08 & 0.009 & 0.003 & N \\
0.01 & 0.04 & $3 \times 10^{4}$ & 0.05 & 0.0094 & 0.003 & N \\
0.01 & 0.025 & $3 \times 10^{4}$ & 0.03 & 0.009 & 0.003 & PW \\
0.001 & 0.02 & $2 \times 10^{4}$ & 0.03 & 0.002 & 0.003 & N \\
\br
\end{tabular}
\item[] $^{\rm a}$ [N]: Newtonian; [PW]: Pac\'{z}y{n}ski \& Wiita \cite{paczynski80}
\end{indented}
\end{table}

\subsection{Stability and Convection}\label{sec:stability}

An important question which cannot be addressed fully through
steady-state calculations is that of stability. Several factors need
to be considered, we address each in turn. The ``runaway radial''
instability discovered by Abramowicz et al. \cite{abramowicz83}
applies to configurations with constant specific angular momentum as a
function of radius. This is fully an effect of General Relativity,
dependent on the existence of an inner Roche lobe for disk accretion,
akin to the $L_{1}$ Lagrange point in binary stellar
evolution. Essentially, if the fluid overflows this lobe, two things
occur: (i) the black hole mass increases, pushing this critical point
outward and (ii) material of the same angular momentum value is now in
a position to be accreted. The effect is a runaway, in which the disk
is accreted in a matter of a few dynamical time scales. It turn out
that even a small positive gradient of the specific angular momentum,
$\ell(R)$, is enough to stabilize this condition, because matter with
greater angular momentum will not be easily accreted, and this damps
the instability quite rapidly. However, none of the post--merger
accretion disks obtained through merger calculations have angular
momentum distributions that are even remotely close to being constant,
and so this turns out to be most likely irrelevant for the evolution
of GRB central engines. Likewise, the self--gravity of the disk is
probably unimportant, particularly if one considers the results for
disk masses from General Relativistic calculations to be more
realistic. This has been raised as a possibility concerning the
late--time X--ray flares in a few of the SGRBs discovered recently,
and will be addressed in the following section.

As for the standard photon--cooled thin disk solution, neutrino cooled
flows were shown to be thermally unstable if radiation pressure
dominates \cite{narayan01,kohri02}. Our numerical solutions which
fully resolve the vertical as well as radial structure of the disk in
time--dependent fashion show this to be the case as well. The
difference is that the disks' response to this condition is a
substantial thickening in the opaque regions, because of (i) the rise
in pressure and (ii) the suppression of cooling. Rather than having a
thin disk, the scale height $H = |P/(dP/dz)| \simeq R$.

\begin{figure} 
\centering \includegraphics[width=3.5in]{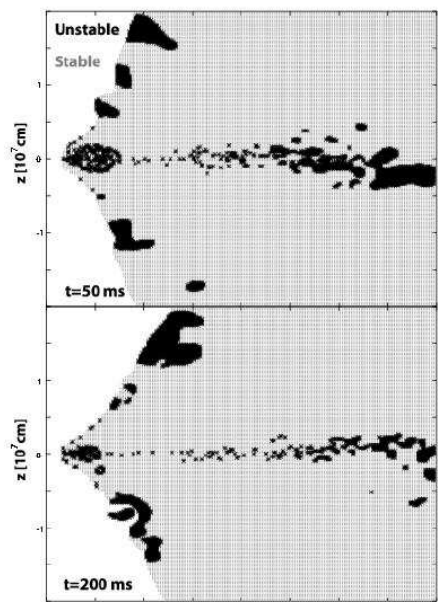} 
\caption{Convective stability of a neutrino--cooled accretion flow
around a black hole with $M_{\rm BH}\simeq 4 M_{\odot}$: The dark
(light) areas represent regions where the Solberg--Hoiland criterion
implies instability (stability). The largest part of the unstable
region clearly coincides with the optically thick region of the
flow. As the disk drains into the black hole, the disk becomes more
transparent and this region becomes smaller. The filament--like
regions of instability close to the equator are due to the
irregularities in the flow. These particular snapshots are taken from
run a2M in \cite{lee05a}, with a Shakura--Sunyaev viscosity
coefficient $\alpha=0.01$ and 0.3~M$_{\odot}$ in the disk initially.}
\label{hoiland} 
\end{figure}

We have also recently found that a convective instability is present
in neutrino cooled accretion flows at high densities (or equivalently,
accretion rates). This is analogous to what occurs above
proto--neutron stars following core collapse \cite{janka96}, and is
directly related to the fact that at $\rho \simeq 10^{11}$g~cm$^{-3}$
the fluid becomes opaque to neutrinos (i.e., the equivalent optical
depth, $\tau_{\nu}$, is of order unity). A further rise in density is
thus accompanied by an increase in entropy per baryon, $s_{b}$, and in
equilibrium electron fraction $Y_{e}$. The classical Solberg--Hoiland
requirement for convective stability can be written as two
simultaneous conditions (e.g., \cite{tassoul78}):
\begin{equation} 
\frac{1}{R^{3}} \frac{d \ell ^{2}}{dR}-\left(
\frac{\partial T}{\partial P}\right)_{s} \nabla P \cdot \nabla s > 0,
\end{equation} 
and 
\begin{equation} -\frac{1}{\rho} \frac{dP}{dz}
\left[ \frac{d\ell ^{2}}{dR} \frac{ds}{dz} - \frac{d\ell ^{2}}{dz}
\frac{ds}{dR}\right] > 0, 
\end{equation} 
where $P$ is the pressure, $T$ is the temperature and $s$ is the
specific entropy.

We find that the first of these is satisfied over most of the disk
volume in dynamical calculations, and marginally so in the inner,
opaque regions. Recall that established convection will erase the
conditions which led to its occurrence in a characteristic turnover
time $t_{\rm con}\simeq l_{\rm con}/v_{\rm con}$, unless some external
condition tries to maintain the instability, in which case (for
efficient convection) marginal stability will ensue (see e.g.,
\cite{quataert00}). The second condition is clearly {\it not}
satisfied in the optically thick regime (see Figure~\ref{hoiland}) and
drives vigorous motions continuously, as can be seen from inspection
of the corresponding velocity field (Figure~4 in
\cite{lee05a}). Although some circulations are also apparent in the
optically thin region, these are of much smaller strength, and are
essentially driven by the overshooting of fluid elements out of the
convective region and into the outer disk, and their subsequent
damping.\\

\newpage

%section 5
\section{Prolonged Engine Activity}\label{sec:prolonged}

\subsection{Motivation}\label{sec:motivation}

It is possible that the mass and accretion rate shape many of the
engine activity's visible manifestations. If we are to distinguish
those properties of the activity which can be regulated by the engine
itself and those which may varied independently, we must look for
guidance from observations of the emitted radiation. A crucial
observational development which must be noted is the discovery of late
time X-ray flaring in a large number of bursts, both long and short
\cite{burrows05,nousek06}. This has proven difficult to interpret in
terms of refreshed shocks, because of their rapid rise and decay. A
likely possibility is that it reflects renewed activity (although see
\cite{giannios06} for an alternative view) in the central engine itself
\cite{merloni01,zhang06,liang06}.

For short bursts, with durations of the order of half a second, these
flares occur tens, or even hundreds of seconds after the main
burst. There is also independent support that X-ray emission on these
time scales is detected when light curves of many bursts are stacked
\cite{lazzati01,montanari05}. If the GRB itself is powered by a
hypercritical accretion flow around a stellar mass black hole, the
dynamical time scale is only milliseconds, while the viscous time
scale, which is usually related to the burst duration, is at most a
few seconds if the effective viscosity is equivalent to $\alpha \sim
10^{-3}$. It is thus in principle a problem to account for a
resurgence of activity ten to one hundred viscous time scales (or
equivalently, up to ten thousand dynamical time scales) later.

Over the past year, a number of specific suggestions have been made
concerning the production of flares, both for long and short events -
in some case, suggesting also that their occurrence in both types of
events should indicate a common origin: interaction with a binary
companion \cite{macfadyen06,dermer06}; fragmentation of a rapidly
rotating core through non-axisymmetric instabilities \cite{king05};
magnetic regulation of the accretion flow \cite{proga06};
fragmentation of the accretion disk through gravitational
instabilities and subsequent accretion \cite{perna06}; differential
rotation in post--merger millisecond pulsars \cite{dai06,gao06} and
magnetar--like activity.

Despite all these suggestions, it remains unclear how exactly the
regulation (in the case of magnetic mechanisms) or the fragmentation
(in the case of self gravity) would come about. Magnetic regulation
was actually addressed by van Putten \& Ostriker \cite{vanputten01} as
a possible way to produce both long and short bursts from the same
central engine, depending only on the spin of the black
hole. Slow--spinning holes would be unable to halt accretion and thus
a short burst would ensue. In rapidly spinning holes, the transfer of
angular momentum through electromagnetic torques would stop or delay
accretion, giving a long event. There are tantalizing clues in
numerical simulations indicating that this might actually
occur. Krolik et al. \cite{krolik05} find a large difference in the
mass accretion rate onto the black hole in MHD simulations depending
on the black hole spin. They report a calculation with a dimensionless
Kerr parameter of $a=0.998$, in which the accretion rate is strongly
suppressed due to electromagnetic stresses, and no stationary state is
achieved through the end of the computation (although they note that
the magnetic field geometry is fundamentally different from that
suggested by van Putten \& Ostriker). This effect is absent for lower
values of the Kerr parameter. It is not clear how to extrapolate these
results to the time scales involved in SGRB flares, however, since
they occur at much later times (the calculations by Krolik et al. span
about half a second for a black hole of four solar masses).

\subsection{Regulating the Accretion Flow}

One of the basic unknowns of accretion disk theory is the physical
mechanism ultimately responsible for angular momentum transport and
energy dissipation in the disk. It is well known that classical
hydrodynamical viscosity cannot drive accretion at the rates inferred
from observations in almost every astrophysical context where
accretion disks are thought to play a crucial role. The usual way to
overcome this difficulty is to assume that transport is dominated by
some {\it anomalous} viscous phenomenon, possibly related to collective
instabilities in the disk, and to give some adhoc parameterizations for
its magnitude.

It has been recently recognized that accretion disks threaded by a
weak magnetic field are subject to magneto hydrodynamic instabilities
(see Balbus \& Hawley \cite{balbus98} and references therein), which
can induce turbulence in the disk and thereby transport angular
momentum, promoting the accretion process.  A possible alternative
source of transport in cold disks is provided by gravitational
instabilities \cite{lin87} - although the outcome strongly depends on
the thermodynamics of the disk. In particular, the fragmentation of a
gravitationally unstable disk requires that the disk be able to cool
very efficiently \cite{gammie01}, with a cooling time
\begin{equation}
t_{\rm cool} < 3\Omega^{-1} 
\end{equation}
where $\Omega$ is the local angular velocity in the
disk. Gravitational instabilities would then lead to a self-regulating
process: if the disk is initially cold, in the sense that $kT \ll
GM/R$, then gravitational instabilities would heat it up on the short
dynamical time-scale, bringing it toward stability; on the other hand,
if the disk is hot enough to begin with, radiative cooling will drive
it toward an unstable configuration. As a result of these competing
mechanisms, the {\it switch} associated with the onset of
gravitational instabilities will act as a thermostat. For
self-regulated disks, a simple relationship holds between the disk
aspect ratio $H/R$ and the mass ratio $M_{\rm disk}/M_{\rm BH}$. The
disk is in fact marginally stable when its temperature is small enough
that
\begin{equation}
M_{\rm disk} > \left({H \over R}\right) M_{\rm BH}. 
\end{equation}
Of course, this relationship can only hold in the limit where $M_{\rm
disk}/M_{\rm BH}\ll 1$.

Disk evolution calculations of neutrino cooled flows have so far
failed to show that the disk is close to instability
\cite{kohri02,lee04,lee05a,setiawan04,setiawan05}, but this eventually
may be the case when enough time has elapsed to allow the disk to
significantly cool. It is not clear how to extrapolate these results
to the time scales involved in SGRB flares, since current calculations
span at most a few seconds. It remains to be discussed under which
conditions the development of a gravitational spiral structure would
be indeed able to transport angular momentum in the disk efficiently,
hence favoring accretion, and whether this might be able to account
for flares. There is an additional concern with the the use of a
viscous formalism in self-gravitating accretion disks: Balbus \&
Papaloizou \cite{balbus99} have shown that in general the energy
transport provided by gravitational instabilities contains global
terms, associated with wave energy transport, that cannot be directly
associated with an effective viscosity. Their relative importance, and
the conditions under which they may play a significant role are as yet
poorly understood.

\subsection{Flares from Tidal Tails}\label{sec:tailflares}

\begin{figure}
\centering \includegraphics[width=4.5in]{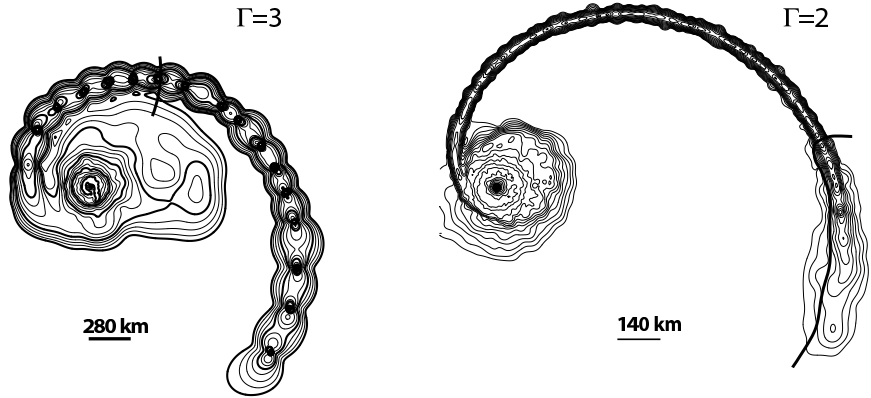}
\caption{During the merger of a black hole and a neutron star, long
one--armed tidal tails of material stripped from the star are
formed. Their structure is plotted here for two values of the
adiabatic index $\Gamma$ of the neutron star fluid, with the scale
indicated in each panel. The thick black line across the tail for
$\Gamma=2$ divides material that is bound to the central black hole
from that which has enough energy to escape the system. Note the
condensations at regular intervals in the tail for the stiff equation
of state, produced by gravitational instability.}
\label{mergertails}
\end{figure}

In this section we explore the possibility that the long tidal tails
formed during compact object mergers and/or collisions may provide the
prolonged mass inflow necessary for the production of late flares in
SGRBs. The general idea of extended emission from compact object
mergers has been advanced before \cite{barthelmy05}, relying on the
gradual disruption of the neutron star core over a time scale that is
much longer than the dynamical one \cite{davies05}. As we have argued
above, we do not believe this will actually occur, but find that the
possibility of injecting matter into the central engine at late times
remains, in a modified form.

\begin{figure}
\centering \includegraphics[width=5.5in]{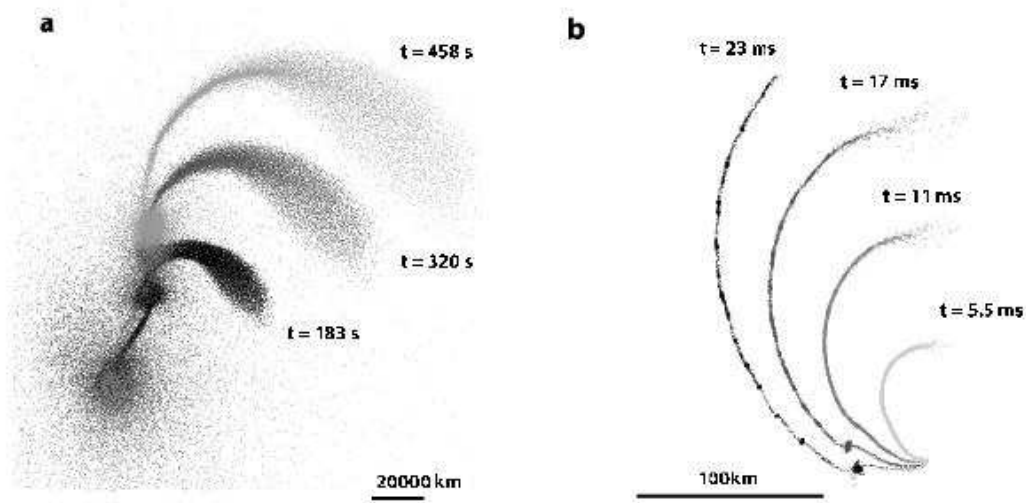}
\caption{During parabolic collisions of compact objects, similar
structures as those found in mergers occur. (a) For a white
dwarf--black hole collision, a large disk and tail form in a few
minutes. The core of the white dwarf will return to the vicinity of
the black hole in $10^3$ s. (b) The collision of a neutron star with a
black hole (the initial mass ratio is $q=0.31$), using the
pseudo-potential of Pac\'{z}y\'{n}ski \& Wiita \cite{paczynski80},
reveals the formation of a single large tail, with condensations due
to self gravity appearing at late times (see
Figure~\ref{mergertails}). Note the different spatial and temporal
scales between the two cases.}
\label{coltails}
\end{figure}

We noted already that large scale tidal tails are a common feature
formed during mergers and collisions between compact objects. These
are typically a few thousand kilometers in size by the end of the
disruption event in the case of neutron stars and black holes, and a
thousand times larger for events involving white dwarfs (see
Figures~\ref{mergertails} and \ref{coltails}). The amount of mass
contained in these structures is significant, and typically $M_{\rm
tail} \approx 0.01-0.05 M_{\odot}$. Depending on the details of the
equation of state, a small fraction of this ($10^{-3}-10^{-4}
M_{\odot}$) is actually unbound, and will escape to the surrounding
medium. The rest will eventually fall back onto the leftover disk, and
probably be accreted onto the central mass.

Once the initial dynamical interaction is over, the fluid in the tails
is practically on ballistic trajectories, moving in the potential
dominated by the central mass. It is thus possible to compute the rate
at which this matter will fall back and accrete onto the existing disk
or black hole. To compare with the simple case of a supermassive black
hole mentioned in \S~\ref{sec:colls}, we have also computed the
differential distribution of mass with energy for this fluid:
$dM/d\epsilon$. Although it appears at first glance to be roughly
constant, there are in fact significant deviations from this. The
corresponding accretion rate thus differs from the $t^{-5/3}$ law
derived by Rees \cite{rees88}, and is somewhat shallower, with
$\dot{M}_{\rm fb} \propto t^{-4/3}$. The fall back accretion rate
illustrated in Figure~\ref{mdottails} indicates that the bulk of the
matter in the tail returns to the black hole in roughly one second.

\begin{figure}
\centering \includegraphics[width=4.5in]{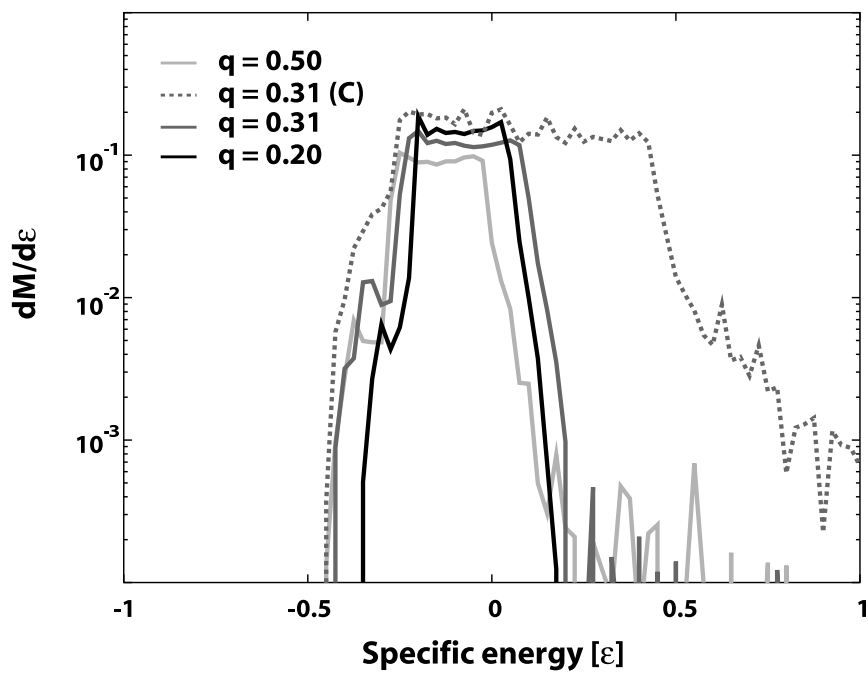}
\caption{The differential mass distribution, $dM/d\epsilon$ with
energy, computed for black hole neutron star mergers and one parabolic
collision (labeled "C"). Each curve is marked with the initial mass
ratio. The material with negative energies will fall back onto the
central mass after following essentially ballistic trajectories. }
\label{dmde}
\end{figure}

\begin{figure}
\centering \includegraphics[width=4.5in]{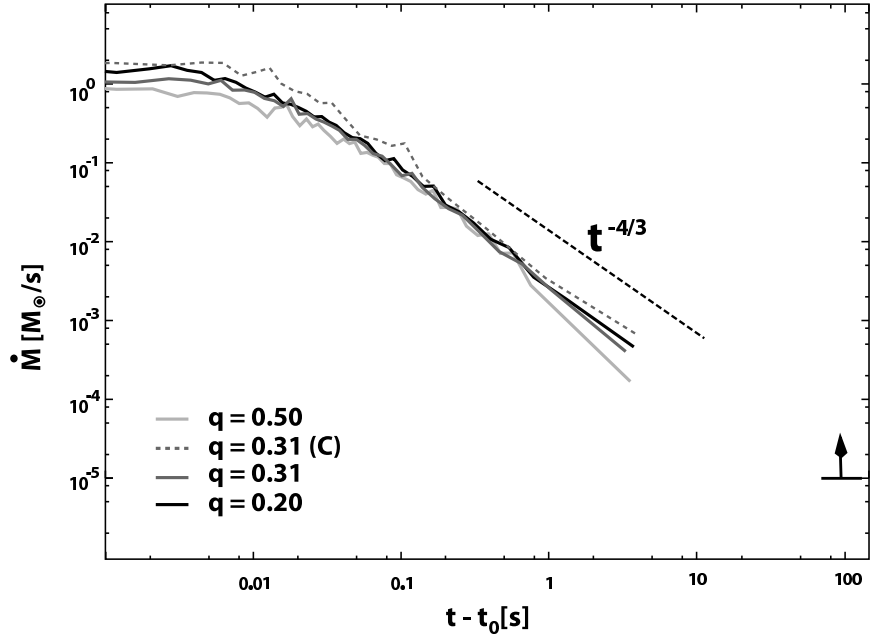}
\caption{The accretion rate onto the central region from the material
in the tails is plotted for the same calculations as shown in
Figure~\ref{dmde}. Initially roughly constant, it rapidly assumes a
power law decay, with index $\approx 4/3$. This is roughly independent
of the type of interaction (although we note that at late times
numerical noise makes an accurate determination of the slope more
difficult). The lower limit indicated at 100 seconds marks the
accretion rate required to account for the typical energy seen in a
late X--ray flare in a SGRB.}
\label{mdottails}
\end{figure}

Now, clearly one second is much too short to account for flares, which
typically occur after 30-100~s. However, the fluid does not directly
fall onto the black hole, since it has considerable angular
momentum. In fact, its circularization radius is comparable to the
size of the disk which was formed around the black hole in the first
place (typically 200-500~km). This is the point at which the gas will
settle if enough of its energy is dissipated, circularizing its
orbit. We thus have a situation in which in a short time scale,
$t_{\rm fb}\approx 1$~s, a mass $M_{\rm tail}\approx 0.01 M_{\odot}$
returns to a radius $R_{\rm circ}\approx 5 \times 10^{7}$~cm. If
dissipation occurs (e.g., through shocks with what is left of the
initial disk), a new ring of matter may form at $R \approx R_{\rm
circ}$. The evolution of this newly injected mass will depend on the
previous history of the original accretion disk, and on how much of it
is left after the fall back time $t_{\rm fb}$. Evidently the component
with the most energy will dominate the behavior of the system, and if
$M_{\rm disk} \gg M_{\rm tail}$ it is hard to see how the infalling
tail could produce a substantial alteration of the overall flow
properties and an accompanying observable signal. If the reverse is
true and $M_{\rm tail} \gg M_{\rm disk}$, however, perhaps the inner
accretion disk and the true accretion rate onto the black hole can be
modified and a secondary episode of energy release be provoked. If we
naively estimate the viscous time scale for the newly formed ring at
$R_{\rm circ}$ as
\begin{equation}
t_{\rm visc} \approx R_{\rm circ}^{2}/10 \nu, 
\end{equation}
where
\begin{equation}
\nu=\alpha c_{s}^{2}/\Omega_{\rm Kep} 
\end{equation}
is the viscosity coefficient, $c_{s}$ is the gas sound speed and
$\Omega_{\rm Kep}$ is the Keplerian angular frequency, then typical
values from the calculations indicate that
\begin{equation}
t_{\rm visc} = 100 \left( \frac{M_{\rm BH}}{4 M_{\odot}} \right)^{1/2}
\left(\frac{R_{\rm circ}}{2 \times 10^{7}{\rm cm}} \right)^{1/2}
\left( \frac{\alpha}{10^{-2}} \right)^{-1} \left(
\frac{c_{s}}{10^{8}\;\mbox{cm~s$^{-1}$}} \right)^{-2} \, \mbox{s}.
\end{equation} 

Viscous time scales of $10-10^3$ seconds are thus in principle
plausible for $10^{-3} \leq \alpha \leq 10^{-1}$, and are much longer
than the injection interval itself. This is akin to the instantaneous
injection of a discrete amount of matter at a given radius, and will
produce an accretion episode with a delay proportional to its
duration, $t_{\rm delay} \propto t_{\rm acc}$ (e.g., \cite{frank92}),
naturally accounting for this observational fact in SGRB flares. 

An additional constraint comes from consideration of the total energy
release in the observed flares. Taking neutrino emission as an
example, the accretion efficiency is
\begin{equation}
\epsilon_{\rm acc} =L_{\nu}/\dot{M}_{\rm BH}c^{2} \approx 0.1, 
\end{equation} 
based on our own calculations \cite{lee04,lee05a} and those reported
by other groups \cite{setiawan04,setiawan05}, where $\dot{M}_{\rm BH}$
is the actual accretion rate that feeds the central black hole, and is
reduced from the rate at which the disk itself is fed by a factor
$\zeta \approx 1/2$, the remainder going into outflows (see also
\cite{narayan01}).  The accretion efficiency is predominantly
dependent upon the flow being optically thin to its own emission, and
can thus be extrapolated down to the average accretion rate produced
by the fall back material, 
\begin{equation}
\dot{M}_{\rm fb} \approx M_{\rm tail}/t_{\rm visc}= 2 \times
10^{-4}M_{\odot}~{\rm s}^{-1}.
\end{equation} 
Thus the neutrino luminosity would be 
\begin{equation}
L_{\nu} \approx \epsilon_{\rm acc} \dot{M}_{\rm BH} c^{2}= 3 \times
10^{49}\;{\rm erg~s}^{-1}.
\end{equation} 
Accounting for a total fluence $L_{\rm flare} \approx 3 \times
10^{46}$erg~s$^{-1}$ and a total energy release $E_{\rm flare} \approx
3 \times 10^{48}$~erg over $t_{\rm flare}\approx 100$~s (typical
numbers for observed events) would require a $\nu \overline{\nu}$
annihilation efficiency of 10$^{-3}$. This is close to being
unrealistically high at this luminosity level, and would thus argue
against neutrinos as an ultimate source for the flaring behavior. On
the other hand, magnetic energy extraction could presumably also
operate, since the injection can deposit enough mass in the disk to
anchor sufficiently strong magnetic fields. It is hard to see,
however, how either option could account for extremely delayed flares
occurring after $10^3$ seconds, but perhaps those can be due to
refreshed shocks \cite{panaitescu06}.

\subsection{Instabilities and Fragmentation in Fluid Tails}\label{sec:knots}

The hydrodynamic stability of the ejected tail is a question that
deserves special comment. The fluid moves globally and to a good
approximation in the potential of the central object. It is,
nevertheless, prone to an instability due to self-gravity, which may
affect its structure on small scales. This is the {\it varicose} or
{\it sausage} instability, and it can be shown (see Ch. XII in
ref.~\cite{chandra61}) that for a cylinder of {\it incompressible}
fluid, gravitational instability sets in for perturbations with
wavelength 
\begin{equation}
\lambda \ge \lambda^{*} =2 \pi R_{\rm cyl}/x^{*}, 
\end{equation}
where $R_{\rm cyl}$ is the radius of the cylinder and $x^{*}\approx
1$. The fastest growing mode, which determines the size of the
fragments upon manifestation of this effect has $x=0.58$,
corresponding to a wavelength $\lambda = 10.8 R_{\rm cyl}$, and a
growth time 
\begin{equation}
\tau = {4 \over \sqrt{4 \pi G \rho}} 
\end{equation}
Binary merger
calculations do not employ incompressible fluids, but for high enough
adiabatic indices one can see that this behavior does indeed occur.

The left panel in Figure~\ref{mergertails} plots the large scale
structure formed during a black hole-neutron star merger in a
simulation with $\Gamma=3$. The knots in the tail are spaced at
regular intervals, with wavelength $\lambda \approx 2.5 \times 10^{7}
\; {\rm cm} \approx 10 R_{\rm tail}$, as predicted by the
incompressible analysis. The estimated time scale for their formation
is $\tau \approx 20$~ms, in good agreement with what is seen in the
calculations.  Binary neutron star mergers with stiff polytropic
pressure--density relations performed by Rasio et al. \cite{rasio94}
show exactly the same structures. Once fragmentation occurs, the
individual clumps continue to move in the overall gravitational
potential, and those that are bound to the central object will return
to its vicinity at discrete intervals to inject matter. For lower
values of the adiabatic index this behavior is no longer
seen. Instead, the fluid expands smoothly over large volumes, and the
tails are less well defined. At the densities encountered in these
structures, one would not expect a compressibility low enough to allow
this instability to operate, and indeed it is not seen in merger
simulations using realistic equations of state
\cite{rosswog05}. Nevertheless, it is worthy to consider their
possible formation and subsequent evolution as a general feature. Note
that if one substitutes the neutron star for a quark (strange) star,
such droplets are a generic feature upon tidal disruption
\cite{lee01b}.\\

\newpage

%Section 6
\section{Summary and Future Prospects}\label{sec:discussion}
In this final section we present a short summary of what we have
learned so far about the physics of SGRB sources.  Though the field is
far from being mature, sufficient progress has been made in
identifying some of the essential ingredients. We also describe the
observational and theoretical prospects for the near future.

\subsection{Summary}\label{sec:summary}
The progress in our understanding of SGRBs has been a story of
consolidation and integration, and there is every indication that this
progression will continue. Until recently, SGRBs were known
predominantly as bursts of $\gamma$-rays, largely devoid of any
observable traces at any other wavelengths.  However, a striking
development in the last several months has been the measurement and
localization of fading X-ray signals from several SGRBs, making
possible the optical and radio detection of afterglows, which in turn
enabled the identification of host galaxies at cosmological
distances. The presence in old stellar populations e.g., of an
elliptical galaxy for GRB 050724, rules out a source uniquely
associated with recent star formation. In addition, no bright
supernova is observed to accompany SGRBs, in distinction from most
nearby long-duration GRBs. It is now clear that short and long events
are not drawn from the same parent stellar population. Even with a
handful of SGRBs detected to date, it has become apparent that they
are far from standard, both in their energetics and cosmological
niche. This hints at the underlying possibility that the progenitor
itself may be quite different from burst to burst, and not entirely
restricted to the most discussed scenario involving the merger of
compact binaries such as the Hulse-Taylor pulsar (although see
\cite{rosswogrr03}) .

The most fundamental problem posed by SGRB sources is how to generate
over $ 10^{50}$ erg in the burst nucleus and channel it into
collimated plasma jets. The cumulative evidence insistently suggests
that the more powerful SGRBs must have ``processed'' upwards of
$10^{-3}M_\odot$ through a compact entity - only this hypothesis
accounts for the high luminosity and compactness inferred from
$\gamma$-ray variability. We believe that accretion onto a compact
object, be it a neutron star or a stellar mass black hole, offer the
best hope of understanding the ``prime mover'' in all types of SGRB
sources although a possible attractive exemption includes a rapidly
spinning neutron star with a powerful magnetic field. Consequently,
most theoretical work has been directed towards describing the
possible formation channels for these systems, and evaluate those
which are likely to produce a viable central engine. The only way to
cool the resulting hypercritical accretion flow (other than through
direct advective transport of the energy into a gravitational sink
hole on a dynamical time scale) is by neutrino emission, circumventing
the classical Eddington limit for photons and allowing for the
conversion of gravitational binding energy into outflowing
radiation. Consideration of the associated energy release thus leads
us down the path of detailed thermodynamical and microphysical
processes unlike those encountered in most areas of astrophysics,
except supernovae, where the physical conditions are quite similar in
terms of density, entropy and internal energy.

As of this writing, it is fairly clear from the concerted efforts of
many groups working on different aspects of these problems that there
is in principle no problem in accounting for the global energy budget
of a typical SGRB from the class of systems here considered. The devil
is in the details, of course, and the actual modes of energy
extraction have yet to be worked out carefully. It would appear,
however, that neutrino emission is more confined to be a competitive
energy source in the early stages of the dynamical evolution, while
magnetically powered events may be able to offer longer staying power.

Still, various alternative ways of triggering the explosions
responsible for SGRBs remain: NS-NS, NS-BH, BH-WD or WD-WD binary
mergers, recycled magnetars, spun-down supra-massive NS and accretion
induced collapse of a NS. Can we decide between the various
alternatives? The progenitors of SGRBs are essentially masked by
afterglow emission, largely featureless synchrotron light, which
reveals little more than the basic energetics and micro physical
parameters of relativistic shocks. In the absence of a supernova-like
feature, the interaction of burst ejecta with a stellar binary
companion \cite{macfadyen06} or with its emitted radiation
\cite{ramirez-ruiz04} may be the only observable signature in the
foreseeable future shedding light on the identity of the progenitor
system. A definitive understanding will, however, come with the
observations of concurrent gravitational radiation or neutrino signals
arising from the dense, opaque central engine.

Our understanding of SGRBs has come a long way since their discovery
almost forty years ago, but these enigmatic sources continue to offer
major puzzles and challenges. SGRBs provide us with an exciting
opportunity to study new regimes of physics. As we have described, our
rationalization of the principal physical considerations combines some
generally accepted features with some more speculative and
controversial ingredients. When confronted with observations, it seems
to accommodate their gross features but fails to provide us with a
fully predictive theory. What is more valuable, though considerably
harder to achieve, is to refine models like the ones advocated here to
the point of making quantitative predictions, and to assemble, assess
and interpret observations so as to constrain and refute these
theories.  What we can hope of our present understanding is that it
will assist us in this endeavour.

\subsection{Observational Prospects}\label{sec:obspros}
High energy astrophysics is a young field.  It owns much of the
remaining unexplored ``discovery space'' in contemporary astronomy.
Two examples of this discovery space are at the extremes of
observation of the electromagnetic spectrum.  At the high end, there
are already a few tens of TeV sources, while at the low end of $\leq
50$~MHz radio astronomy there are essentially no sources.  Neutrino
astronomy claims only two cosmic sources so far, the sun and
SN1987a. Finally, as many of the most interesting high energy sources
are ultimately black holes and neutron stars, the exciting field of
gravitational wave astronomy --- perhaps a decade away from birth ---
is inextricably linked to high energy astrophysics. These are the next
frontiers and we consider the role of SGRBs within them in turn.

Atmospheric Cerenkov techniques are being used to detect $\gamma$-rays
in the GeV--TeV range. These are important as both sources and as
probes.  Persistent sources have been identified with pulsars, blazars
and supernova remnants and in each case are likely to provide the best
approach we have to understanding the fundamental nature of these
sources. A tentative $\geq 0.1$ TeV detection of a long GRB has been
reported with the water Cherenkov detector Milagrito \cite{atkins00}.
Here the big question is to determine whether the SGRB jets comprise
ultrarelativistic protons, that interact with either the radiation
field or the background plasma, or if they are $\e^\pm$ pairs.  TeV
sources should be far more plentiful in the latter case.  The
combination of GLAST and telescopes like HESS and VERITAS ought to be
able to sort this out.

The same shocks which are thought to accelerate the electrons
responsible for the non-thermal $\gamma$-rays in SGRBs should also
accelerate protons \cite{waxman04,waxman06,dermer05}.  Both the
internal and the external reverse shocks are mildly relativistic, and
are expected to lead to relativistic protons \cite{waxman04apj}. The
maximum proton energies achievable in SGRB shocks are $E_p\sim
10^{20}$ eV, comparable to the highest energies measured with large
cosmic ray ground arrays \cite{hayashida99}. For this, the
acceleration time must be shorter than both the radiation or adiabatic
loss time and the escape time from the acceleration region
\cite{waxman95}. The accelerated protons can interact with the
fireball photons, leading to charged pions, muons and neutrinos.  For
internal shocks producing observed 1 MeV photons this implies $\geq
10^{16}$ eV protons, and neutrinos with $\sim 5\%$ of that energy,
$\epsilon_\nu\geq 10^{14}$ eV \cite{waxmanbahcall97}. Another copious
source of target photons in the UV is the afterglow reverse shock, for
which the resonance condition requires higher energy protons leading
to neutrinos of $10^{17}-10^{19}$ eV \cite{waxmanbahcall99}. Whereas
photon-pion interactions lead to higher energy neutrinos and provide a
direct probe of the shock proton acceleration as well as of the photon
density, inelastic proton-neutron collisions may occur even in the
absence of shocks, leading to charged pions and neutrinos
\cite{derishev99} with lower energies than those from photon-pion
interactions. The typical neutrino energies are in the $\sim$ 1-10 GeV
range, which could be detectable in coincidence with observed
SGRBs. This is the province of projects like AMANDA, IceCube and
ANTARES. Success in the former will suggest that ultrarelativistic
outflows comprise mainly protons.  Neutrino astronomy has the
advantage that we can see the universe up to $\sim$~EeV energies. By
contrast, the universe becomes opaque to $\gamma$-rays above $\sim$
TeV energies through absorption by the infrared background.

Finally the last and most challenging frontier is that of
gravitational radiation, which is largely unknown territory. A
time-integrated luminosity of the order of a solar rest mass ($\sim
10^{54}$ erg) is predicted from merging NS-NS and NS-BH models, while
that from collapsar models is less certain, but estimated to be
lower. Ground-based facilities, like LIGO, TAMA and VIRGO, will be
seeking such stellar sources. The observation the associated
gravitational waves would be facilitated if the mergers involve
observed SGRB sources; and conversely, it may be possible to
strengthen the case for (or against) NS-NS or NS-BH progenitors of
SGRBs if gravitational waves were detected (or not) in coincidence
with some bursts. The technical challenge of achieving the
sensitivities necessary to measure waves from assured sources should
not be understated. It may well take more than another decade to reach
them.  Neither, however, should the potential rewards. Gravitational
waves offer the possibility of observing in an entirely {\em new}
spectrum, not merely another window in the same (electromagnetic)
spectrum. Furthermore, they will by their very nature tell us about
events where large quantitites of mass move in such small regions that
they are utterly opaque and forever hidden from direct electromagnetic
probing, and are the only way (except perhaps for neutrinos) through
which we can learn about them. There have been few regions of the
electromagnetic spectrum where observation conformed to previous
expectation. So it would be indeed remarkable if gravitational wave
astronomy, or any of the other frontiers, turned out to be as we have
described. Indeed, if past experience is any guide, they will surely
provide us with new surprises, challenges and potential loose threads
through which we can unravel another piece of the fabric of the
universe.

\subsection{Theoretical Prospects}\label{sec:thep}

Although some of the features now observed in GRB sources (especially
afterglows) were anticipated by theoretical discussions, the recent
burst of observational discovery has left theory lagging behind. There
are, however, some topics on which we do believe that there will be
steady work of direct relevance to interpreting observations.

One of the most important is the development and use of hydrodynamical
codes for numerical simulation of SGRB sources with detailed physics
input. Existing two and three dimensional codes have already uncovered
some gas-dynamical properties of relativistic flows unanticipated by
analytical models (e.g., \cite{mckinney06}), but there are some key
questions that they cannot yet address. In particular, higher
resolution is needed because even a tiny mass fraction of baryons
loading down the outflow severely limits the maximum attainable
Lorentz factor. We must wait for useful and affordable three
dimensional simulations before we can understand the nonlinear
development of instabilities. Well-resolved three dimensional
simulations are becoming increasingly common and they rarely fail to
surprise us. The symmetry-breaking involved in transitioning from two
to three dimensions is crucial and can lead to qualitatively new
phenomena. A particularly important aspect of this would be to link in
a self-consistent manner the flow within the accretion disk to that in
the outflowing gas, allowing for feedback between the two
components. The key to using simulations productively is to isolate
questions that can realistically be addressed and where we do not know
what the outcome will be, and then to analyse the simulations so that
we can learn what is the correct way to think about the problem and to
describe it in terms of elementary principles. Simulations in which
the input physics is so circumscribed that they merely illustrate
existing prejudice are of limited value!

A second subject ready for a more sophisticated treatment is related
to the intensity and shape of the intrinsic spectrum of the emitted
radiation.  Few would dispute the statement that the photons which
bring us all our information about the nature of GRBs are the result
of particle acceleration in relativistic shocks (e.g.,
\cite{granot06}).  Since charged particles radiate only when
accelerated, one must attempt to deduce from the spectrum {\it how}
the particles are being accelerated, {\it why} they are being
accelerated, and to identify the macroscopic source driving the
microphysical acceleration process.

Collisionless shocks are among the main agents for accelerating ions
as well as electrons to high energies whenever sufficient time is
available (e.g., \cite{blandford87,achterberg01}). Particles reflected
from the shock and from scattering centres behind it in the turbulent
compressed region have a good chance of experiencing multiple
scattering and acceleration by first-order Fermi acceleration when
coming back across the shock into the turbulent upstream
region. Second-order or stochastic Fermi acceleration in the broadband
turbulence downstream of collisionless shocks will also contribute to
acceleration. In addition, ions may be trapped at perpendicular
shocks. The trapping is a consequence of the shock and the Lorentz
force exerted on the particle by the magnetic and electric fields in
the upstream region. With each reflection at the shock the particles
gyrate parallel to the motional electric field, picking up energy and
surfing along the shock surface. All these mechanisms are still under
investigation, but there is evidence that shocks play a most important
role in the acceleration of cosmic rays and other particles to very
high energies.

There is no in situ information available from astrophysical plasmas.
So one is forced to refer to indirect methods and analogies with
accessible plasmas, found only in near-Earth space.  Actually, most of
the ideas about and models of the behaviour of astrophysical plasmas
have been borrowed from space physics and have been refitted to
astrophysical scales. However, the large spatial and long temporal
scales in astrophysics and astrophysical observations do not allow for
the resolution of the collisionless state of the plasmas. For
instance, in the solar wind the collisional mean free path is of the
order of a few AU. Looked at from the outside, the heliosphere, the
region which is affected by the solar wind, will thus be considered
collision dominated over time scales longer than a typical propagation
time from the Sun to Jupiter. On any smaller and shorter scales this
is wrong, because collisionless processes govern the solar wind
here. Similar arguments apply to stellar winds, molecular clouds,
pulsar magnetospheres and the hot gas in clusters of galaxies. One
should thus be aware of the mere fact of a lack of small-scale
observations. Collisionless processes generate anomalous transport
coefficients. This helps in deriving a more macroscopic
description. However, small-scale genuinely collisionless processes
are thereby hidden. This implies that it will be difficult, if not
impossible, to infer anything about the real structure, for instance,
of collisionless astrophysical shock waves.  The reader is refer to
\cite{baumjohann96}, and \cite{treumann97} for an excellent
presentation of the basic kinetic collisionless (space) plasma theory.

The most interesting problem remains, however, in the nature of the
central engine and the means of extracting power in a useful
collimated form. In all observed cases of relativistic jets, the
central object is compact, either a neutron star or black hole, and is
accreting matter and angular momentum. In addition, in most systems
there is direct or indirect evidence that magnetic fields are present
-- detected in the synchrotron radiation in galactic and extragalactic
radio sources or inferred in collapsing supernova cores from the
association of remnants with radio pulsars. This combination of
magnetic field and rotation may be very relevant to the production of
relativistic jets (e.g., \cite{blan02}). Much of what we have
summarized in this respect is conjecture and revolves largely around
different prejudices as to how three dimensional flows behave in
strong gravitational fields. There are serious issues of theory that
need to be settled independently of the guidance we obtain from
observations of astrophysical black holes.\\

\ack We are indebted to a considerable number of colleagues for their
help and advice.  We acknowledge, in particular, the contributions of
M. Aloy, E. Berger, K. Belczy{\'n}ski, L. Bildsten, D. Burrows,
D. Fox, C. Fryer, J. Fynbo, N. Gehrels, D. Guetta, J. Hjorth, P. Hut,
Th. Janka, C. Kouveliotou, S. Kulkarni, E. Nakar, P. M\'{e}sz\'{a}ros,
T. Piran, D. Pooley, M. Prakash, F. Rasio, M.Ruffert, A. Soderberg,
V. Usov, G. van de Ven, D. Watson, E. Waxman, S. Woosley and
Z. Zheng. Our views on the topics discussed here have been clarified
through discussions with J. Bloom, J. Granot, W. Klu\'{z}niak,
A. MacFadyen, D. Page, X. Prochaska, M. Rees, A. Socrates and
especially S. Rosswog. We are particularly grateful to Y. Kaneko for
providing the time histories and spectra of BATSE SGRBs. Major
portions of this review were written at DARK Cosmology Centre,
Copenhagen; Institute for Advanced Study, Princeton; Institute of
Astronomy, Cambridge; Instituto de Astronom\'{\i}a, M\'{e}xico. We
thank the directors of these institutions for their generous
hospitality. Financial support for this work was provided in part by
NSF (PHY0503584), CONACyT (36632E,45845E) and PAPIIT
(110600,119203).\\

\end{document}